\documentclass[longauth]{aa}  
\usepackage[english]{babel}
\usepackage{multirow}
\usepackage{xcolor}
\usepackage{graphicx}
\usepackage{natbib} 
\usepackage{gensymb}
\usepackage{ftnxtra}
\usepackage{fnpos}
\usepackage{txfonts}
\begin{document}

   \title{The CARMENES search for exoplanets around M dwarfs}

   \subtitle{Detection of a mini-Neptune around LSPM~J2116+0234 and refinement of orbital parameters of a super-Earth around GJ~686 (BD+18 3421)}

   \author{S.~Lalitha \inst{1}\fnmsep\thanks{S. Lalitha and D. Baroch contributed equally to this work},
          D.~Baroch \inst{2,3}\fnmsep\footnotemark[1],
          J.~C.~Morales \inst{2,3},
          V.~M.~Passegger \inst{4},
          F.~F.~Bauer \inst{5},  
          C.~Cardona~Guill\'en \inst{6,7}, 
          S.~Dreizler \inst{1},
          M.~Oshagh \inst{1},
          A.~Reiners \inst{1},
          I.~Ribas \inst{2,3},
         J.~A.~Caballero\inst{8},
         A.~Quirrenbach \inst{9},
         P.~J. Amado \inst{5},
          V.~J.~S.~B\'ejar \inst{6,7},
          J.~Colom\'e \inst{2,3},
          M.~Cort\'es-Contreras \inst{8,10},
          D.~Galad\'i-Enr\'iquez \inst{11},
          L. Gonz\'alez-Cuesta \inst{6,7},
          E.~W.~Guenther \inst{12},
          H.-J.~Hagen \inst{4},
          T.~Henning \inst{13},
          E.~Herrero \inst{2,3},
          T.-O.~Husser \inst{1}, 
          S.~V.~Jeffers \inst{1},
          A.~Kaminski \inst{9},
          M.~K\"urster \inst{13},
          M.~Lafarga \inst{2,3},
          N.~Lodieu \inst{6,7},
          M.J.~L\'opez-Gonz\'alez\inst{5},
          D.~Montes \inst{14},
          M.~Perger \inst{2,3},
          A.~Rosich \inst{2,3},
          E.~Rodr\'iguez \inst{5}, 
          C.~Rodr\'iguez-L\'opez\inst{5},
          J.~H.~M.~M.~Schmitt \inst{4},
          L.~Tal-Or \inst{15,1},
          M.~Zechmeister \inst{1}
           }

   \institute{
              Institut f\"ur Astrophysik, Georg-August-Universit\"at,
              Friedrich-Hund-Platz 1, D-37077 G\"ottingen, Germany\\
              \email{lalitha.sairam@uni-goettingen.de}
         \and
              Institut de Ci\`encies de l'Espai (ICE, CSIC),
              Campus UAB, C/ de Can Magrans s/n, E-08193 Bellaterra (Cerdanyola del Vall\`es), Barcelona, Spain\\
              \email{baroch@ice.cat}
         \and
              Institut d'Estudis Espacials de Catalunya (IEEC),
              C/ Gran Capit\`a 2-4, E-08034 Barcelona, Spain
          \and
              Hamburger Sternwarte,
              Gojenbergsweg 112, D-21029 Hamburg, Germany
         \and
              Instituto de Astrof\'isica de Andaluc\'ia (IAA-CSIC),
              Glorieta de la Astronom\'ia s/n, E-18008 Granada, Spain
         \and
              Instituto de Astrof\'isica de Canarias,
              V\'ia L\'actea s/n, E-38205 La Laguna, Tenerife, Spain
         \and
              Departamento de Astrof\'isica, Universidad de La Laguna,
              E-38026 La Laguna, Tenerife, Spain
         \and
              Centro de Astrobiolog\'ia (CSIC-INTA), ESAC,
              Camino Bajo del Castillo s/n, E-28692 Villanueva de la Ca\~nada, Madrid, Spain
         \and
              Landessternwarte, Zentrum f\"ur Astronomie der Universt\"at Heidelberg,
              K\"onigstuhl 12, D-69117 Heidelberg, Germany
         \and
              Spanish Virtual Observatory (SVO)
         \and
              Centro Astron\'onomico Hispano Alem\'an, Observatorio de Calar Alto, Sierra de los Filabres, E-04550 Gergal, Spain
         \and
              Th\"uringer Landesstenwarte Tautenburg,
              Sternwarte 5, D-07778 Tautenburg, Germany
         \and
              Max-Planck-Institut f\"ur Astronomie,
              K\"onigstuhl 17, D-69117 Heidelberg, Germany
         \and
              Departamento de F\'{i}sica de la Tierra y Astrof\'{i}sica and IPARCOS-UCM (Intituto de F\'{i}sica de Part\'{i}culas y del Cosmos de la UCM), Facultad de Ciencias F\'{i}sicas, Universidad Complutense de Madrid, E-28040, Madrid, Spain
         \and
              Department of Geophyics, Raymond and Beverly Sackler Faculty of Exact Sciences, Tel Aviv University, Tel Aviv, IL-6997801, Israel
             }
             
\titlerunning{The CARMENES search for exoplanets around M dwarfs.}

\authorrunning{Lalitha et al.}

\date{Received ; accepted }

 
  \abstract{Although M dwarfs are known for high levels of stellar activity, they are ideal targets for the search of low-mass exoplanets with the radial velocity method. We report the discovery of a planetary-mass companion around LSPM~J2116+0234 (M3.0\,V) and confirm the existence of a planet orbiting  GJ~686 (BD+18~3421; M1.0\,V). The discovery of the planet around LSPM~J2116+0234 is based on CARMENES radial velocity observations in the visual and near-infrared channels. We confirm the planet orbiting around GJ~686 by analyzing the radial velocity data spanning over two decades of observations from CARMENES VIS, HARPS-N, HARPS, and HIRES. 
  We find planetary signals  at 14.44 and 15.53\,d  in the radial velocity data for LSPM~J2116+0234 and GJ~686, respectively. 
Additionally, the radial velocity, and photometric time series, as well as various spectroscopic indicators, show hints of variations of 42\,d for LSPM~J2116+0234 and  37\,d for  GJ~686, which we attribute to the stellar rotation periods. 
The orbital parameters of the planets are modeled with Keplerian fits together with correlated noise from the stellar activity. A mini-Neptune with a minimum mass of 11.8\,M$_{\oplus}$ orbits LSPM~J2116+0234  
producing a radial velocity semi-amplitude of 6.19\,m\,s$^{-1}$, while a super-Earth of mass 6.6\,M$_{\oplus}$ orbits GJ~686 
and produces a radial velocity semi-amplitude of 3.0\,m\,s$^{-1}$. 
  Both LSPM~J2116+0234 and GJ~686 have planetary companions populating the regime of exoplanets with masses lower than 15\,M$_{\oplus}$ and orbital periods <20\,d. }


   \maketitle
%
\section{Introduction}
Nearly $75\%$ of the stellar population of our galaxy consists of M-type stars, making them the most common potential planetary hosts \citep{henry_2006, Dressing_2015, Gaidos_2016}. Since the discovery of the first exoplanet around a main-sequence star in 1995 \citep{Mayor1995}, an important goal has been the discovery and characterization of terrestrial planets located inside the habitable zone. M-dwarf stars are in the focus of ongoing surveys for habitable planets for two main reasons. Firstly, the induced radial velocity (RV) amplitude is inversely proportional to the mass of the star \citep{newton1687}, increasing the detection probability of lower-mass planets around them \citep{ Marcy_1998, Udry_2007, Bonfils_2013}. Secondly, due to the low luminosities of M dwarfs, their habitable zones are located closer to the host star with relatively shorter orbital periods.

The downside of surveying M-dwarf stars with high-resolution spectrographs is their high levels of magnetic activity. Large inhomogeneities such as dark spots and bright faculae are produced on the stellar surface due to the activity which in turn affects the spectral line profile and induces a Doppler-shift in the spectrum \citep{vogt1983}. Consequently, the stellar activity can induce large-amplitude RV variations which may have periodicities close to the stellar rotation period, and therefore, they can be misinterpreted as planetary signals. Several techniques have been developed over the past years to disentangle activity-induced variations and planetary signals. Some of these techniques are the study of correlations between activity indicators and RVs \citep[see e.g.][]{Queloz_2001,Boisse_2011,Oshagh2017,zechmeister_2018}, the selection of individual spectral lines less affected by activity \citep{Dumusque_2018}, and the use of Bayesian statistical models such as Gaussian Processes \citep{haywood2014,Rajpaul_2015,faria_2016,Jones_2017}.
 
 The spot- and facula-induced RV amplitude generally tends to decrease toward longer wavelengths \citep{Desort_2007}. This is a consequence of the lower temperature-contrast between heterogeneities and the quiet surface at longer wavelengths \citep{barnes_2011, jeffers_2014}. 
However, in stars with a strong magnetic field the relative importance of the Zeeman effect increases with wavelength \citep{Reiners_2013}.
In addition, M dwarfs emit the bulk of their spectral energy at wavelengths redward of 1\,$\mu$m 
(\citealt{Reiners_2010}, and refernces therein). Hence, in theory, observations at wavelengths around 700--900\,nm are ideal for both reducing the effect of stellar activity on RVs and minimizing the exposure time when surveying M dwarfs  \citep{Reiners2018b}.

\begin{table}
\centering
\caption{Basic properties of the host stars.}
\label{tab:tab1}
\begin{tabular}{lcccc} 
\hline\hline
\noalign{\smallskip}
Parameters & LSPM~J2116+0234 & GJ~686   & Ref.\\
\noalign{\smallskip}
\hline
\noalign{\smallskip}
Karmn\tablefootmark{a} & J21164+025 & J17378+185 &   \\
$\alpha$ (J2000) & 21:16:27.28& 17:37:53.35 & {\it Gaia}\\
$\delta$ (J2000) &+02:34:51.40&+18:35:30.16&{\it Gaia}\\
$d$ [pc]&17.64$\pm$0.02&8.16$\pm$2$\times10^{-3}$&{\it Gaia}\\
$G$ [mag] & 10.86$\pm$8$\times10^{-4}$&8.74$\pm$ 6$\times10^{-4}$&{\it Gaia}\\
$J$ [mag]&8.219$\pm$0.032&6.360$\pm$0.023&2MASS\\
Sp. Type & M3.0\,V&M1.0\,V&PMSU,\\
& & &L\'ep13\\
$J$ [mag]&8.219$\pm$0.032&6.360$\pm$0.023&2MASS\\
$T_{\rm eff}$ [K]&3475$\pm51$&3654$\pm$51&Pas18\\
$\log{g}$ [cgs]&4.95$\pm$0.07&4.88$\pm$0.07 &Pas18\\

[Fe/H] [dex]&-0.05$\pm$0.16&-0.22$\pm$0.16&Pas18\\
$L$ [$L_{\odot}$]&$0.02\pm3.2\times10^{-4}$&0.03$\pm$7.3$\times10^{-4}$&Sch19\\
$R$ [$R_{\odot}$]&0.431$\pm$0.015&0.427$\pm$0.017&Sch19\\
$M_{\star}$ [$M_{\odot}$]&0.430$\pm$0.031&0.426$\pm$0.033&Sch19\\
pEW (H$\alpha$) [$\AA$] & +0.004$\pm$0.005&-0.128$\pm$0.03&Jef18\\
$v\sin{i}$ [km\,s$^{-1}$] &$<$2&<2&Rei18\\
$\log R'_{HK}$&$\cdots$ &$-5.42\pm0.05$ &  Sua18\\
U [km\,s$^{-1}$] & -23.99$\pm$0.21 & -33.56$\pm$0.28 & {\it Gaia} \\
V [km\,s$^{-1}$] & -18.12$\pm$0.29 & 35.40$\pm$0.25 & {\it Gaia} \\
W [km\,s$^{-1}$] & -5.24$\pm$0.21 & -21.20$\pm$0.17 & {\it Gaia} \\
\noalign{\smallskip}
\hline
\end{tabular}

\footnotesize{{\bf References:} 2MASS: \cite{skrutskie_2006}; Cor16: \cite{Cortes2016}; {\it Gaia}: \cite{Gaia2016,gaia_2018}; Jef18: \cite{jeffers_2018}; Rei18:\cite{Reiners2018b}; PMSU: \cite{hawley_1996}; L\'ep13: \cite{Lepine2013}; Pas18: \cite{passegger_2018}; Sch19: \cite{schweitzer_2019}; Sua18: \cite{Suarez_2018}; UCAC4: \cite{zacharias_2013}. {\bf Notes.} $^{(a)}$ CARMENES identifier.
}
\end{table}


The high-resolution spectrograph CARMENES installed at the 3.5\,m telescope at the Calar Alto Observatory (Almer\'ia, Spain) is specifically designed to cover a wide wavelength range \citep{Quirrenbach_2016, Quirrenbach2018}. CARMENES extends further into the near-infrared (NIR) than most high-precision spectrographs to cover the range where M dwarfs emit the bulk of their spectral energy.
The instrument consists of two cross-dispersed \'echelle spectrographs covering visible (VIS) wavelengths (0.52--0.96\,$\mu$m, R$\sim$94\,600) and NIR wavelengths (0.96--1.71\,$\mu$m, R$\sim$80\,400) \citep{Quirrenbach_2014}. Since beginning its operation in January 2016, CARMENES has been regularly monitoring  M dwarfs pre-selected from the CARMENES input catalog \citep[Carmencita;][]{alonso_2015, caballero_2016a, Reiners2018b}. Several planetary systems around them have already been confirmed or discovered  \citep{Trifonov_2018,Reiners_2018a,Sarkis_2018,Kaminski_2018,Luque_2018,Nagel_2019,perger2019}, including a planetary companion orbiting Barnard's star \citep{Ribas_2018}. 

In this paper, we analyze the RV data of LSPM~J2116+0234 and GJ~686, monitored as part of the CARMENES Guaranteed Time Observations (GTO) M dwarf survey.
The data reveal the presence of a mini-Neptune around LSPM~J2116+0234 with a period of 14.4\,d. Furthermore, we used CARMENES data to refine the orbital parameters of GJ~686b reported by \citet{affer2019} (hereafter Aff19), a super-Earth with a period of 15.5\,d. In \S\ref{sec:targets}, we introduce the basic properties of the host stars, and we describe the spectroscopic and photometric data in \S\ref{sec:data}. In \S\ref{sec:analysis}, we present our results from the analysis of RVs, photometry and activity indicators. 
We model activity as correlated noise using a Bayesian framework in order to find the orbital parameters and discuss the stability of the signals through time and wavelength. We conclude and summarize our work in \S\ref{sec:summary}.  

\section{Targets} \label{sec:targets} 

A summary of the basic stellar properties of both targets is presented in Table~\ref{tab:tab1}. The photospheric parameters such as the effective temperature $T_{\rm eff}$, surface gravity $\log{g}$, and metallicity [Fe/H] of the targets were determined in the CARMENES framework by \cite{passegger_2018} using the PHOENIX-ACES model grid \citep{Husser2013}. The stellar masses and radii were determined based on the photospheric parameters and a mass-radius relation.  

\object{LSPM~J2116+0234} (Karmn~J21164+025) is an M3.0\,V star at a distance $\sim$17.64\,pc \citep{2016AJ....151..160F,Gaia2016,gaia_2018}. It was discovered by \citet{2005AJ....129.1483L} as a northern star with a proper motion larger than 250\,mas\,yr$^{-1}$, and was characterized photometrically and spectroscopically (\citealt{2011AJ....142..138L};  \citealt{Lepine2013}; \citealt{2014MNRAS.443.2561G}). LSPM~J2116+0234 has been identified as a nearby potential target for planet searches \citep{2013MNRAS.435.2161F,2014AJ....148..119F}, activity analyses \citep{2017ApJ...834...85N,jeffers_2018}, and determination of photospheric stellar parameters \citep{passegger_2018}.

\begin{figure}[!t]
\begin{center}
\includegraphics[width=1.05\columnwidth,clip]{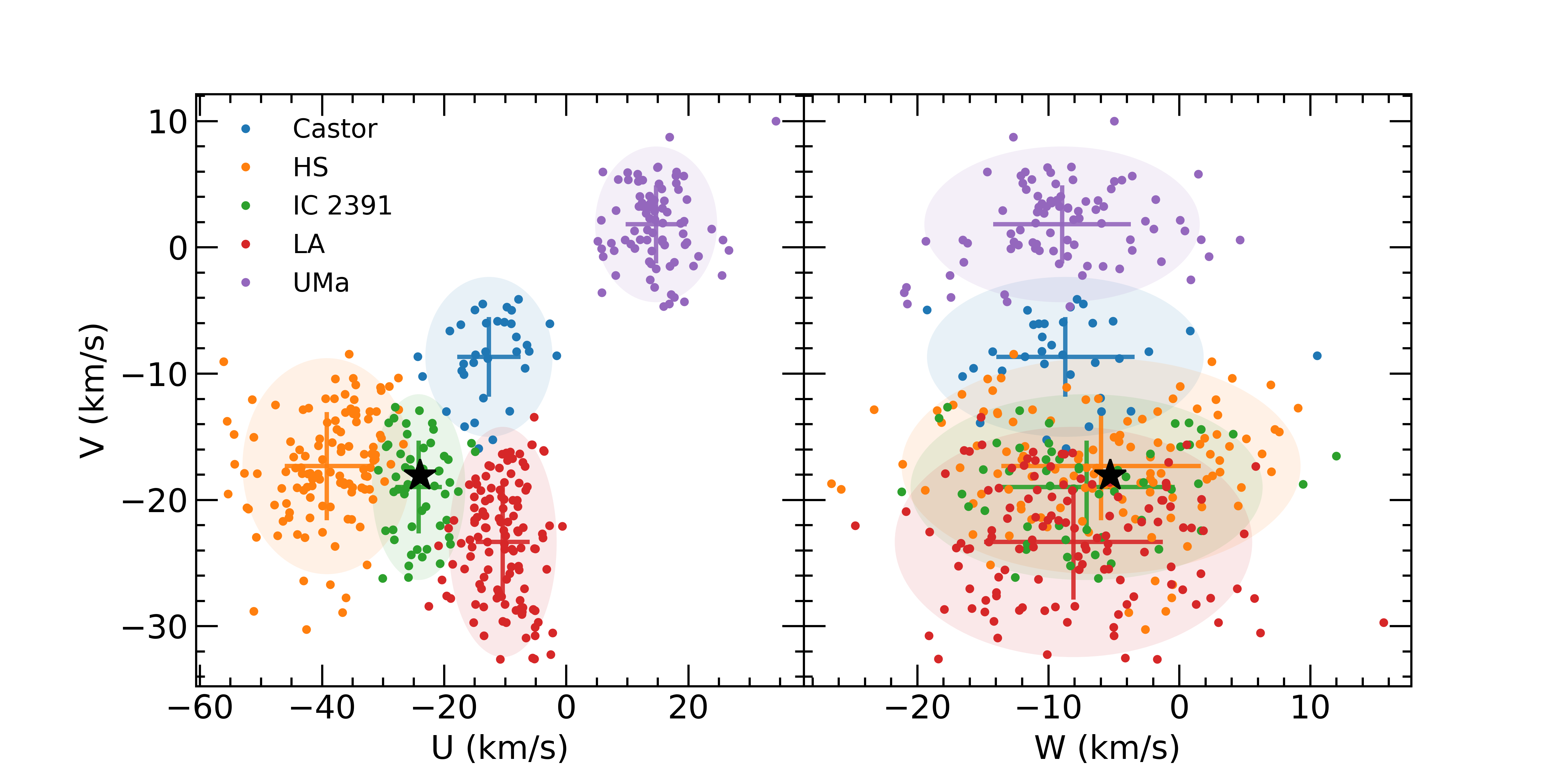}
\caption{\label{uvw}  The UVW velocities of young moving groups and LSPM J2116+023 (star symbol). The crosses are the 1-sigma value for each moving group while the ellipses represents a 2-sigma value.}
\end{center}
\end{figure}

\citet{Cortes2016} computed Galactocentric space velocities consistent with membership to the young disk population and the IC~2391 stellar kinematic group. 
We evaluate the membership of LSPM J2116+0234 to the young disk population and the IC~2391 supercluster using the {\em Gaia} second data release (GDR2) astrometric data, and the RV as published by \cite{Reiners2018b}. 
In Figure~\ref{uvw}, we compare the UVW velocities of LSPM J2116+0234 with known members of young moving groups from \cite{montes_2001}. 
However, the lack of H$\alpha$ feature in emission (pEW(H$\alpha$) = +0.004 $\pm$ 0.005\,{\AA}; \citealt{jeffers_2018}) seems to be inconsistent with the mean age of the supercluster. The H$\alpha$ line can be seen in all the members with similar spectral type of the IC~2391 open cluster, which is thought to be the birthplace of the supercluster \citep{eggen_1995} and, hence, to have a similar age (50 $\pm$ 5 Myr, \citealt{barrado_2004}). 
Besides, we do not detect any X-ray emission based on {\em ROSAT} All Sky Survey data. Therefore, we estimate the upper-limit luminosity $L_{\rm X}$< 10$^{27}$\,erg\,s$^{-1}$, which is lower than the expected value of $L_{\rm X}$< 10$^{29}$\,erg\,s$^{-1}$ found for the members with similar spectral type of the IC~2391 and other older young open clusters such as the Pleiades or Hyades \citep{patten_1996}. Furthermore, the tentative rotational period of about $\sim$42\,d, obtained in this work (see Section~\ref{subsec:lspm}), is longer than those of the IC~2391 cluster members \citep{patten_1996} and those of  the members of the older Pleiades open cluster (110--120\,Myr, \citealt{dahm_2015}), which have typical rotational periods of less than 10 days \citep{rebull_2016}. Therefore, although LSPM J2116+0234 shares the same kinematics as the IC~2391 supercluster, all the activity indicators and the rotation period indicate that the object is older than 50\,Myr. In fact, from the gyrochronologic relation in \cite{Barnes2017}, we estimate an age for this star of $\sim2$\,Gyr, using its rotation period and $B-V$ color.

\object{GJ~686} (BD+18~3421, Karmn~J17378+185) is an M1.0\,V star, located in the Hercules constellation, at only $d\sim$8.2\,pc \citep{Gaia2016,gaia_2018}. Because of its earlier spectral type and closer heliocentric distance, GJ~686 is also brighter than LSPM~J2116+0234. In particular, its bright visual magnitude  $V$ $\approx9.6$ \,mag \citep{Koen2010,zacharias_2013} made the star to be tabulated in the "Bonner Durchmusterung des s\"udlichen Himmels" by \citet{Bonner1886}, and its parallax to be measured more than 100 years ago \citep{Barnard1913,Adams1926, Osvalds1957}. 
It was one of the first late-type stars for which RV and metallicity were measured \citep{1953GCRV..C......0W,1990PAZh...16...52T} and one of the first M-type standard stars \citep{1994AJ....108.1437H}. Later, GJ~686 took more relevance with the investigation of its moderate activity level \citep{1986ApJS...61..531S,1986AJ.....92..139S,1987AJ.....94..150H,1989A&A...219..239R,1993A&AS..100..343P}. 
Its moderate activity level has been confirmed by more recent, comprehensive studies \citep{1998A&A...331..581D,2004ApJS..152..261W,2010ApJ...725..875I,jeffers_2018}, and is consistent with its kinematic classification as a thin-disk star \citep{Cortes2016}. In the past 20 years several spectra were taken with HIRES on the Keck-I telescope to search for extrasolar planets around it. Using these data, \cite{butler_2017} found a signal at $15.5303\pm0.0030$\,d and an amplitude of $3.46\pm0.56$\,m~s$^{-1}$, which they listed as a signal requiring confirmation.  Recently, Aff19 analysed high-precision RV data from HIRES together with HARPS and HARPS-N spanning over 20 years, yielding the detection of a 
super-Earth orbiting  GJ~686. The planetary companion was reported to have a  minimum mass of 7.1$\pm$0.9\,M$_{\oplus}$, orbiting its host star with a period of 15.5321\,d and a semi-major axis of 0.091\,au. 

Furthermore, they also analyzed the activity indicators of HIRES, HARPS and HARPS-N, from which they estimated a rotation period of 37\,d and an activity cycle of $\sim$2000\,d.

\begin{table}
\centering
\caption{Basic information of archival and CARMENES observations.}
\label{tab:Nobs}
\resizebox{\columnwidth}{!}{
\begin{tabular}{lccccc} 
\hline\hline
\noalign{\smallskip}
Target & Instrument & $N_{\rm obs}~(N_{\rm used})$  & $\Delta$t & $rms$ & $\overline{\sigma}$\\
 & & [$\#$] &[d]& [m\,s$^{-1}$] & [m\,s$^{-1}$]\\
 \noalign{\smallskip}
\hline
\noalign{\smallskip}
\multirow{2}{*}{LSPM~J2116+0234} & CARM-VIS & 72(70) & 882 & 5.30 & 1.48\\
 & CARM-NIR & 57(55) & 823 & 6.89 & 6.54 \\
 \noalign{\smallskip}
\hline
\noalign{\smallskip}
\multirow{4}{*}{GJ~686} & HIRES & 114(112) & 5947 & 4.09 & 1.85\\
 & CARM-VIS & 100(96) & 987 & 3.11 & 1.71\\
 & HARPS-N & 64(61) & 1347 & 3.02 & 0.71\\
 & HARPS & 20(19) & 2299 & 2.42 & 0.69\\
\noalign{\smallskip}
\hline
\end{tabular}
}
\end{table}

\begin{table*}
\centering
\caption{Photometric observation log of GJ~686.}
\label{tab:obslog}
\begin{tabular}{lcccc}
\hline\hline
\noalign{\smallskip}

\multirow{1}{*}{}& \multirow{1}{*}{Monet-S} & \multirow{1}{*}{SNO} & \multirow{1}{*}{TJO} &\multicolumn{1}{c}{LCO}\\
&Sutherland& Granada & Lleida & Global network\\
\noalign{\smallskip}
\hline
\noalign{\smallskip}
Latitude & -32$^{\circ}$22'44"&+37$^{\circ}$03'51"&+42$^{\circ}$03'05"&\\
Longitude & +20$^{\circ}$48'39"&-03$^{\circ}$23'05"&+00$^{\circ}$43'46"&\\
Altitude [m] & 1700&2896&1568&\\
\hline
\noalign{\smallskip}
Filters [Johnson]  &  $B$ & $V$,\,$R$& $R$& $B$,\,$V$\\
FOV [arcmin]& 12.6$\times$12.6 &13.2\,$\times$\,13.2&12\,$\times$\,12&19\,$\times$\,29\\
Exposure time [s] & 2\,($B$) & 25\,($V$), 10\,($R$) & 20\,($R$) &72\,($B$), 17\,($V$)\\
Number of nights [d]& 26 &26&38&69\\
Time Span [d] & 65 &58&87&93\\
Observation period (2018)& Jul--Sep &Jul--Sep &Jul--Oct &Jul--Oct\\
\noalign{\smallskip}
\hline
\end{tabular}

\footnotesize{\textbf{Notes: } For LCO node positions see {\tt https://lco.global/observatory/sites/}.  \\Epochs with long duration observation with LCO was exposed to 5s in $V$ filter.}
\end{table*}

\section{Observations} \label{sec:data}
\subsection{Spectroscopic data}
High-resolution spectroscopic observations were obtained with the VIS and NIR channels of the CARMENES spectrograph. 
The wavelength calibration of both channels is done with hollow-cathode lamps (U-Ar, U-Ne, Th-Ne) and temperature-pressure stabilized Fabry-P\'erot etalons \citep{schafer_2018} to interpolate the wavelength solution and monitor any instrumental drift during observations \citep{Bauer_2015}. Reduction of raw spectra is automatically performed using the CARACAL \citep[CARMENES Reduction And Calibration,][]{caballero_2016a} pipeline, which corrects for bias, flat-field, and cosmic rays.

\begin{figure}[!t]
\begin{center}
\includegraphics[width=0.95\columnwidth,clip]{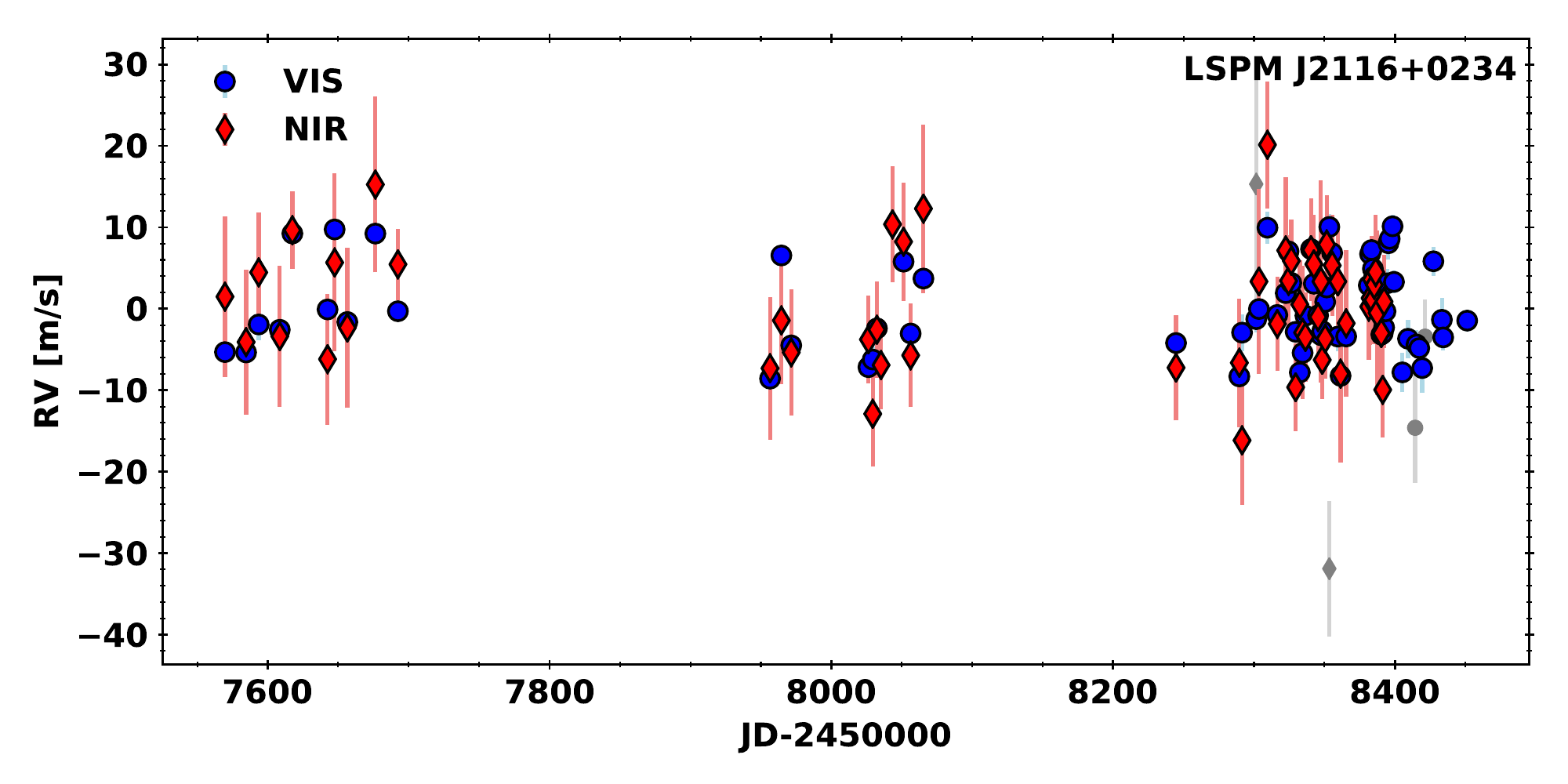}
\includegraphics[width=0.95\columnwidth,clip]{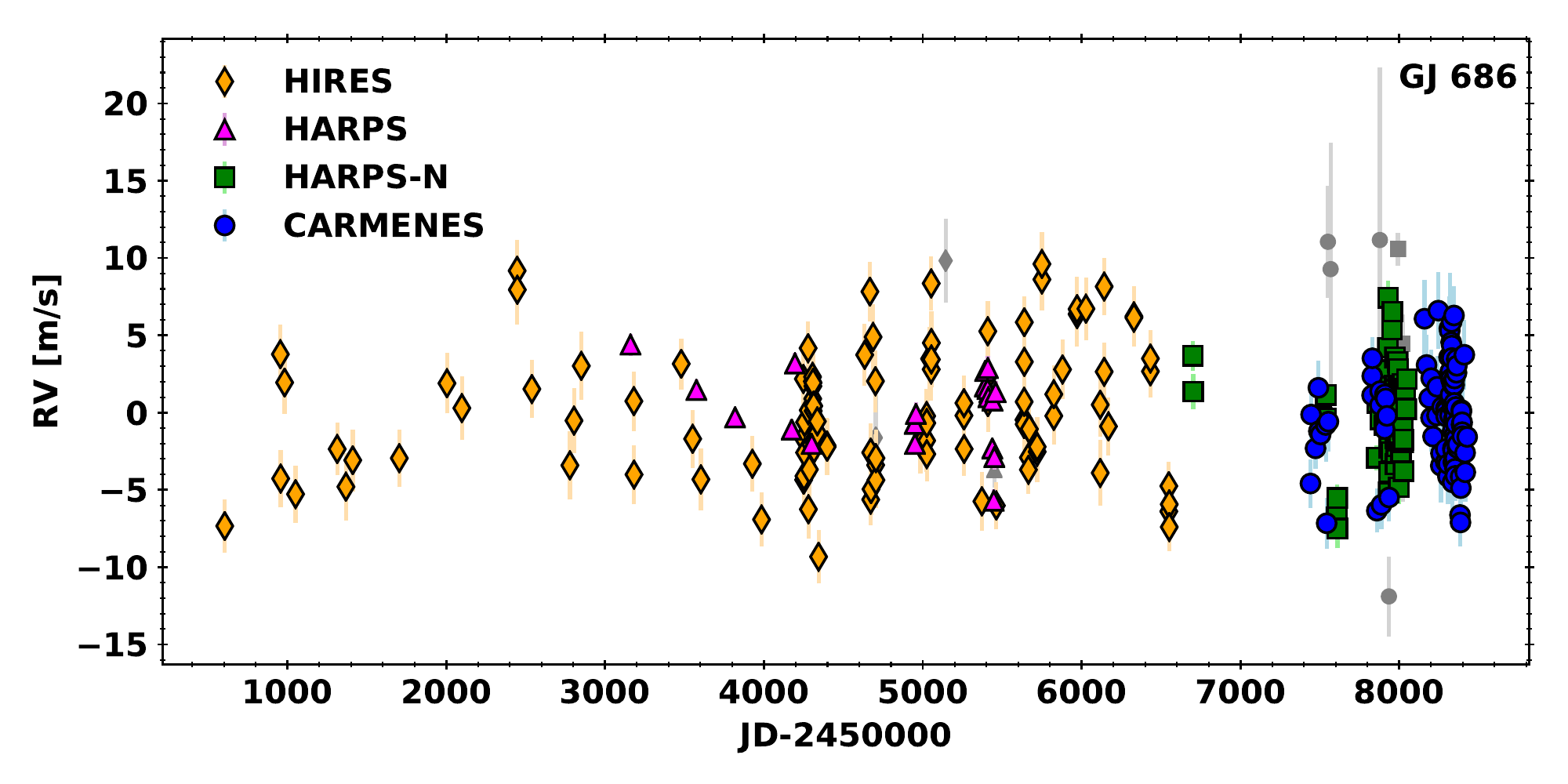}
\includegraphics[width=0.95\columnwidth,clip]{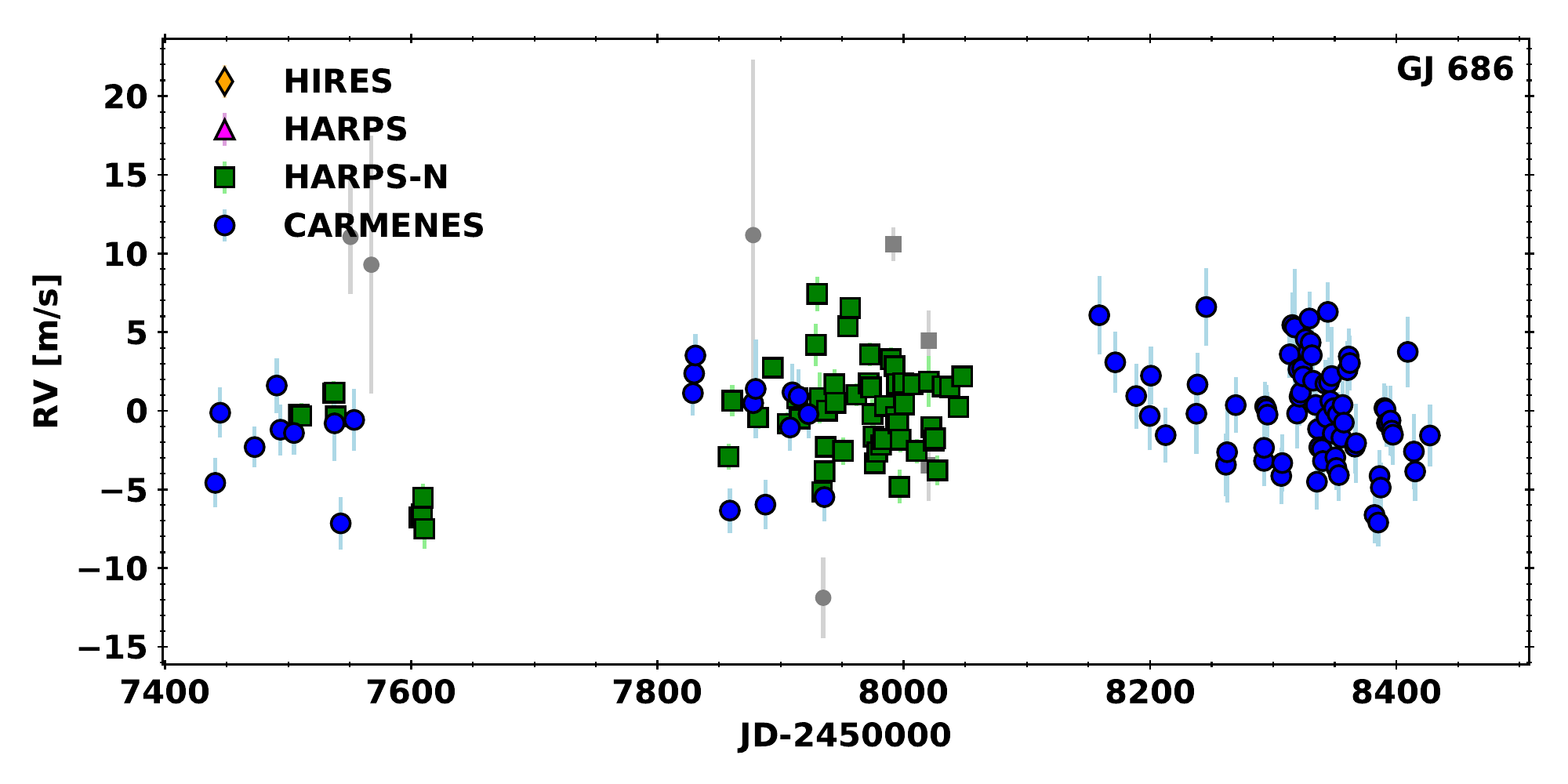}
\caption{\label{RVdataset} RV time series of the M-dwarf stars LSPM~J2116+0234 ({\it top}), GJ~686 ({\it middle}) and zoom into the data with JD$>$2457000 of GJ~686 ({\it bottom}). The grey symbols correspond to the clipped RVs.}
\end{center}
\end{figure}

 High-precision RVs are routinely computed by the CARMENES SERVAL pipeline \citep{zechmeister_2018}, using an algorithm based on a least-square fitting of the RV shifts of the individual spectra against a high signal-to-noise ratio (S/N) template, which is constructed by co-adding all available spectra of the target \citep[see also][]{anglada_2012}. A nightly zero-point correction is applied to the RVs to track remaining systematics of the instrument and/or pipeline \citep[for more details see][]{Trifonov_2018, lev_2019}. These nightly zero-points are calculated using all the CARMENES GTO stars with an RV standard deviation lower than 10\,m\,s$^{-1}$. The median magnitude of these corrections are 1.79\,m\,s$^{-1}$ and 1.78\,m\,s$^{-1}$ for the VIS and NIR channel data of LSPM~J2116+0234, respectively, and 1.86\,m\,s$^{-1}$ for the VIS channel data of GJ~686.

Telluric contamination and unmasked detector defects can lead to systematic RV errors in spectral orders with low RV content. Therefore, we carefully select the orders to exclude from our computation of the NIR RVs. This process is done iteratively to minimize the sample RMS of the entire CARMENES M-dwarf sample. 

The SERVAL pipeline also provides information about stellar activity such as line indices for a number of spectral features (e.g.  H$\alpha$, Na \textsc{i} D and Ca \textsc{ii} IRT), the differential line width (dLW), and the chromatic index (CRX), as defined in \cite{zechmeister_2018}. Furthermore, for each CARMENES spectrum, the cross-correlation function (CCF) is computed using a weighted mask of co-added stellar spectra. The CCFs are fitted with a Gaussian function to determine the contrast, the full width at half maximum (FWHM), and the bisector velocity span (BIS). A detailed description on CCF computation methodology is given by \cite{Reiners_2018a}.

LSPM~J2116+0234 was monitored between 30 Jun 2016 and 29 Nov 2018, obtaining 72 and 57 high-resolution spectra from the CARMENES VIS and NIR channels, respectively. In total, the observations cover a time span of 882\,d, with typical exposure times of 1800 seconds. 
In Table \ref{tab:Nobs}, we provide a summary of the total number of available RVs, the time span of the data, standard deviation and median internal uncertainty $\overline{\sigma}$.

For GJ~686, 100 CARMENES spectra from the VIS channel are available, which were obtained between 22 Feb 2016 and 29 Nov 2018, covering 987\,d.  Besides, as outlined in Aff19, other instruments have monitored GJ~686 during the past 21 years, adding an additional 198 precise RVs.  

To avoid using RV epochs contaminated by flares, or spectra with a low S/N, we applied a 3$\sigma$ clipping to both the RVs and errors of each individual dataset, removing a total of 10 RVs from GJ~686 ($3.5\%$) and 4 from LSPM~J2116+0234 ($3.1\%$).  Since the internal RV precision in the NIR is larger than the expected RV signal for GJ~686, we decided not to use the NIR RVs. In Fig.~\ref{RVdataset}, we show the RV time series of both targets. The radial velocities for LSPM~J2116+0234 and GJ~686 are given in  Table~\ref{tab:lspmrvs} and \ref{tab:gj686rvs}, respectively.

\subsection{Photometric data}

Several potential exoplanet candidates from CARMENES are monitored photometrically by ground-based telescopes to constrain the stellar rotation \citep{diez2019} as well as to search for planetary transits.  LSPM~J2116+0234 was not monitored by our photometric follow-up program, therefore, we searched through the archival surveys such as All-Sky Automated Survey \citep[ASAS \footnote{{\tt http://www.astrouw.edu.pl/asas/}},][]{Pojmanski_1997} and Catalina Sky Survey \footnote{{\tt https://catalina.lpl.arizona.edu}} \citep{Drake_2009}. These survey data were used to investigate the stellar rotation period.

\begin{figure}
\begin{center}
\includegraphics[width=\columnwidth]{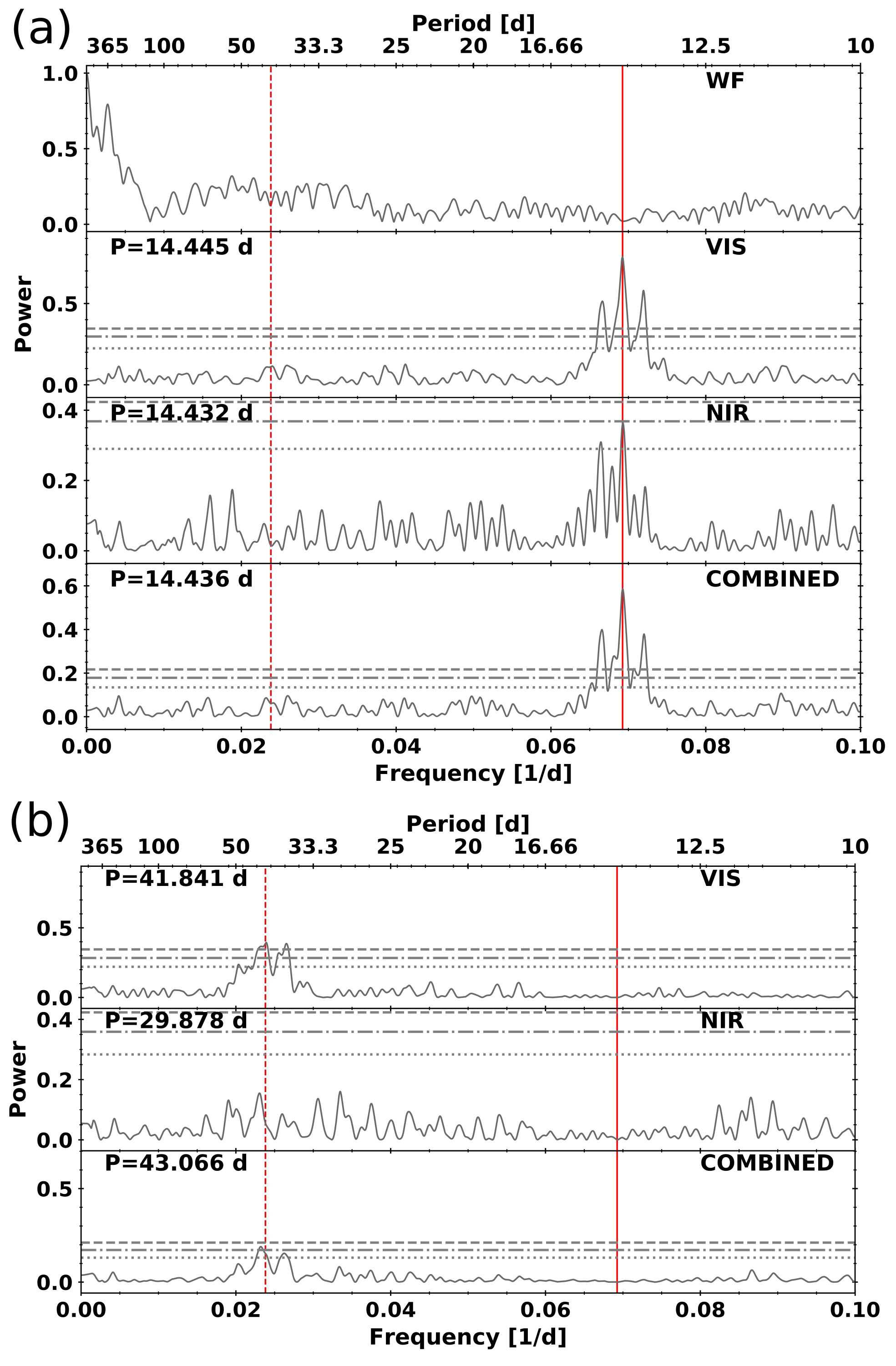}
\caption{\label{LSPMperiodogram_all}(a) GLS periodograms of LSPM~J2116+0234 RV data. The top panel shows the WF of the combined dataset. The next two panels correspond to the VIS and the NIR CARMENES channels, respectively, while the bottom panel shows the periodogram of the combined data set. The horizontal lines represent bootstrapped FAP levels of 10, 1 and 0.1\%. The periods reported in each panel refer to the highest peak. The vertical solid and dashed red lines indicate the period of the proposed planet and the estimated stellar rotation period at 14.44 and $\sim$42\,d, respectively. Although we have inspected the periodogram for significant signals at frequencies up to 1\,d$^{-1}$, for visual purposes, we only show the region from 0 to 0.1\,d$^{-1}$ in all the periodograms. (b) GLS periodograms of the RV residuals after removing a sinusoid with the period found in (a). }
\end{center}
\end{figure}

Along with the photometry from the ASAS database, we have monitored GJ~686 with the following facilities: 
\begin{itemize}
\item MONET: The MONET 1.2\,m telescope located at the Sutherland station of the South African Astronomical Observatory (SAAO). It is equipped with a 2k\,$\times$\,2k CCD with a plate scale of 0.36\,arcsec per pixel.
\item SNO: The T90 telescope located at Sierra Nevada Observatory, Spain is a 0.9\,m Ritchey-Chrétien telescope. It is equipped with a CCD camera VersArray 2k\,$\times$\,2k with a plate scale of 0.38\,arcsec per pixel \citep{Rodriguez_2010}.  
\item TJO: The Joan Or\'o telescope is located at the Montsec Astronomical Observatory (OAdM), Spain. It is a fully robotic 0.8\,m Ritchey-Chr\'etien telescope with an FLI PL4240 2k\,$\times$\,2k camera and a plate scale of 0.36\,arcsec per pixel.
\item LCOGT: The Las Cumbres Observatory Global Telescope is a network of robotic telescopes deployed at several sites around the globe. The observations were performed using the 0.4\,m telescopes in Haleakala, Hawai'i (kb27 and kb82 SBIG CCDs), the Teide Observatory in Tenerife (kb23 and kb99 SBIG CCDs), the McDonald Observatory in Texas (kb92 SBIG CCDs), the South African Astronomical Observatory (kb96 SBIG CCDs) and the Cerro Tololo Interamerican Observatory (kb 81 SBIG CCDs). The telescopes have a plate scale of 0.57\,arcsec per pixel. 
\end{itemize}

In Table~\ref{tab:obslog}, we give the detailed photometric observation log for GJ~686. The MONET, TJO and SNO photometric data were reduced and analyzed with standard packages and tasks of the Image Reduction and Analysis Facility (IRAF\footnote{{IRAF is distributed by the National Optical Astronomy Observatory, which is operated by the Association of Universities for Research in Astronomy under a cooperative agreement with the National Science Foundation. \tt http://iraf.noao.edu/}}). The LCOGT images were reduced by the BANZAI pipeline \citep{banzai}. The differential photometry was performed by dividing the flux of GJ~686 by the combined flux of all comparison stars.

\begin{figure}[!t]
\begin{center}
\includegraphics[width=\columnwidth]{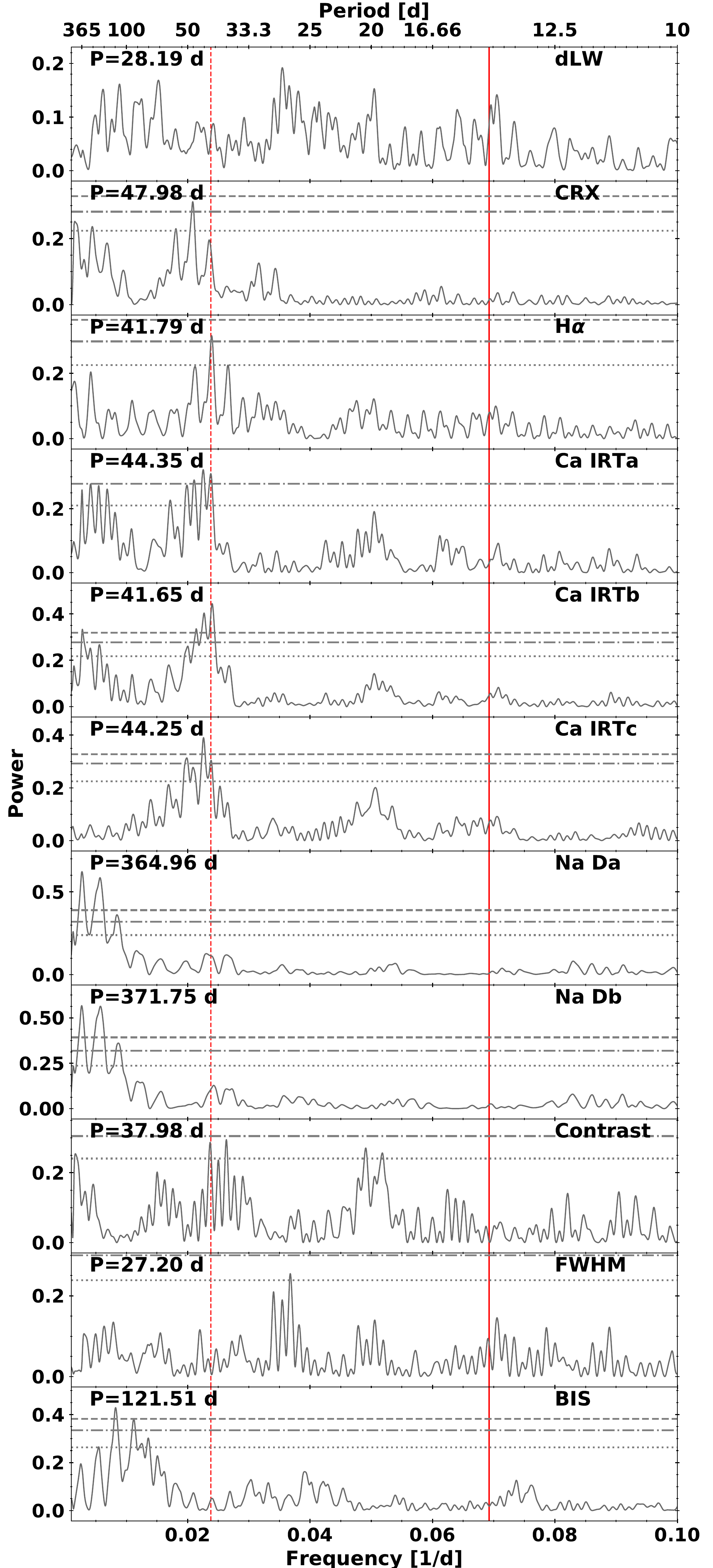}
\caption{\label{periodogram_activity_lspm}GLS periodograms of the activity indicators of LSPM~J2116+0234. The vertical red solid line indicates the period of the suggested planet, while the vertical red dotted indicates the stellar rotation period. The periods reported in each panel refer to the highest peak. Horizontal lines represent the bootstrapped 10, 1 and 0.1$\%$ FAP levels.}
\end{center}
\end{figure}

\begin{figure}[!t]
\begin{center}
\includegraphics[width=0.98\columnwidth]{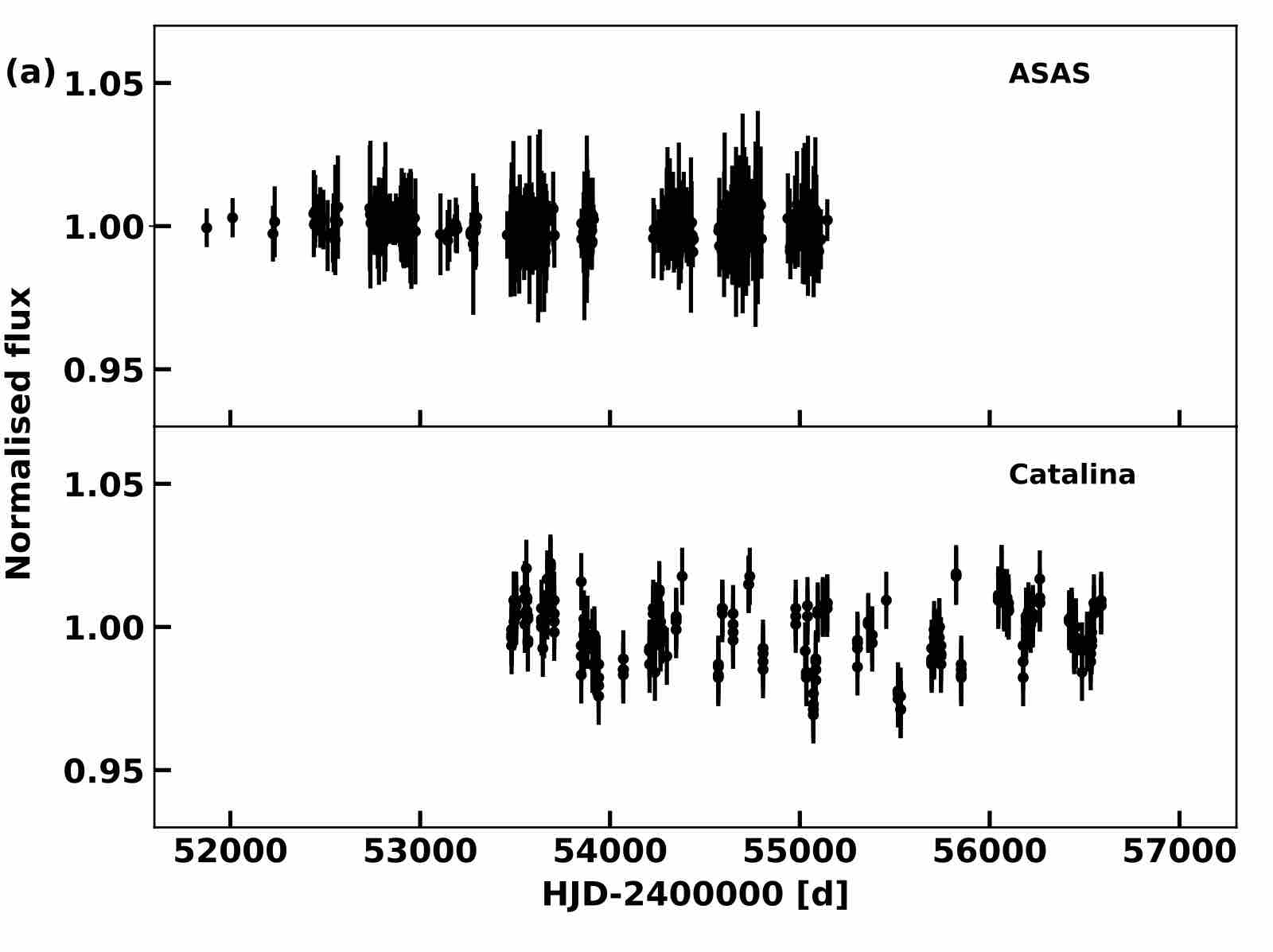}
\includegraphics[width=1.0\columnwidth]{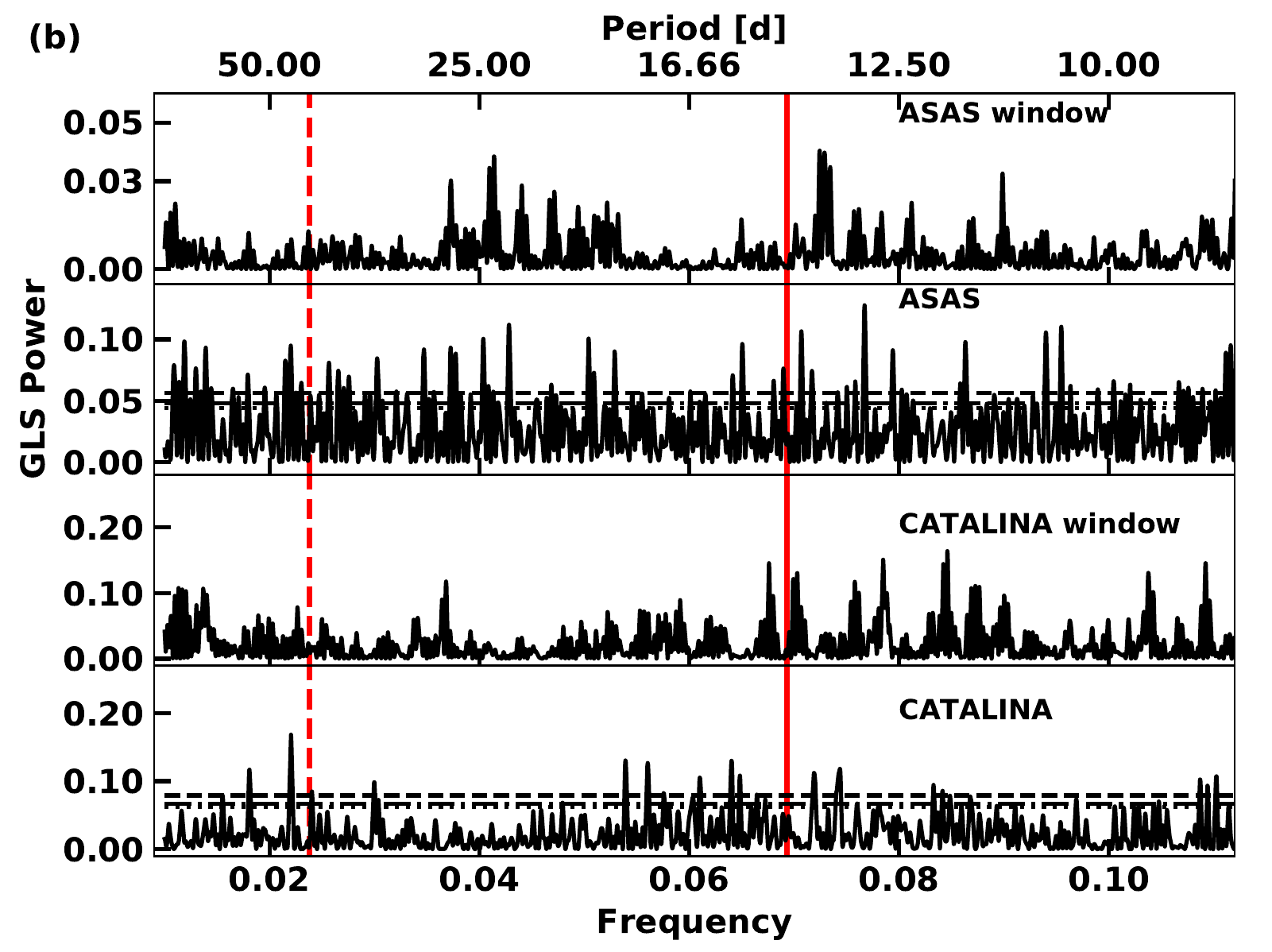}
\caption{\label{lspmlc} (a) Photometric time series of LSPM~J2116+0234 observed with ASAS and Catalina surveys. (b) GLS periodograms of the photometric time series of LSPM~J2116+0234 observed with ASAS and Catalina surveys. The WF for each of the surveys are plotted. The vertical solid and dashed lines indicate the planetary period and the estimated stellar rotation period at 14.44\,d and $\sim$42\,d, respectively. The horizontal lines represent 10, 1 and 0.1\% bootstrapped FAP levels, respectively. }
\end{center}
\end{figure}

\section{Analysis and results} \label{sec:analysis}

\subsection{LSPM~J2116+0234}\label{subsec:lspm}

To investigate the RV variability of LSPM~J2116+0234, we computed the Generalized Lomb-Scargle (GLS) periodogram \citep{zechmeister_2009} of the RVs from the individual VIS and NIR CARMENES channels, as well as of the combined dataset. 
In Figure~\ref{LSPMperiodogram_all}a, the resultant periodograms are plotted together with the window function (WF) of the combined dataset (top panel). We computed the false alarm probability (hereafter, FAP) levels of 10, 1 and 0.1$\%$ using 10\,000 bootstrap randomizations of the input data. 
 The signal is considered significant if it reaches a FAP level $<0.1\%$. The VIS data periodogram (Fig. \ref{LSPMperiodogram_all}a, second panel) displays a significant and isolated signal at 0.0692\,d$^{-1}$ (14.445\,d), accompanied by the two yearly aliases at $\pm$0.0027\,d$^{-1}$ ($\sim$365\,d) from the central peak, as suggested by the WF. The NIR data, shown in the third panel of Fig. \ref{LSPMperiodogram_all}a also displays a signal at 14.44\,d just above the 1$\%$ FAP level. Due to the larger errors of the NIR RVs, the periodogram of the combined dataset (Fig. \ref{LSPMperiodogram_all}a, bottom panel) is very similar to the periodogram of the VIS channel. The combined periodogram also has a very significant peak at 14.436\,d. Furthermore, we also found another significant signal at a period of 0.9308\,d$^{-1}$ (1.07\,d), which is the expected period of a daily alias of the 14.436\,d signal.

We analyzed the periodograms of the residuals after subtracting the main signal from the RV periodograms, to investigate if there are significant RV variations remaining. The residual periodograms are shown in Fig. \ref{LSPMperiodogram_all}b. They show a significant peak in the VIS data at 0.0230\,d$^{-1}$ (41.8\,d), and 0.0232\,d$^{-1}$ (43.1\,d) at the 1$\%$ FAP level in the combined dataset. However, the NIR data do not show any signal above the 10$\%$ FAP level. 

Furthermore, we carried out a periodogram analysis of the activity indicators to investigate if the signals at $\sim$14.44\,d and $\sim$42\,d in the RV data may have a stellar origin. 
We show the resulting GLS periodograms of the activity indicators  and the properties of the CCFs in Fig. \ref{periodogram_activity_lspm}. None of the periodograms show any significant peak at 14.436\,d (vertical solid line). The calcium triplet lines show significant peaks between 41 and 44\,d (vertical dashed line). These peaks are close to the significant peak in the residual RV periodogram. At $1\%$ FAP level, the H$\alpha$ index shows a peak at 41.8\,d, and the CRX at 48.0\,d, which has a higher power than its yearly alias at 42.4\,d. We interpret this yearly alias as the real signal due to its consistency with the signals found on the other indicators. On subtraction of these signals, the residuals do not show any significant peaks in their periodograms (see Fig. \ref{LSPMGLSresidualsactivity} in the appendix). Thus, we attribute the signals at $\sim$42\,d in the RV residuals and the activity indices to be related to the rotation period of the star. 

We investigated if there is any significant correlation between the RVs and 
various activity indicators (Figure \ref{correlation_activity_lspm}). None of the indicators show a correlation above the significance limit, which we set as a p-value of 0.05 or lower, except for one of the Ca \textsc{ii} IRT lines with a p-value of 0.025. Further, we can see a color gradient in the correlation plot of this line, indicating that the correlation is indeed caused by the rotation period. We also observe that there are indicators that despite having significant signals in the periodogram, they show no correlation with the RV, which could be indicative of shift in phase between indicator and RV (0 or $\pi$).

The analysis of the CARMENES RVs was complemented with the archival photometric data for LSPM~J2116+0234. Figure~\ref{lspmlc}a shows the observations from ASAS spanning $\sim$9 years and Catalina spanning $\sim$8.5 years with an overlap of $\sim$ 4.5 years.
The GLS periodograms of the photometric data are depicted with bootstraped FAP levels (Fig. \ref{lspmlc}b). We do not find prominent signals around the planetary period or the activity indicator peaks found in the RV data. However, the Catalina dataset shows a significant period near 45\,d. After removing this signal, no prominent period can be found in the data. In contrast, we find a significant signal close to the 14\,d signal in the ASAS data at $\sim$13.03\,d. However, the ASAS WF shows a significant period of about 13.80\,d. After removing the signal at 13.80\,d, the signal at 13.03\,d also disappears. Furthermore, the remaining signals do not reach the $10\%$ FAP level, with a second signal at $\sim$45\,d. 

We note that between HJD$\sim$2453400 - 2455100\,d there are overlapping observations by ASAS and Catalina. We analyzed the dataset during this overlapping period together applying an offset to both the datasets. However, we do not find any significant period.

To summarize, based on CARMENES VIS and NIR RV data we identify a strong signal at $\sim$14.44\,d with no counterpart in the activity indicators, which we attribute to a planetary origin. We also find significant signals at $\sim$42\,d in the residuals of the RV data. This signal is also significant in some activity indicators, and thus we relate it to the rotation period of the star.

\begin{table*}
     \caption{Best-fit parameters to different models of the planetary system LSPM~J2116+0234b.}

     \label{tab:lspmparams}

     \centering %
\resizebox{\textwidth}{!}{\begin{tabular}{lcccccccc}
       \hline \hline 
       \noalign{\vskip 1mm}
       \multirow{3}{*}{} & \multicolumn{8}{c}{LSPM~J2116+0234~b} \\ 
       \cline{2-9}
       \noalign{\vskip 1mm}
      & VIS+NIR & \multicolumn{2}{c}{VIS} & NIR & \multicolumn{3}{c}{VIS+NIR} & \multirow{2}{*}{Prior} \\
      \cline{3-4} \cline{6-8}
 	  \noalign{\smallskip}
      & \multirow{2}{*}{null model} & \multirow{2}{*}{Keplerian} & GP + Keplerian & \multirow{2}{*}{Keplerian} & \multirow{2}{*}{Keplerian} & GP + Keplerian & GP + Keplerian & \\
      &  &  & \texttt{George} & &  &  \texttt{George}& \texttt{Celerite}\\
   
 	   \hline 

	  {\bf Planetary parameters}\\
	   \hline
	   \noalign{\vskip 1mm}
 	  $P$ [d] & $\cdots$ & $14.4432^{+0.0080}_{-0.0086}$ & $14.4433^{+0.0079}_{-0.0086}$  & $14.425^{+0.030}_{-0.029}$ & $14.4399^{+0.0078}_{-0.0087}$ &  $14.4410^{+0.0076}_{-0.0088}$&$14.451^{+0.012}_{-0.010}$ & $\mathcal{U}(10,20)$\\
 	  \noalign{\vskip 1mm}
 	  $T_0$ [JD-2457000] & $\cdots$ & $573.36^{+1.05}_{-0.99}$ & $573.27^{+0.94}_{-0.87}$   & $574.7^{+1.3}_{-1.3}$ & $573.6^{+1.1}_{-1.1}$ & $573.34^{+0.87}_{-0.90}$ & $572.52^{+0.75}_{-0.88}$ & $\mathcal{U}(550,590)$\\
 	  \noalign{\vskip 1mm}
 	  $K$ [m\,s$^{-1}$] & $\cdots$ & $6.43^{+0.45}_{-0.42}$ & $6.31^{+0.44}_{-0.43}$   & $5.1^{+1.3}_{-1.3}$ & $6.26^{+0.41}_{-0.39}$ & $6.19^{+0.38}_{-0.40}$&$6.29^{+0.25}_{-0.29}$  & $\mathcal{U}(0,20)$\\
 	  \noalign{\vskip 1mm}
 	  $e\sin{\omega}$ & $\cdots$ & $-0.069^{+0.068}_{-0.070}$ & $-0.081^{+0.059}_{-0.053}$ & -0.069 (fixed) & $-0.066^{+0.066}_{-0.067}$ & $-0.084^{+0.054}_{-0.050}$&$0.015^{+0.013}_{-0.014}$ & $\mathcal{U}(-1,1)$ \\
 	  \noalign{\vskip 1mm}
 	  $e\cos{\omega}$ & $\cdots$ & $0.168^{+0.067}_{-0.066}$ &  $0.170^{+0.064}_{-0.065}$  & 0.168 (fixed) & $0.157^{+0.065}_{-0.065}$ & $0.164^{+0.062}_{-0.060}$&$0.159^{+0.061}_{-0.064}$ & $\mathcal{U}(-1,1)$ \\
 	  \noalign{\vskip 1mm}
 	  $a$ [AU] & $\cdots$ & $0.0876^{+0.0022}_{-0.0021}$ &  $0.0876^{+0.0021}_{-0.0021}$   &$0.0876^{+0.0022}_{-0.0021}$ & $0.0876^{+0.0022}_{-0.0021}$ & $0.0876^{+0.0021}_{-0.0020}$&$0.0876^{+0.0020}_{-0.0021}$ &$\cdots$\\
 	  \noalign{\vskip 1mm}
 	  $m_{\rm p}$ $\sin i$ [M$_{\oplus}$] & $\cdots$& $13.6^{+1.1}_{-1.1}$ &  $13.4^{+1.1}_{-1.1}$    & $10.8^{+2.9}_{-2.8}$ & $13.3^{+1.0}_{-1.1}$ & $13.1^{+1.0}_{-1.0}$&$13.56^{+0.54}_{-0.62}$ &$\cdots$\\
 	  \noalign{\vskip 1mm}
	  \hline
	  {\bf RV offsets and jitter}\\
	   \hline
	   \noalign{\vskip 1mm}
 	  $\gamma_{\rm VIS}$ [m\,s$^{-1}$] & $0.27^{+0.64}_{-0.60}$ & $0.41^{+0.30}_{-0.30}$ &  $0.00^{+0.55}_{-0.63}$  & $\cdots$ & $0.41^{+0.30}_{-0.29}$ & $-0.09^{+0.56}_{-0.62}$&$0.22^{+0.30}_{-0.26}$ & $\mathcal{U}(-100,100)$\\
 	  \noalign{\vskip 1mm}
 	  $\gamma_{\rm NIR}$ [m\,s$^{-1}$] & $0.12^{+0.91}_{-0.88}$ & $\cdots$ &  $\cdots$  & $-0.02^{+0.92}_{-0.89}$ & $-0.02^{+0.90}_{-0.90}$ & $0.03^{+0.93}_{-0.92}$&$0.15^{+0.35}_{-0.42}$ & $\mathcal{U}(-100,100)$\\
 	  \noalign{\vskip 1mm}
 	  $\sigma_{\rm jit, VIS}$ [m\,s$^{-1}$] & $5.1^{+0.53}_{-0.43}$ & $1.83^{+0.31}_{-0.28}$ &  $0.42^{+0.59}_{-0.29}$  & $\cdots$ & $1.87^{+0.31}_{-0.28}$ & $0.35^{+0.36}_{-0.25}$&$0.67^{+0.31}_{-0.28}$ & $\mathcal{U}(0,10)$\\
 	  \noalign{\vskip 1mm}
 	  $\sigma_{\rm jit, NIR}$ [m\,s$^{-1}$] & $1.8^{+1.5}_{-1.2}$ & $\cdots$  &  $\cdots$  & $1.08^{+1.17}_{-0.77}$ & $1.28^{+1.36}_{-0.88}$ & $1.06^{+1.19}_{-0.74}$&$0.96^{+0.51}_{-0.28}$ & $\mathcal{U}(-10,10)$\\
      \noalign{\vskip 1mm}
      \hline
     {\bf Hyper-parameters}\\
	   \hline
	   \noalign{\vskip 1mm}
 	  $K_{\rm QP, VIS}$ [m\,s$^{-1}$] & $\cdots$ & $\cdots$  &  $1.80^{+0.48}_{-0.53}$  & $\cdots$ & $\cdots$ & $1.86^{+0.49}_{-0.40}$&$\cdots$ & $\mathcal{U}(0.001,10)$ \\
      \noalign{\vskip 1mm}
 	  $K_{\rm QP, NIR}$ [m\,s$^{-1}$] & $\cdots$ & $\cdots$  &  $\cdots$   & $\cdots$ & $\cdots$ & $0.04^{+0.54}_{-0.04}$&$\cdots$ & $\mathcal{U}(0.001,10)$ \\
      \noalign{\vskip 1mm} 
 	  $\lambda_{\rm QP}$ [d] & $\cdots$ & $\cdots$  &  $125^{+140}_{-72}$  & $\cdots$  & $\cdots$ & $102^{+111}_{-55}$&$\cdots$ & $\mathcal{U}(5,500)$ \\
      \noalign{\vskip 1mm}
 	  $w_{\rm QP}$ & $\cdots$ & $\cdots$  &  $0.28^{+0.18}_{-0.16}$   &$\cdots$ & $\cdots$ & $0.30^{+0.19}_{-0.11}$&$\cdots$ & $\mathcal{U}(0,1)$ \\
      \noalign{\vskip 1mm}
 	  $P_{\rm QP}$ [d] & $\cdots$ & $\cdots$   &  $42.1^{+2.5}_{-2.0}$   & $\cdots$ & $\cdots$ & $42.0^{+2.0}_{-1.5}$&$\cdots$ & $\mathcal{U}(28,56)$ \\
 	  \noalign{\vskip 1mm}
	  P$_0$ [d]& $\cdots $&$\cdots $ &  $\cdots $  &$\cdots$ & $\cdots$ &  $\cdots$&44.5$^{+4.6}_{-7.0}$ & $\mathcal{U}(1,1500)$\\
	  \noalign{\vskip 1mm}
    $\tau$ [d]& $\cdots$ &$\cdots$  &  $\cdots $  &$\cdots$ & $\cdots$ & $ \cdots$&$480^{+390}_{-412}$ & $\mathcal{U}(1,1500)$\\
    \noalign{\vskip 1mm}
    S$_0$& $\cdots$ & $\cdots$ &  $\cdots $  &$\cdots$ & $\cdots$ &$\cdots$  &0.46$^{+1.59}_{-0.39}$ & $\mathcal{U}(-15,15)$\\

	     \noalign{\vskip 1mm}
	  \hline

	  {\bf Fit quality}\\
	   \hline
	   \noalign{\vskip 1mm}
 	  $\sigma_{\rm O-C}$ [m\,s$^{-1}$] & 4.64 & 2.44 &  1.31  & 5.49 & 4.14 & 3.85&3.93 &$\cdots$\\
 	  \noalign{\vskip 1mm}
 	  $ \ln L$ & -400.0 & -160.4 &  -144.36  &-174.4 & -336.1 & -319.6 & -322.3 &$\cdots$\\ 
 	  \noalign{\vskip 1mm}
 	  $\Delta \ln L$ & 0 & $\cdots$ &  $\cdots $  & $\cdots$ & 63.9 & 80.4& 77.7 &$\cdots$\\
 	  \noalign{\vskip 1mm}
 
 	  \hline
     
     \end{tabular}}
     
   \end{table*}

\subsubsection{Keplerian modeling} \label{subsubsec:lspmkep}

Assuming that the strong signal at 14\,d has planetary origin, we determined the orbital parameters of the signal by fitting a Keplerian model with semi-amplitude (\emph{K}), eccentricity (\emph{e})\footnote{$e^2=(e \sin{\omega})^2 + (e \cos{\omega})^2$}, orbital period (\emph{P}), longitude of periastron ($\omega$), time of periastron passage\footnote{For a circular orbit, we define it as the time of maximum RV.} ($T_0$) and an RV offset for each channel ($\gamma_{\rm INS}$) as free parameters. Furthermore, we also allowed an adjustable RV jitter for each set of RVs ($\sigma_{\rm jit;INS}$) in the fit, as defined by \cite{Baluev_2009}.

We computed the uncertainties and final orbital parameters by running the Markov Chain Monte Carlo (MCMC) sampler \texttt{emcee} \citep{Foreman_2013}, with the natural logarithm of the model likelihood as the objective function. We run 500 chains of 15000 steps each, with a burn-in of 10000. The uncertainties were derived from the 1$\sigma$ (68.3\%) confidence interval of the posterior parameter distribution. We chose uniform priors as those shown in the last columns in Tables \ref{tab:lspmparams} and \ref{tab:gjparams}. 

The best model parameters for the VIS, NIR and the combined dataset can be found in Table \ref{tab:lspmparams}. To test the consistency of the signals in both datasets, we modeled a Keplerian orbit for each one separately. The Keplerian model parameters of the NIR channel, listed in the fifth column in Table \ref{tab:lspmparams}, resulted in an $e\sin{\omega}$ and $e\cos{\omega}$ compatible with zero, so we fixed their values to those obtained with the VIS channel, shown in the third column of Table \ref{tab:lspmparams}. All the orbital parameters are compatible within their respective uncertainties. The jitter in the VIS channel is slightly higher than in the NIR channel, indicating that the unaccounted errors are larger (e.g. from stellar variability).  

The best fit parameters of the combined dataset correspond to a planet with a minimum mass of $13.3^{+1.0}_{-1.1}$\,M$_{\oplus}$ orbiting its host star every $14.4399^{+0.0078}_{-0.0087}$\,d with an eccentricity of $0.183^{+0.062}_{-0.063}$ causing a RV semi-amplitude of 6.26$^{+0.41}_{-0.39}$\,m\,s$^{-1}$.

\subsubsection{Model comparison and signal stability} \label{subsubsec:lspmod}

To evaluate the statistical significance of our model, we computed the improvement in the natural logarithm of the likelihood ($\Delta \ln L$). The likelihood function is the probability distribution of the data fitting the model and depends on the adopted noise model \citep[see e.g.][]{baluev2013, Ribas_2018}.

\noindent Here we consider two different noise models: 
\begin{itemize}
    \item A white-noise model, which assumes that all the measurements are statistically independent from each other (null model). 
    \item A correlated-noise model using a Gaussian process (GP), which parametrizes the covariance function correlating all the measurements. \cite{rasmussen2005} describe many different covariance functions with different properties, among which the quasi-periodic harmonic oscillator has been widely used to disentangle planetary signals from stellar activity signals \citep{haywood2014, mortier2018, perger2019} or even to infer stellar rotation periods \citep{angus2018}.
\end{itemize}

\begin{figure}[!t]
\begin{center}
\includegraphics[width=\columnwidth]{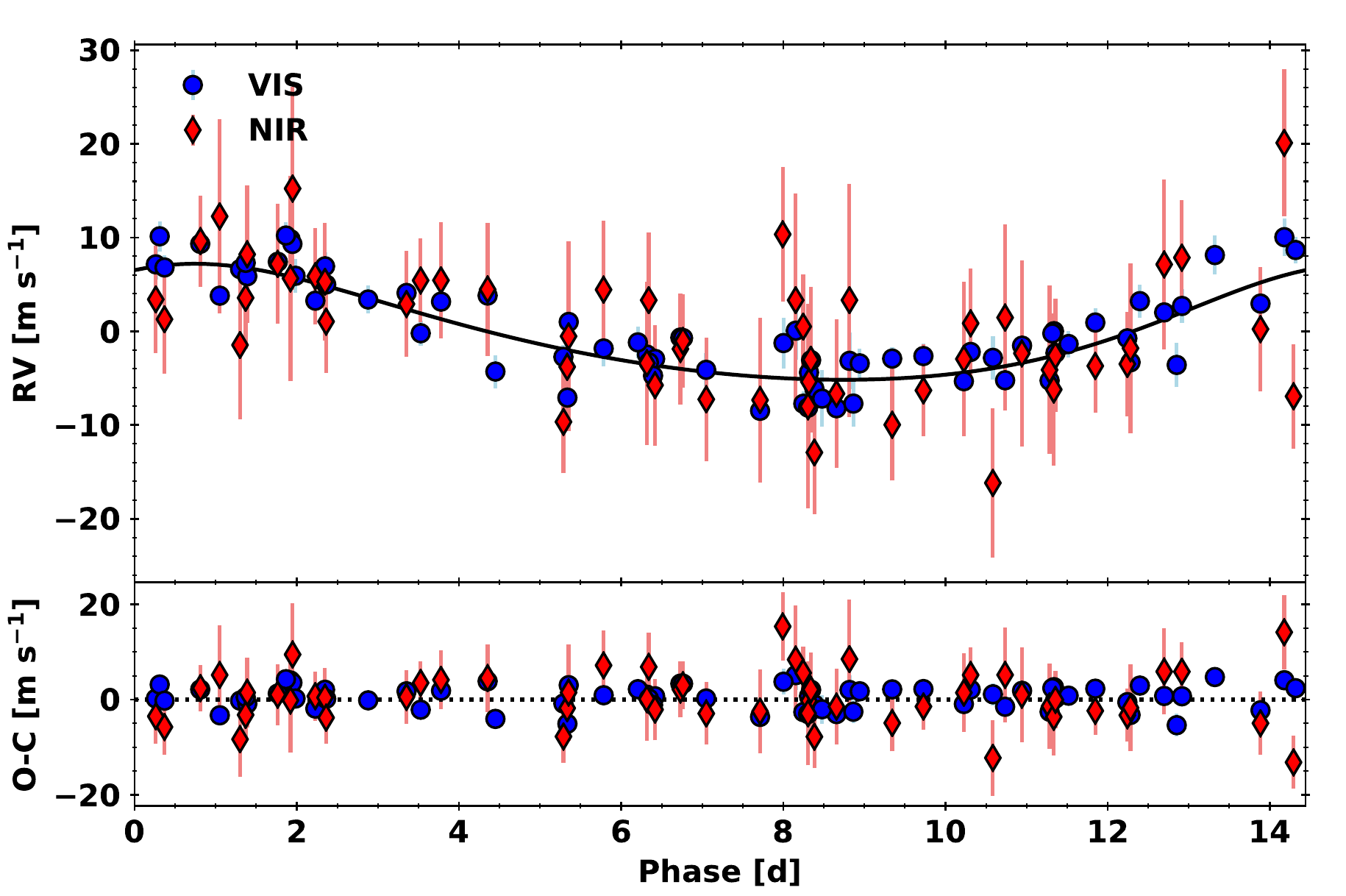}
\includegraphics[width=\columnwidth]{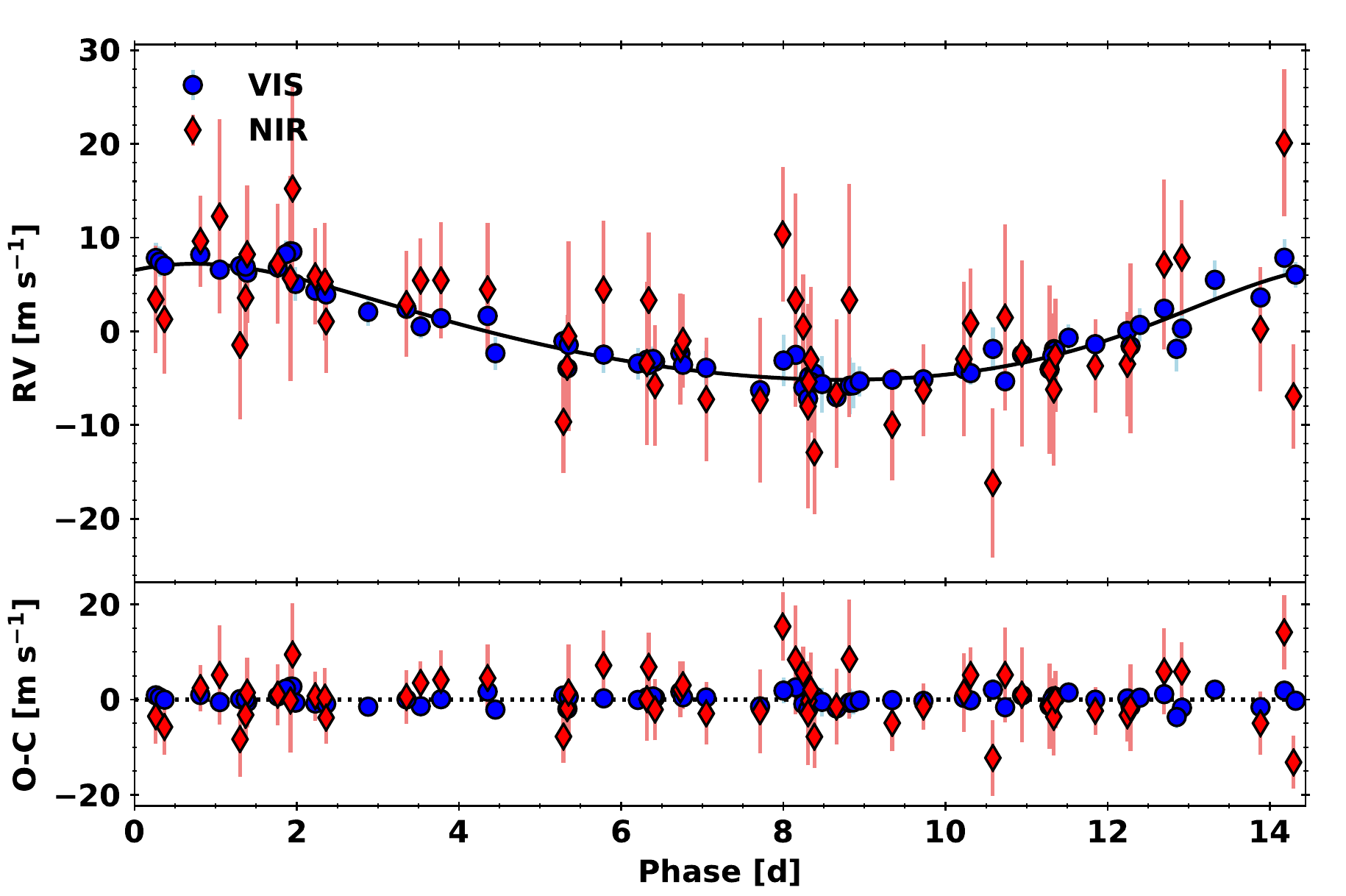}

\caption{\label{phasefoldlspm} LSPM~J2116+0234 RVs of the CARMENES VIS and NIR channels phase-folded with the best Keplerian fit (\textit{top}) and with the best Keplerian + Gaussian process fit (\textit{bottom}), with a 14.441\,d period.}
\end{center}
\end{figure}

For the null model, we assume that the data variation is produced by uncorrelated noise, and thus we only fit the offset and a jitter term for each different dataset. We computed the final parameters and uncertainties running the MCMC sampler \texttt{emcee}. The null model solution is shown in the second column of Table \ref{tab:lspmparams}, which gives jitter terms with higher values than for the other models, especially in the VIS, pointing towards an extra RV variability that is unaccounted by the measurement uncertainity. We obtain a best-fit $\ln L=-400.0$, which we take as the base value to compute the $\Delta \ln L$ of the other models in Table \ref{tab:lspmparams}.

We investigate the influence of the activity-induced RV variations on the determination of the orbital parameters by modeling together a Keplerian orbit and an activity term with a GP, which uses the quasi-periodic function as the covariance matrix. This function is characterized by four hyperparameters: (1) the output-scale amplitude $K_{\rm QP}$, which contains the amplitude of the RV variations due to the activity, (2) the decay time $\lambda_{QP}$, which is related to the lifetime of the active regions, (3) the smoothing parameter $w_{QP}$, which controls the high-frequency noise, and may be related to the number of spots and/or facule in the photosphere and (4) the periodicity of the correlations $P_{\rm QP}$, which is usually interpreted as the rotation of the star. We model the data of each instrument with separate GPs sharing all the parameters except the amplitude $K_{\rm QP}$, which should be different for instruments working in different wavelength ranges. This model is implemented using the \texttt{george} python library \citep{ambikasaran_2015}. We modeled the Keplerian orbit with the same parameters as those used in Section \ref{subsubsec:lspmkep}, including also a different RV offset and a jitter term for each instrument. All the parameters have been optimized simultaneously, and their solutions and uncertainties have been computed from the MCMC posterior distribution. We consider a model as tentative or as statistically significant over the null model if it reaches a FAP level of 1$\%$ or 0.1$\%$, respectively. These values corresponds to $\Delta \ln L$=15.1 and 18.7, which are computed from a bootstrap randomization of 5000 permutations of the datapoints.

\begin{figure}[!t]
\begin{center}
\includegraphics[width=\columnwidth,clip]{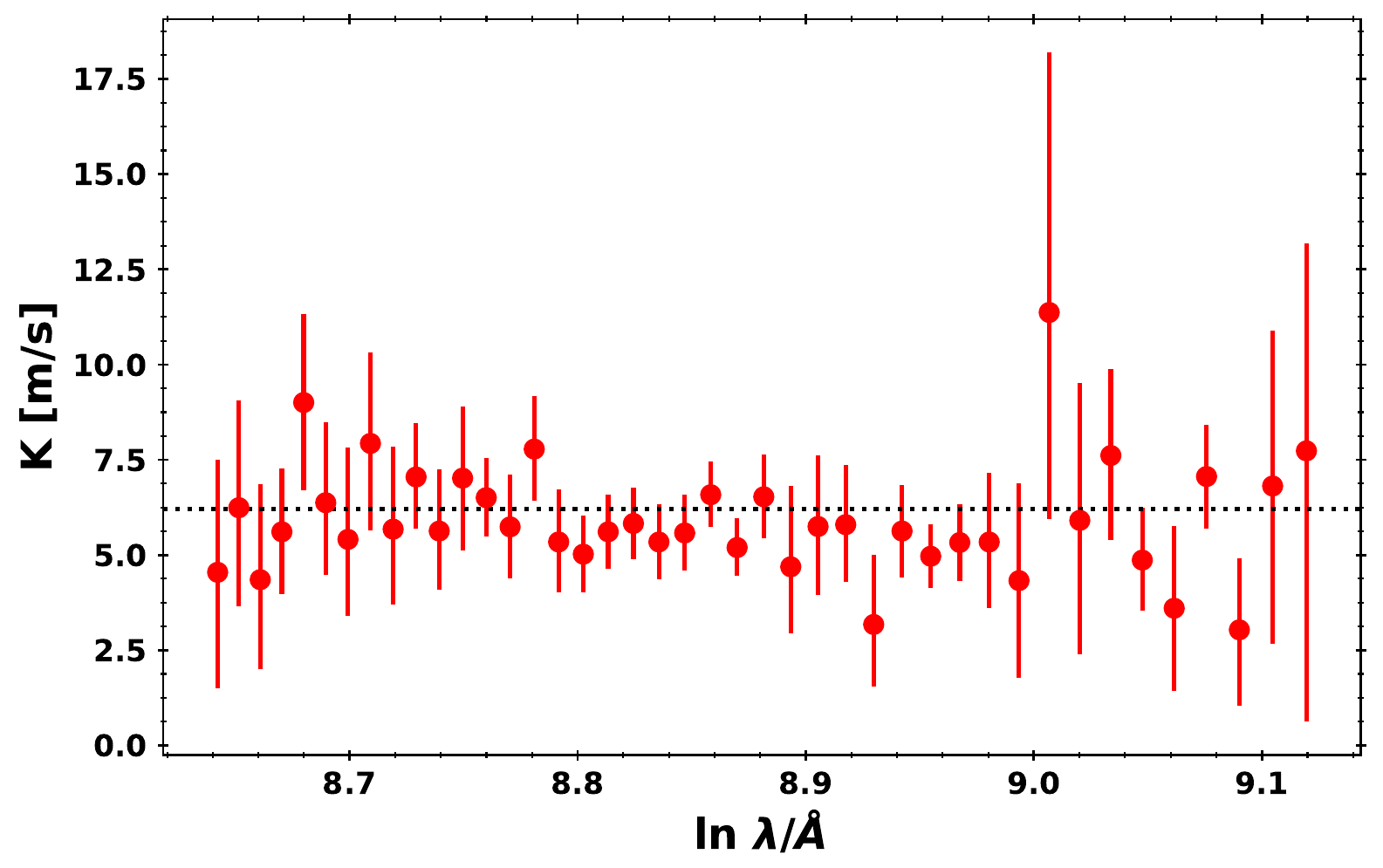}
\caption{\label{KstabLSPM} Stability of the semi-amplitude of the signal as a function of the logarithm of the wavelength at the center of the order used for LSPM~J2116+0234. The red circles show the semi-amplitudes of a Keplerian orbit fit using the velocities of each order individually. The values and uncertainties are computed from the MCMC posterior distribution. The black dashed line indicates the semi-amplitude found when using the combined RVs.}
\end{center}
\end{figure}

The parameters of the solution for the activity plus Keplerian modeling of the combined dataset of LSPM~J2116+0234 are shown in the seventh column of Table \ref{tab:lspmparams}. The orbital parameters are very similar to the values obtained with only a Keplerian model. We note that all the parameters agree within 1$\sigma$ uncertainties, except for the jitter term in the VIS channel, which is smaller. This is expected since we are adding an extra term modeling the activity which was included as part of the jitter term previously. As for the GP hyperparameters, we found a periodicity of P$_{\rm GP}=42.0^{+2.0}_{-1.5}$\,d, which is very similar to the periodicities found in the activity indicators. Hence, we consider the period of $\sim$42\,d as the rotation period of the star. The amplitude of the activity RV term variations is $1.86^{+0.49}_{-0.40}$\,m\,s$^{-1}$ in the VIS channel, while it is nearly zero in the NIR. This is in agreement with the expected decrease of the activity signal toward longer wavelengths, but could also be produced by the larger uncertainties of this channel. Note that the GP model with activity term is favored over the null model and over the Keplerian model, with an increase of the logarithm of the likelihood of 80.4 and 16.5, respectively.  

We investigate if the NIR RVs are modifying significantly the VIS RV solution. We fitted a Keplerian plus an activity term to the VIS RVs alone, whose best parameters are shown in the fourth column in Table \ref{tab:lspmparams}. All the parameters are compatible within one sigma. Thus, although the NIR RVs have internal uncertainties higher than the planet signal does not affect significantly the orbital solution given by the VIS RVs.

\begin{figure}
\begin{center}
\includegraphics[width=\columnwidth]{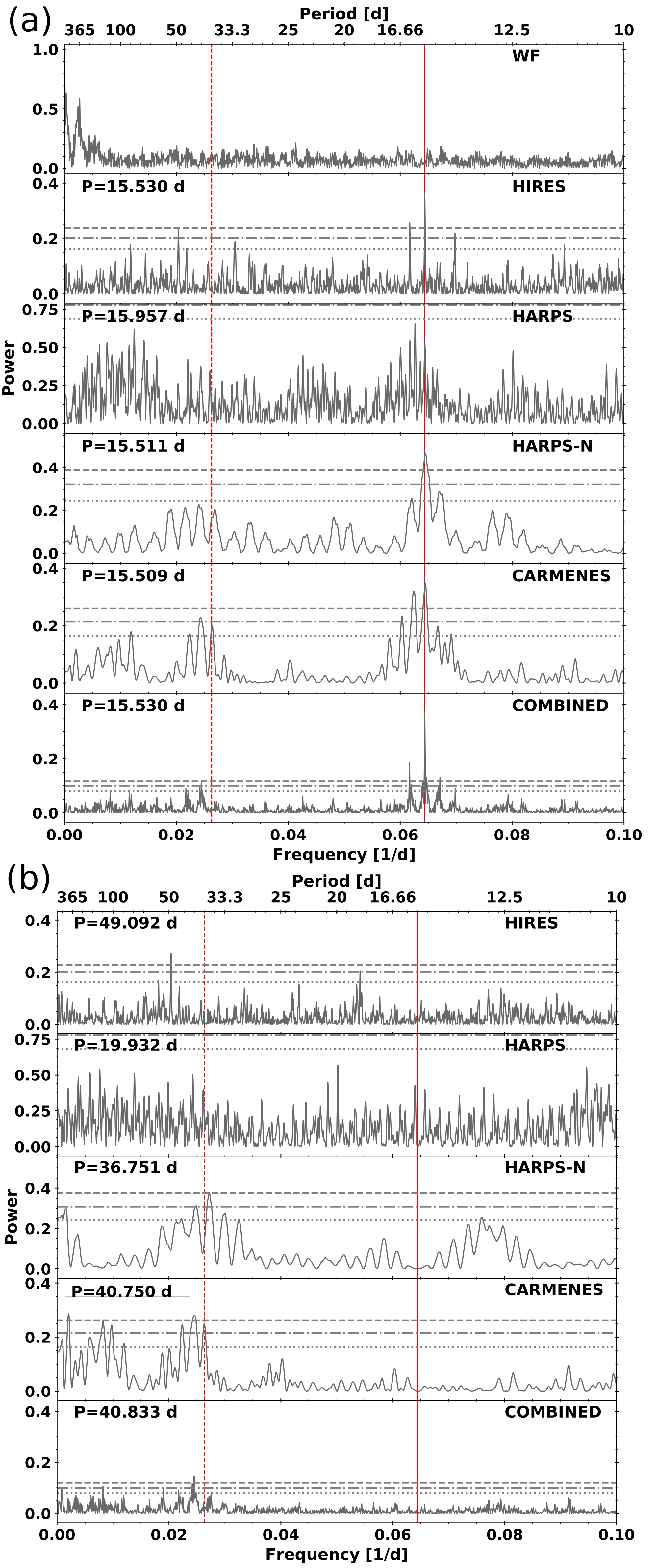}
\caption{\label{periodogram_all_gj686}(a) GLS periodograms of GJ~686 RV data. The top panel shows the WF of the combined dataset. The next four panels represent the HIRES, HARPS, HARPS-N and CARMENES data, respectively, and the bottom panel shows the periodogram of the combined dataset. The periods reported in each panel refer to the highest peak. Horizontal lines represent the bootstrapped FAP levels of 10, 1 and 0.1$\%$. The vertical solid and dashed red lines indicate the period of the proposed planet and estimated stellar rotation period at 15.53 and $\sim$38\,d, respectively. (b) GLS periodograms of the RV residuals after removing a sinusoid with the period found in (a). }
\end{center}
\end{figure}

As a consistency check between GP models, we also used \texttt{celerite}, the fast and scalable GP regression package \citep{celerite}, which uses as covariance matrix the model of a stochastically driven simple harmonic oscillator \citep{Kaminski_2018, Ribas_2018}. This model is characterized by the damping time $\tau$, the oscillator frequency P$_0$, and the height of the peak S$_0$. The parameter S$_0$ scales with the power of the associated frequency. The eighth column in Table \ref{tab:lspmparams} shows the best-fit parameters using this model, which gives compatible periods for the planet and rotation of the star.

\begin{figure}[!ht]
\begin{center}
\includegraphics[width=0.95\columnwidth,clip]{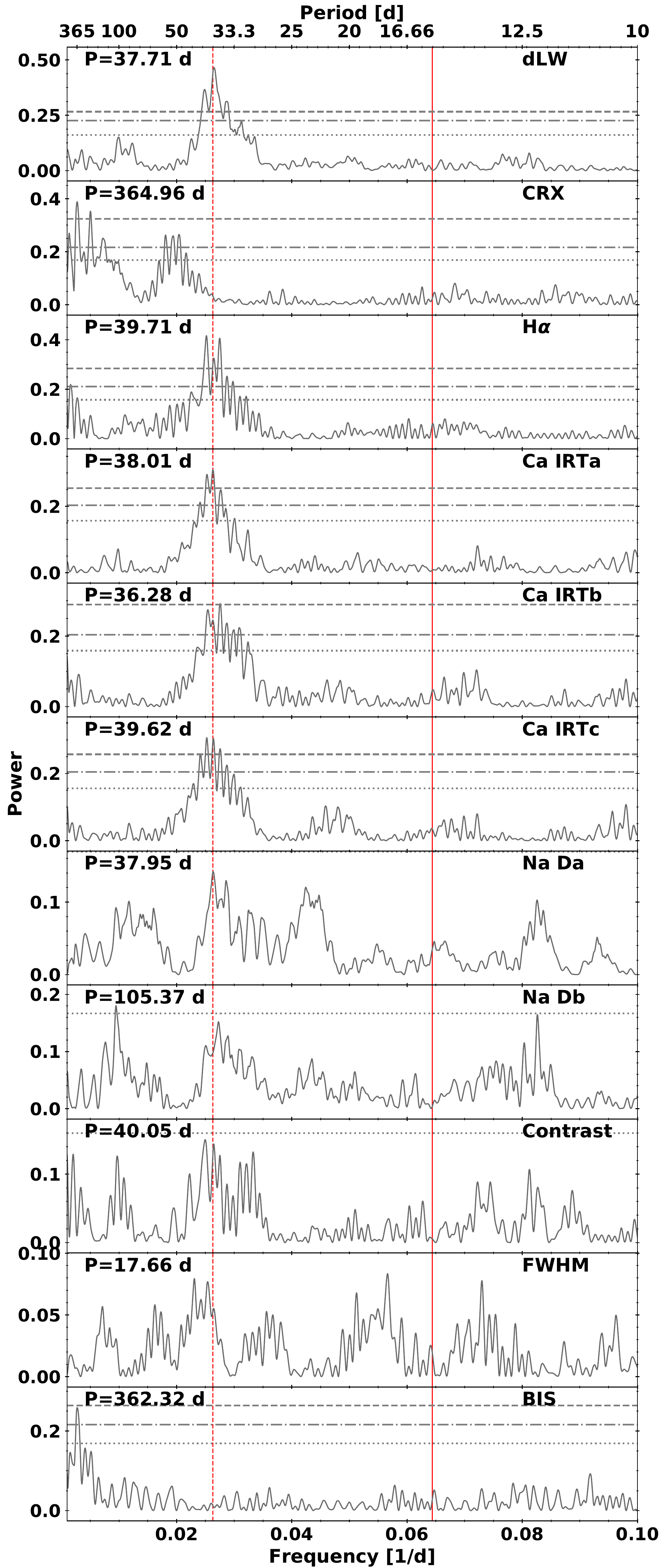}
\caption{\label{periodogram_activity_gj686} GLS Periodograms of the CARMENES activity indicators of GJ~686. The vertical solid line indicates the period of the suggested planet, while the vertical red dotted line marks the period attributed to the rotation. The periods reported in each panel refer to the highest peak. Horizontal lines represent the bootstrapped 10, 1 and 0.1$\%$ FAP levels.}
\end{center}
\end{figure}

In Fig.~\ref{phasefoldlspm}, we show the phased RV data with the best fit Keplerian model (top panel) and  Keplerian+GP (bottom panel). The posterior distribution of the parameters and their correlations of the best-model solution are depicted in Fig.~\ref{cornerlspm} in the appendix. The histograms suggest that all the orbital parameters follow a well behaved normal distribution, except for the expected correlation between $T_0$ and $e\sin{\omega}$ and $e\cos{\omega}$ due to their proximity to the degenerate solution at zero eccentricity.

Finally, we checked the stability of the signals throughout the wavelength range covered, to provide more evidence against a potential activity-induced origin of the signals. The RVs from each spectral order of CARMENES were used to compute the orbital parameters of a circular orbit and their uncertainties from the final posterior distribution of an MCMC sample of 500 walkers and 1000 steps. In Fig. \ref{KstabLSPM}, we show the resulting semi-amplitude of the circular orbit as a function of the logarithm of the wavelength at the center of each CARMENES order. All the values are consistent within 2$\sigma$ of the semi-amplitude found with the RVs of all the orders combined. Further, we do not see a decrease in amplitude towards longer wavelength, as it would be expected if the signal is activity-induced. Therefore, we conclude that the signal at the $\sim$14.44\,d period in LSPM~J2116+0234 is 
consistent with the planet hypothesis.

\subsection{GJ~686}
To investigate the RV variability of GJ~686, we computed the GLS periodograms of 
the HIRES, HARPS, HARPS-N and CARMENES measurements.  In Fig.~\ref{periodogram_all_gj686}a (top to bottom), we show the WF of the combined dataset, the periodograms of the HIRES, HARPS, HARPS-N and CARMENES RVs and of all data combined. We subtracted the mean value of each RV dataset to compute the periodogram of the combined dataset. The horizontal lines indicate the 10, 1 and 0.1$\%$ bootstrapped FAP levels.

Except for HARPS, all the instruments have the strongest signal at a period of 15.5\,d with FAP $<0.1\%$. The HARPS dataset shows a signal around 16.0\,d just below the 10$\%$ FAP, although we also notice an excess power at $\sim$15.5\,d. The periodogram of the combined dataset has a highly significant signal at $0.06439$\,d$^{-1}$ (15.53\,d). We notice an additional peak at  $0.06165$\,d$^{-1}$ (16.22\,d) with high significance, due to one yearly alias of the main signal, which is clearly observed in the WF.

\begin{figure*}
\begin{center}
\includegraphics[width=0.49\textwidth]{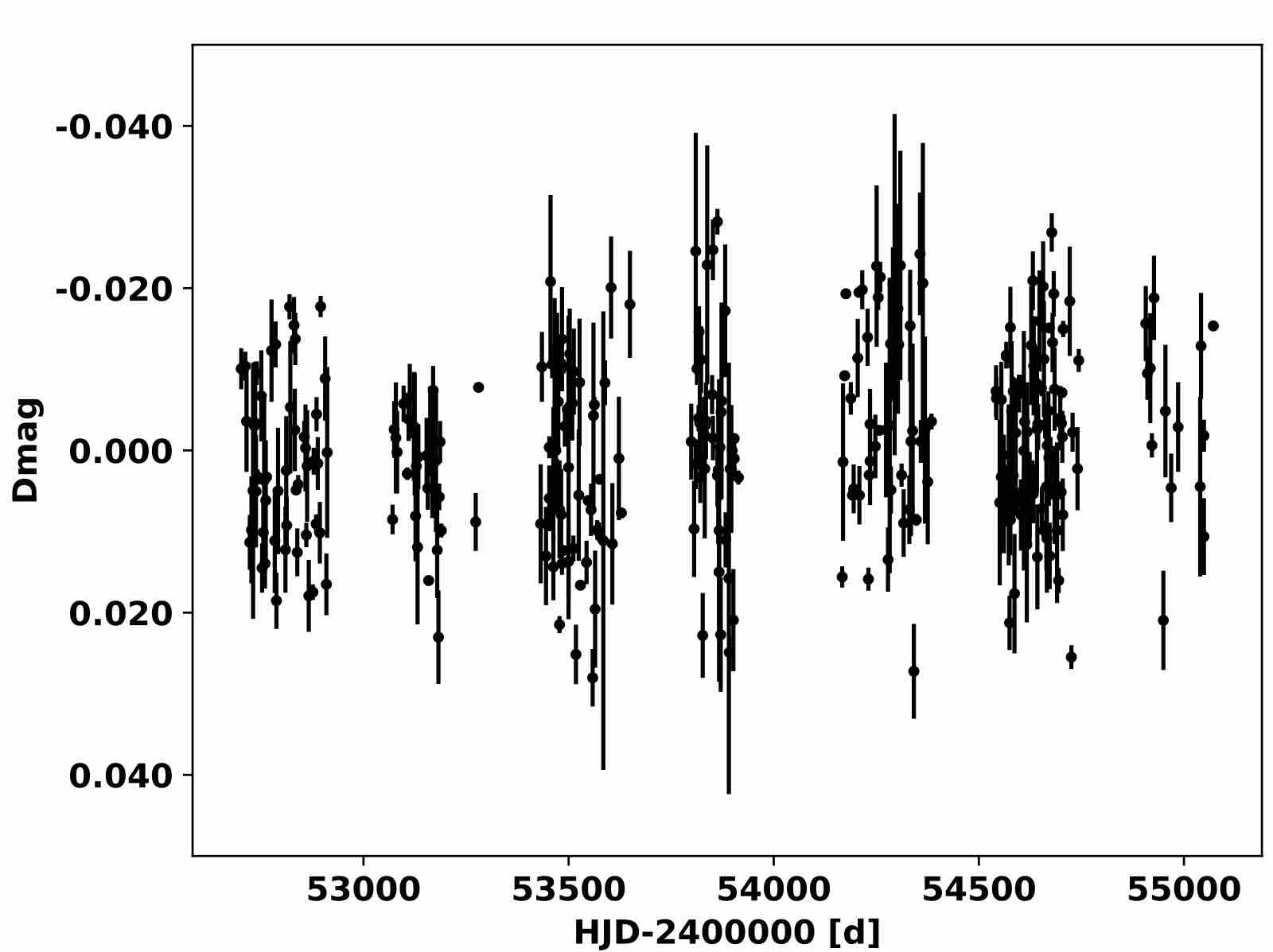}
\includegraphics[width=0.475\textwidth,clip]{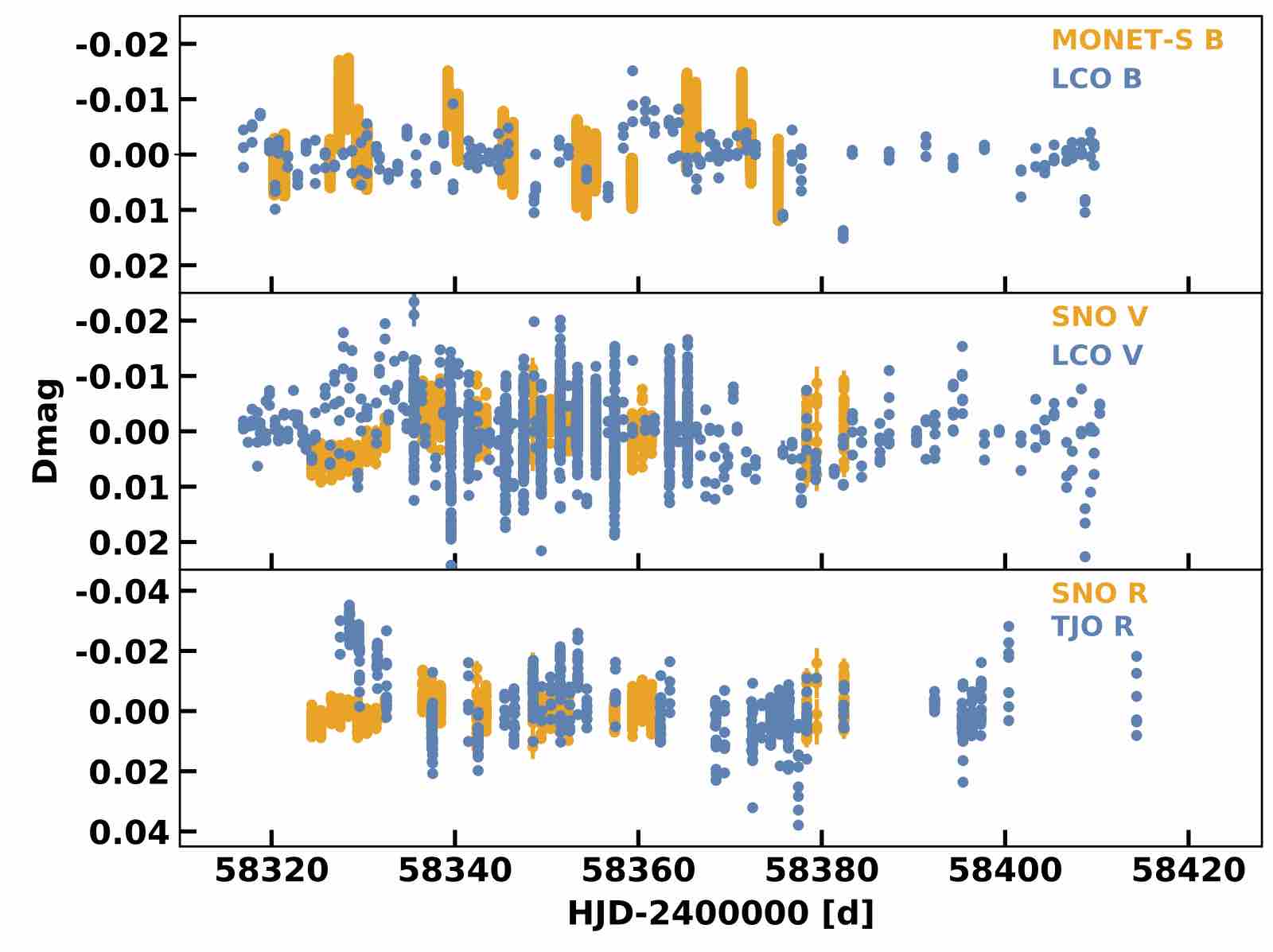}
\caption{\label{photGJ686} \textit{Left panel}: ASAS archival differential photometric data in $V$ filter.\textit{ Right panel}: Differential photometric follow-up of GJ~686 in $B$ filter with MONET-S and LCO (\textit{top panel}), in $V$ filter with SNO and LCO (middle panel) and in $R$ filter with SNO and TJO (\textit{bottom panel}). }
\end{center}
\end{figure*}

\begin{figure}
\begin{center}
\includegraphics[width=\columnwidth]{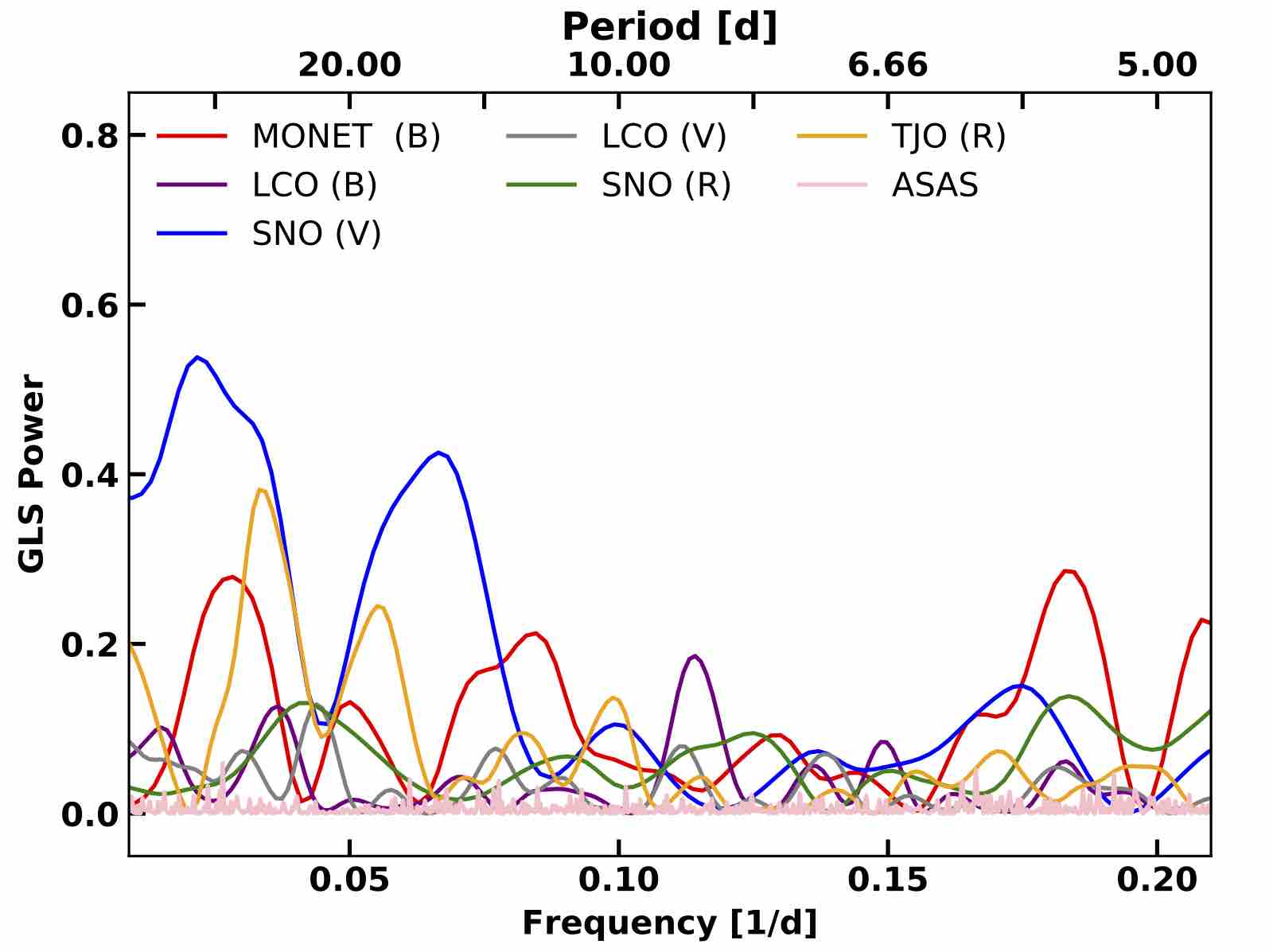}
\caption{\label{photperiodGJ686} GLS periodogram of all the photometric time series data sets of GJ~686. }
\end{center}
\end{figure}

We notice additional peaks with 1$\%$ FAP level at $\sim$49\,d in the HIRES data and at $\sim$41\,d in the CARMENES and combined datasets. This signal becomes significant in the combined dataset once the main signal is removed, as shown in Fig. \ref{periodogram_all_gj686}b, and there is also a signal at $\sim$37\,d in the HARPS-N data just below the 0.1$\%$ FAP. 
Further subtracting the main signal of the residuals, the CARMENES dataset has a significant signal at $\sim$500\,d, and two signals at the $1\%$ level around 1100 and 120\,d are seen in the combined dataset. They could be produced either by a long term activity cycle or an offset mismatch between datasets, or by a long period planet.  After another iteration of subtraction, no other significant signals remain in the residuals. 

Further, we searched for periodic signals in the GLS periodograms of the CARMENES activity indicators and photometric data. We investigated the possibility of the periodic RV variations being produced by stellar activity. As before, we consider a signal to be significant when it reaches a FAP below the $0.1\%$ level. Figure \ref{periodogram_activity_gj686} depicts the GLS periodograms of the dLW, CRX, the H$\alpha$, calcium infrared triplet (Ca IRT a, b and c), and sodium D doublet (Na D a and b) indices, and the contrast, FWHM and bisector span of the CCFs, as introduced in Sect.~\ref{sec:data}. There are several activity indicators with significant signals, among which there is a recurrence of signals between 36 and 40\,d. However, neither of the activity indices show any significant signals at or near 15.53\,d. 
The detected periods are in agreement with the periods found by Aff19 in the activity indicators. In particular, they found significant signals at 37 and 45\,d in the H$\alpha$ data from HARPS-N and HIRES, respectively, and a significant signal at 38\,d 
 in the S-index measured with HIRES. Furthermore, the activity time series during the last $\sim$100\,d of CARMENES observations 
 also shows a modulation of the signals at   $\sim$38\,d (see Fig.~\ref{gj686activitytimeseries} in the appendix).  The modulation may be caused by an epoch of high stellar activity. 
  Additionally, we also see significant signals in the CRX and BIS at 365\,d caused by a combination of the WF and a long-period trend.
 The GLS periodograms of the residuals of the activity indicators (Fig. \ref{GJ686GLSresidualsactivity}, in the appendix), only show significant signals at long periods in H$\alpha$, and a signal just below the 1\% FAP in the CRX. We further investigate the correlations between several activity indicators and the RV.  As seen in Fig.~\ref{correlation_activity_gj}, we find significant correlations (p-values below 0.05) for a  few activity indicators. This can be deduced by the strong modulations seen in Fig.~\ref{gj686activitytimeseries}. We do not find correlations with the CRX or the Na \textsc{i} D lines, which might be only due to shifts in phase \citep{perger2019}. Given that the signals at $\sim$38\,d are present in both RV and activity indicators, we attribute the variability to the stellar rotation period. 
 
We plot the available photometric time series in Fig \ref{photGJ686},  to investigate further the stellar rotation period. The GLS periodogram of the available photometry is depicted in Figure \ref{photperiodGJ686}. The MONET and ASAS photometry data show peaks at around 0.0279\,d$^{-1}$ (35.83\,d) and 0.0264\,d$^{-1}$ (37.87\,d). The $V$ band observations with SNO show a peak at $\sim$45\,d, although the broad amplitude of the peak makes it also compatible with the $\sim$38\,d signal observed in the activity indicators. However, the $R$ band observations with SNO and the $V$ band observations with LCO have peaks around 22\,d. Since they do not have a counterpart in the RV activity indicators, the nature of these peaks is not clear. 
Finally, we note that the TJO $R$ band and LCO $B$ band photometry have signals around $\sim$29\,d, which may be caused by the lunar period. In fact, the S/N of the data is strongly modulated with a period of $\sim$29\,d, supporting this hypothesis. No other significant signals remain after the subtraction of this periodicity.  

As a summary, based on the signals found in MONET and ASAS photometry and the activity indicators, we conclude that the rotation period of the star likely lies within the range 36--40\,d. Further, we have not found any significant signal at $\sim$15.53\,d in the photometry and the activity indicators, and thus, this signal probably has a planetary origin.

\begin{figure}
\begin{center}
\includegraphics[width=\columnwidth]{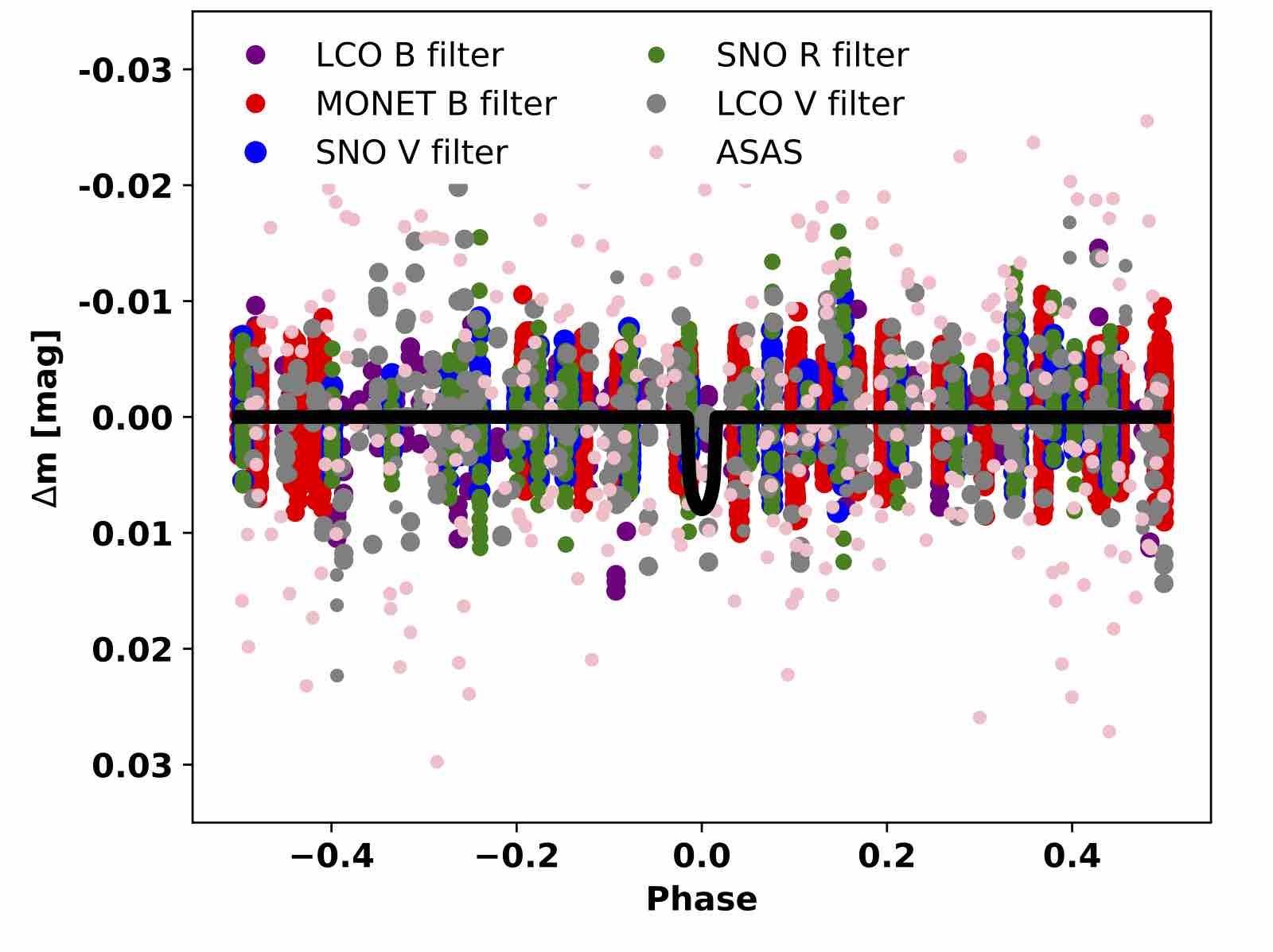}
\caption{\label{transit} Various photometric datasets phase-folded to 15.53\,d, the orbital period of the planet. Depicted in black is the transit model showing the expected signal and maximum transit depth.}
\end{center}
\end{figure}

\subsubsection{Keplerian modeling} \label{subsubsec:gjkep}

Assuming a planetary origin of the 15.5-day signal, we fitted a Keplerian and a sinusoidal model to the combined RVs. We computed the orbital parameters and uncertainties using the MCMC technique to infer the posterior distribution of the fitted parameters. The best fit models and their uncertainties are shown in the fourth and fifth column of Table \ref{tab:gjparams}, respectively. We compared the $\Delta \ln L$ for both models, although the solutions were statistically equivalent ($\Delta \ln L<1$). We adopted a circular model since $e\sin{\omega}$ and $e\cos{\omega}$ are consistent with zero within one sigma. The final parameters of the circular orbit give a planet with a period of $15.5311\pm0.0015$\,d at $0.0917^{+0.0024}_{-0.0023}$\,au, which produces an RV semi-amplitude of $2.83\pm0.22$\,m\,s$^{-1}$. Using the stellar mass given in Table \ref{tab:tab1}, we derive a minimum planet mass of $m_p \sin{i}=6.24^{+0.58}_{-0.59}$\,M$_{\oplus}$.

Using the derived orbital parameters, we investigated in detail all the accumulated photometric data for a possible planetary transit signature. The transit probability of GJ~686~b is 2.1$\%$, and the transit duration would be 2.5\,h. We detrended the photometric time series data and performed the Box-fitting Least Square (BLS) periodogram \citep{kovacs_2002}. We found no significant signal. Furthermore, in Figure~\ref{transit} all photometric data phase-folded to the planetary period of 15.53\,d along with an example of the expected transit signal. 
Since there are gaps in the photometric data, the estimated transit duration is $\approx$2.5\,h and the uncertainties of the transit window are large.


\begin{table*}
\caption{Best-fit parameters to different models of the planetary system Gl~686~b. }
\label{tab:gjparams}
\centering %
\resizebox{\textwidth}{!}{\begin{tabular}{lccccccc}
\hline \hline 
\noalign{\vskip 1mm}
\multirow{4}{*}{} & \multicolumn{6}{c}{Gl~686~b}\\
\cline{2-8}
\noalign{\vskip 1mm}
& Aff19 & \multicolumn{6}{c}{This work}\\
\cline{3-8}
& \multirow{2}{*}{\texttt{George}}  & \multirow{2}{*}{null model} & \multirow{2}{*}{Keplerian} & \multirow{2}{*}{Circular} & GP + Keplerian & GP + Keplerian & \multirow{2}{*}{Prior} \\
&  & & &  &  \texttt{George} &  \texttt{Celerite} & \\
	   	   \hline 
	  {\bf Planetary parameters}\\
	   \hline
	   \noalign{\vskip 1mm}
$P$ [d] & $15.5321^{+0.0017}_{-0.0017}$ & $\cdots$&$15.5311^{+0.0015}_{-0.0017}$ & $15.5311^{+0.0015}_{-0.0015}$ & $15.5314^{+0.0015}_{-0.0014}$& $15.5309^{+0.0017}_{-0.0015}$ & $\mathcal{U}(10,20)$\\
 	  \noalign{\vskip 1mm}
$T_0$ [JD-2450000]& $7805.69^{+0.28}_{-0.28}$ & $\cdots$ & $605.8^{+9.5}_{-11.5}$ & $610.83^{+0.63}_{-0.61}$ & $606.8^{+1.8}_{-2.3}$ & $605.2^{+4.1}_{-3.1}$ & $\mathcal{U}(585,625)$\\
 	  \noalign{\vskip 1mm}
$K$ [m\,s$^{-1}$] & $3.29^{+0.31}_{-0.32}$ & $\cdots$ &$2.85^{+0.21}_{-0.22}$ &$2.83^{+0.22}_{-0.22}$ &$3.02^{+0.18}_{-0.20}$&$3.11^{+0.28}_{-0.29}$ & $\mathcal{U}(0,20)$\\
 	  \noalign{\vskip 1mm}
$e\sin{\omega}$ & $\cdots$ & $\cdots$ & $-0.019^{+0.092}_{-0.098}$ & $\cdots$ & $-0.077^{+0.056}_{-0.058}$ &$0.009^{+0.007}_{-0.006}$ & $\mathcal{U}(-1,1)$\\
 	  \noalign{\vskip 1mm}
$e\cos{\omega}$ & $\cdots$ & $\cdots$ & $-0.012^{+0.070}_{-0.082}$ & $\cdots$ & $0.001^{+0.056}_{-0.064}$ &$0.079^{+0.060}_{-0.051}$ & $\mathcal{U}(-1,1)$\\
 	  \noalign{\vskip 1mm}
$a$ [AU] & $0.091\pm0.004$& $\cdots$ & $0.0917^{+0.0024}_{-0.0023}$ & $0.0917^{+0.0024}_{-0.0023}$ & $0.0917^{+0.0024}_{-0.0023}$&$0.0917^{+0.0023}_{-0.0023}$ &$\cdots$\\
 	  \noalign{\vskip 1mm}
$m_{\rm{p}}$ $\sin i$ [M$_{\oplus}$] & $7.1\pm0.9$ & $\cdots$ & $6.22^{+0.60}_{-0.61}$ & $6.24^{+0.58}_{-0.59}$ & $6.64^{+0.53}_{-0.54}$&$6.89^{+0.89}_{-0.87}$ &$\cdots$\\
 	  \noalign{\vskip 1mm}
	  \hline
	  {\bf RV offsets and jitter}\\
	   \hline
	   \noalign{\vskip 1mm}
$\gamma_{\rm HIRES}$ [m\,s$^{-1}$] & $0.65^{+0.52}_{-0.49}$ & $-0.12^{+0.35}_{-0.37}$ & $-0.05^{+0.32}_{-0.32}$ & $-0.08^{+0.32}_{-0.33}$ & $0.07^{+0.56}_{-0.57}$&$0.05^{+0.45}_{-0.47}$ & $\mathcal{U}(-100,100)$\\
 	  \noalign{\vskip 1mm}
$\gamma_{\rm HARPS}$ [m\,s$^{-1}$] & $-0.33^{+0.60}_{-0.61}$ & $0.15^{+0.56}_{-0.53}$ & $0.11^{+0.46}_{-0.46}$ & $0.12^{+0.40}_{-0.43}$ & $0.59^{+0.63}_{-0.66}$&$0.12^{+0.72}_{-0.80}$ & $\mathcal{U}(-100,100)$\\
 	  \noalign{\vskip 1mm}
$\gamma_{\rm HARPS-N}$ [m\,s$^{-1}$] &$-0.41^{+0.53}_{-0.63}$& $-0.19^{+0.38}_{-0.39}$& $-0.11^{+0.29}_{-0.28}$ & $-0.10^{+0.29}_{-0.28}$ & $-0.41^{+0.68}_{-0.64}$&$-0.33^{+0.41}_{-0.47}$ & $\mathcal{U}(-100,100)$\\
 	  \noalign{\vskip 1mm}
$\gamma_{\rm CARM}$ [m\,s$^{-1}$] &$\cdots$ & $-0.34^{+0.32}_{-0.31}$ &$-0.44^{+0.28}_{-0.28}$ & $-0.43^{+0.26}_{-0.26}$ & $-1.11^{+0.63}_{-0.65}$&$-1.09^{+0.66}_{-0.67}$ & $\mathcal{U}(-100,100)$\\ 	  
\noalign{\vskip 1mm}
$\sigma_{\rm jit, HIRES}$ [m\,s$^{-1}$] &$0.51^{+0.47}_{-0.35}$& $3.68^{+0.31}_{-0.29}$ & $2.84^{+0.29}_{-0.27}$ &  $2.81^{+0.29}_{-0.27}$ &  $0.63^{+0.51}_{-0.44}$&0.55$^{+0.47}_{-0.53}$ & $\mathcal{U}(0,10)$\\
 	  \noalign{\vskip 1mm}
$\sigma_{\rm jit, HARPS}$ [m\,s$^{-1}$] &$0.67^{+0.46}_{-0.41}$& $2.44^{+0.50}_{-0.36}$ &  $1.68^{+0.42}_{-0.33}$ &  $1.66^{+0.38}_{-0.30}$ &  $0.83^{+0.48}_{-0.40}$&$1.30^{+0.31}_{-0.33}$ & $\mathcal{U}(0,10)$\\
 	  \noalign{\vskip 1mm}
$\sigma_{\rm jit, HARPS-N}$ [m\,s$^{-1}$] &$1.44^{+0.29}_{-0.26}$& $2.93^{+0.31}_{-0.26}$& $2.07^{+0.25}_{-0.21}$ &  $2.09^{+0.24}_{-0.21}$ &  $1.04^{+0.23}_{-0.22}$&1.14$^{+0.34}_{-0.44}$ & $\mathcal{U}(0,10)$\\
 	  \noalign{\vskip 1mm}
$\sigma_{\rm jit, CARM}$ [m\,s$^{-1}$] & $\cdots$ & $2.55^{+0.27}_{-0.27}$ &  $1.85^{+0.44}_{-0.32}$ &  $1.82^{+0.27}_{-0.26}$ &  $0.26^{+0.29}_{-0.17}$&1.49$^{+0.39}_{-0.47}$ & $\mathcal{U}(0,10)$\\
\noalign{\vskip 1mm}
      \hline
     {\bf Hyper-parameters}\\
	   \hline
	   \noalign{\vskip 1mm}
$K_{\rm QP, HIRES}$ [m\,s$^{-1}$] & $3.16^{+0.44}_{-0.40}$ & $\cdots$ & $\cdots$ &  $\cdots$ &  $3.24^{+0.50}_{-0.45}$&$\cdots$ & $\mathcal{U}(0.001,10)$\\
\noalign{\vskip 1mm}
$K_{\rm QP, HARPS}$ [m\,s$^{-1}$] & $1.76^{+0.31}_{-0.28}$ & $\cdots$ & $\cdots$ &  $\cdots$ &  $1.72^{+0.35}_{-0.28}$&$\cdots$ & $\mathcal{U}(0.001,10)$\\
\noalign{\vskip 1mm}
$K_{\rm QP, CARM}$ [m\,s$^{-1}$] & $\cdots$ & $\cdots$ &  $\cdots$ &  $\cdots$ &  $2.04^{+0.43}_{-0.34}$&$\cdots$ & $\mathcal{U}(0.001,10)$\\
\noalign{\vskip 1mm}
$\lambda_{\rm QP}$ [d] & $23^{+31}_{-18}$ & $\cdots$ &  $\cdots$ &  $\cdots$ &  $49^{+14}_{-11}$& $\cdots$  & $\mathcal{U}(5,500)$\\
\noalign{\vskip 1mm}
$w_{\rm QP}$ & $0.48^{+0.31}_{-0.18}$ & $\cdots$ &  $\cdots$ &  $\cdots$ &  $0.50^{+0.14}_{-0.10}$& $\cdots$  & $\mathcal{U}(0,1)$\\
\noalign{\vskip 1mm}
$P_{\rm QP}$ [d] & $37.0^{+5.5}_{-14.6}$ & $\cdots$&  $\cdots$ &  $\cdots$ &  $38.4^{+1.6}_{-1.3}$& $\cdots$ & $\mathcal{U}(20,50)$ \\
\noalign{\vskip 1mm}
P$_0$ [d]& $\cdots $&$\cdots $ &$\cdots$ & $\cdots$ &  $\cdots$&39.0$^{+3.2}_{-4.3}$ & $\mathcal{U}(1,1500)$\\
\noalign{\vskip 1mm}
$\tau$ [d]& $\cdots$ &$\cdots$  &$\cdots$ & $\cdots$ & $ \cdots$&820$^{+923}_{-792}$  & $\mathcal{U}(1,1500)$\\
\noalign{\vskip 1mm}
S$_0$& $\cdots$ & $\cdots$ &$\cdots$ & $\cdots$ &$\cdots$  &0.48$^{+0.96}_{-0.88}$  & $\mathcal{U}(-15,15)$\\
\noalign{\vskip 1mm}
	  \hline
	  {\bf Fit quality}\\
	   \hline
	   \noalign{\vskip 1mm}
$\sigma_{O-C}$ [m\,s$^{-1}$] & $\cdots$ & 3.47 & 2.81 & 2.80 & 1.36 & 1.49 &$\cdots$ \\
 	  \noalign{\vskip 1mm}
 $\ln L $ & $\cdots$  & -758.2 & -691.9 & -691.1 & -637.7 & -636.1 &$\cdots$ \\
 $\Delta \ln L $ & $\cdots$  & 0 & 66.3 & 67.1 & 120.5 & 122.1 &$\cdots$ \\
 \noalign{\vskip 1mm}
\hline
     
\end{tabular}}

\end{table*}

\begin{figure}[!t]
\begin{center}

\includegraphics[width=\columnwidth]{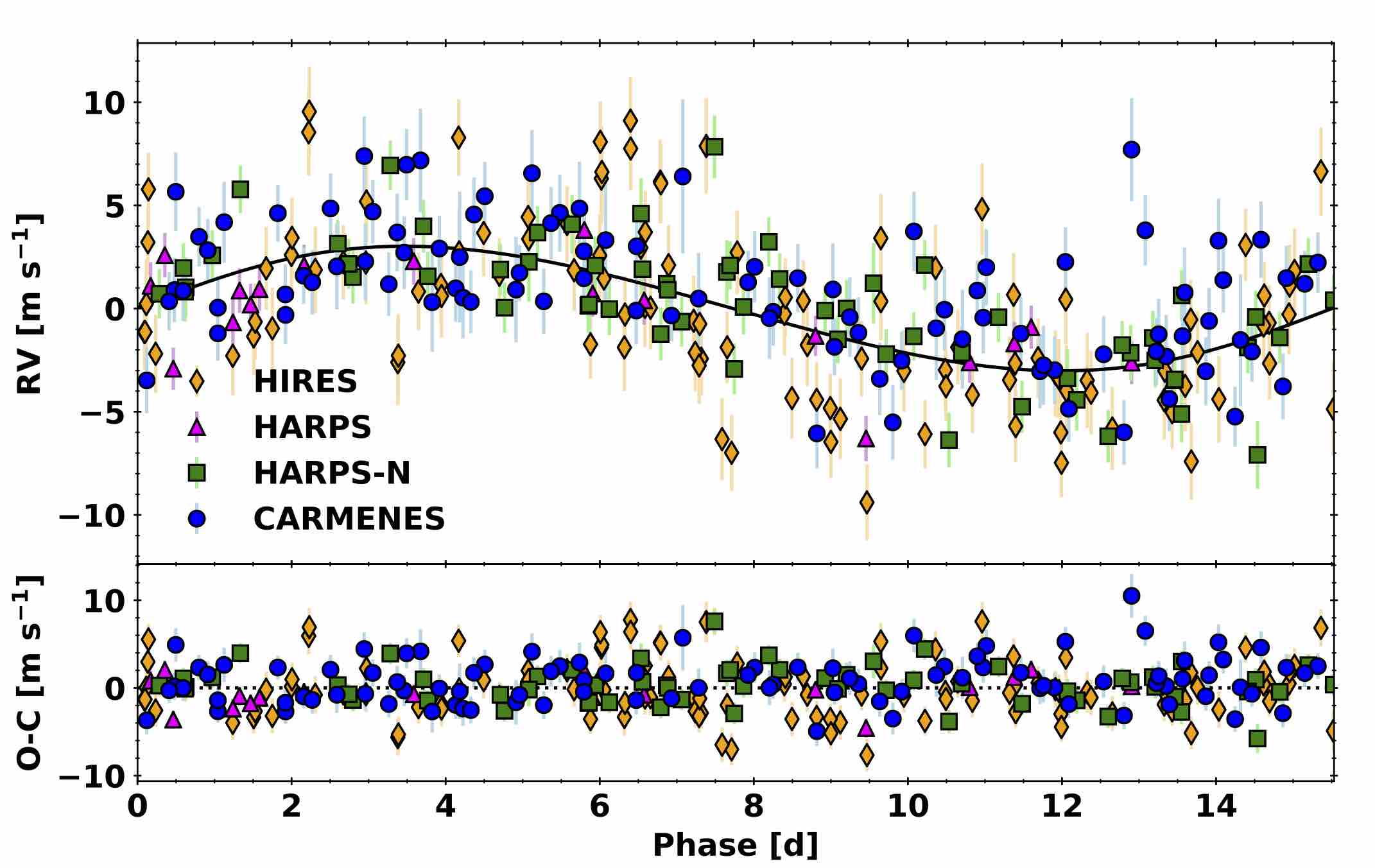}
\includegraphics[width=\columnwidth]{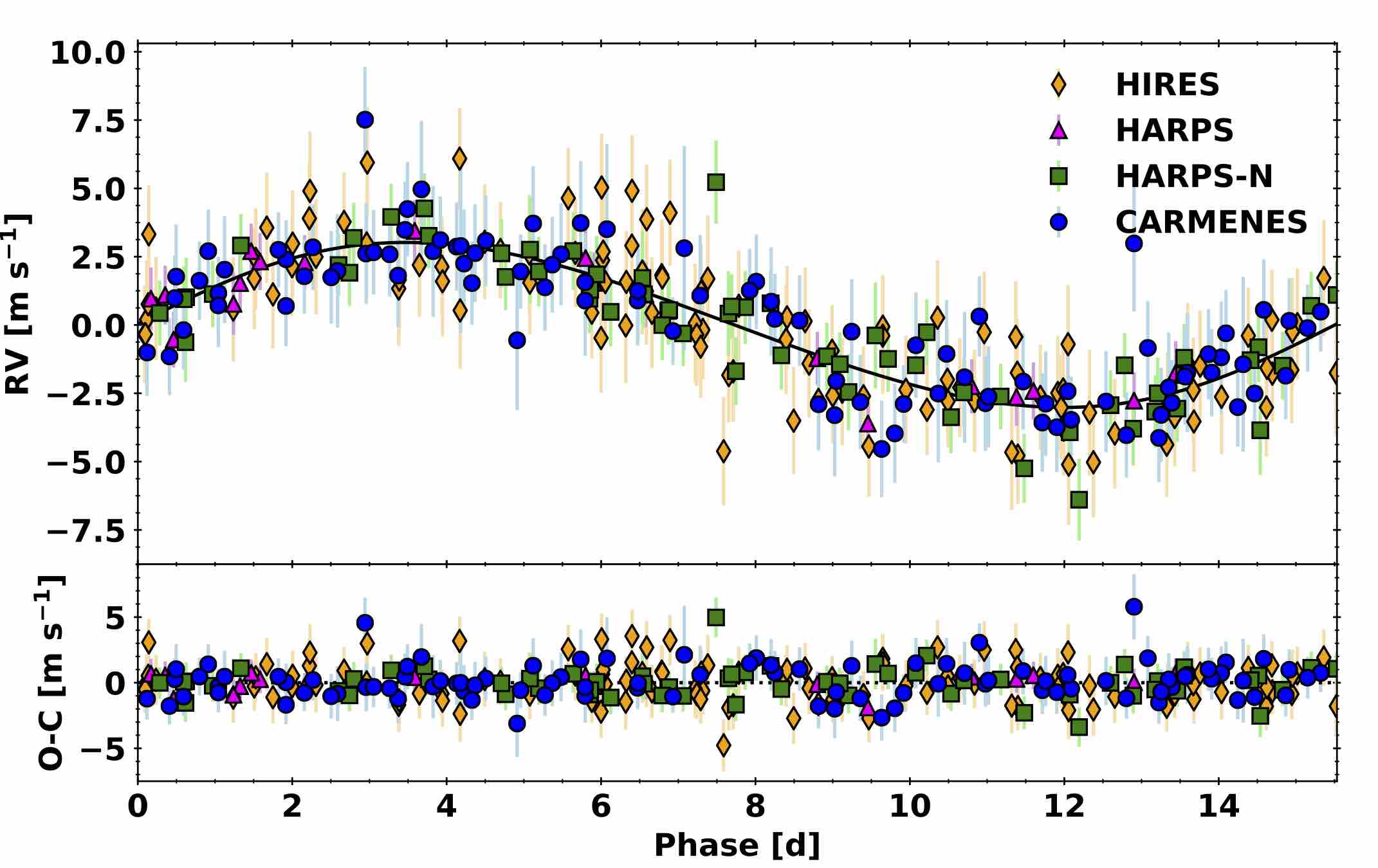}
\caption{\label{phasefoldgj} 
Phase-folded RV measurements with HIRES, HARPS, HARPS-N and CARMENES of GJ~686. 
\textit{Top panel}:  the best-fit Keplerian model with a 15.531\,d period. \textit{Bottom panel}: the best-fit  Keplerian + Gaussian process model.}
\end{center}
\end{figure}

\begin{figure}
\begin{center}
\includegraphics[width=\columnwidth,clip]{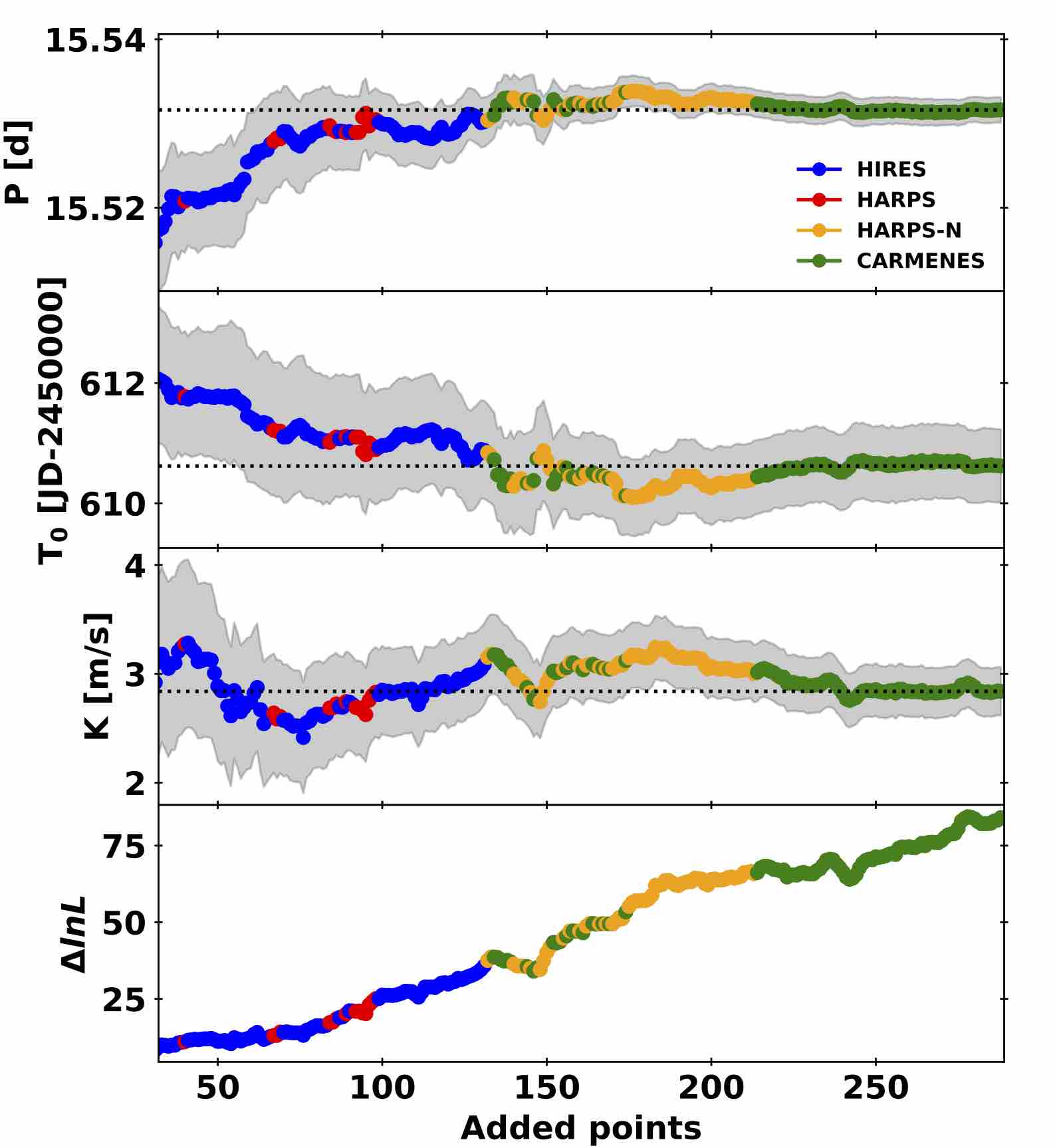}
\caption{\label{GJ686evo} Orbital parameters of a circular orbit and increment in the log likelihood as a function of the number of RV points, which are added chronologically. Each color represents a different instrument, and the gray shaded regions indicate the uncertainties computed from the MCMC posterior distribution. The first three panels show the period, time of maximum RV and semi-amplitude, respectively. The bottom panel shows the increment in log likelihood with respect to a fit to the mean value. 
The black dotted line indicates the orbital parameters obtained from the combined RVs. }
\end{center}
\end{figure}

\subsubsection{Model comparison and signal stability} \label{subsubsec:gjmod}

Using the same approach as for the previous system (see Sec. \ref{subsubsec:lspmod}), we modeled our RV data with a null-model described by white-noise, to evaluate the statistical significance of our models, and with a correlated-noise model that simultaneously fits the activity-induced RV variation and a Keplerian orbit.

The null model solution for GJ~686 is listed in the third column of Table \ref{tab:gjparams}. The best-fit null model yielded a $\ln L=-758.2$, which we used as a reference to compare the $\Delta \ln L$ against other models. For this star, the $\Delta \ln$L corresponding to a FAP of 1\% and 0.1\% are 13.4 and 17.4, respectively, as computed from a bootstrap randomization of 5000 permutations. 

We modeled the planetary signal of GJ~686 and the activity term with a GP using a quasi-periodic function as the covariance matrix. Since HARPS and HARPS-N are working in the same wavelength range, we modeled their RVs with the same amplitude $K_{\rm QP}$. We list the parameter solution of this approach in the sixth column of Table \ref{tab:gjparams}. With a period of $15.5314^{+0.0015}_{-0.0014}$\,d, the planetary signal has a RV semi-amplitude of 3.02$^{+0.18}_{-0.20}$\,m\,s$^{-1}$, which is slightly lower than the amplitude found in Aff19, but consistent within the uncertainties. Consequently, we also derived a smaller minimum mass of $6.64^{+0.53}_{-0.54}$\,M$_{\oplus}$. Unlike the model with only a planetary signal, in this case, we found a non-negligible eccentricity, of 0.077$^{+0.056}_{-0.058}$, computed from $e \sin{\omega}$ and $e \cos{\omega}$.  All the other orbital parameters are consistent within the respective uncertainties. We found a strong periodicity at $38.4^{+1.6}_{-1.3}$\,d in the GP hyperparameters, reducing the uncertainties in Aff19 
by about one order of magnitude. This periodicity is in agreement with the signals found in the activity indicators, therefore, we consider $\sim$38.4\,d as the rotation period of the star. 
Furthermore, we found a large increase in $\ln L$ with respect to the null model and also with respect to the circular orbit model, with $\Delta \ln L$ of 120.5 and 53.4, respectively.

We show the phase-folded RV data for GJ~686 with the best Keplerian fit in Fig.~\ref{phasefoldgj} (top panel), while the best Keplerian + Gaussian process fit is depicted in the bottom panel. Figure~\ref{cornergj} in the appendix shows the posterior distribution of the parameters and their correlations for GJ~686 with the planet + activity model. All the orbital parameters follow a well behaved normal distribution and there are no strong correlations between parameters, except for the correlation between $e\sin \omega$ and $e\cos \omega$ with $T_0$. 

Based on the orbital parameters obtained with the Keplerian plus activity model, we note a small decrease in the RV semi-amplitude of GJ~686 with respect to that found in Aff19, and further decrease when compared to the tentative signal found in \cite{butler_2017}, of $3.46\pm0.56$\,m\,s$^{-1}$. Although they are consistent within their respective uncertainties, the amplitudes of the signals are not directly comparable, since \cite{butler_2017} model only a Keplerian orbit, whereas Aff19 model a circular orbit plus an activity term, and in this work, we model a Keplerian + activity term. Hence, we check the stability of the signal over time using the same model. Here we use a circular orbit, fixing the offsets and jitter terms to the values found in the fifth column of Table \ref{tab:gjparams}. We iteratively add the RV data in chronological order and compute the final parameters and uncertainties from the parameter distribution of an MCMC chain of 1000 steps. The results are shown in Fig. \ref{GJ686evo}, with each color representing the instrument with which the RV measurement was made, and the grey shaded regions indicate the uncertainties. As observed, the amplitude is almost always compatible within uncertainties, and the period and time of periastron passage are stable after the addition of the $\sim$60$^{\rm th}$ measurement.

\begin{figure}
\begin{center}
\includegraphics[width=0.49\textwidth,clip]{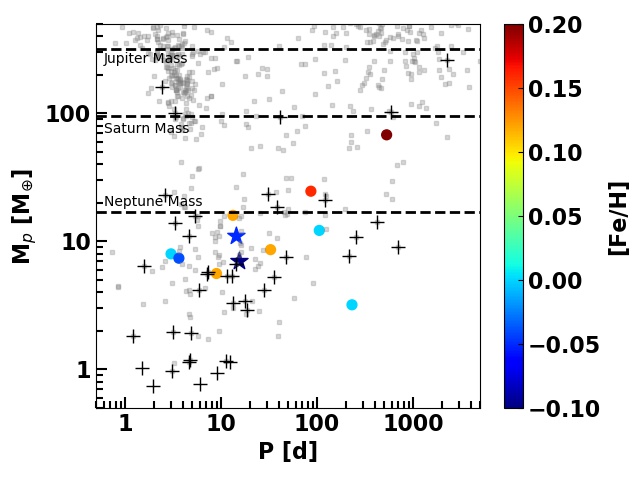}
\includegraphics[width=0.47\textwidth,clip]{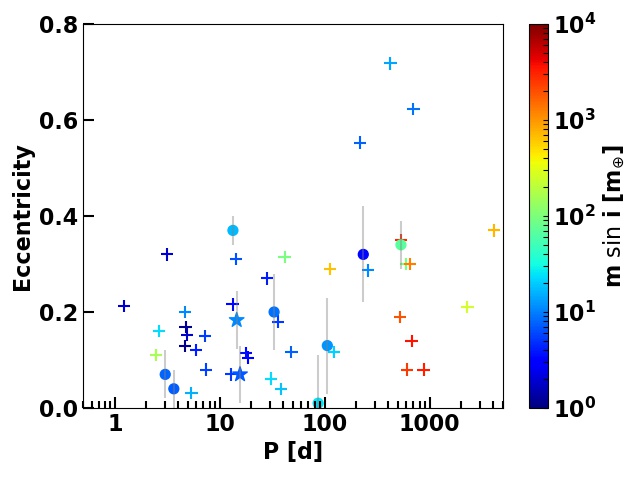}
\caption[\textit{Top panel}: CARMENES measured planetary masses against the planetary orbital periods are represented as color-coded circles using the  stellar metallicity. The two planetary systems discussed in this paper are represented as stars. The scattered grey squares and black crosses are all known exoplanets and planets orbiting M stars, respectively. The horizontal lines represent the mass of solar system planets. \textit{Bottom panel}: Eccentricity against the planetary orbital period of exoplanets around M stars (crosses). CARMENES planets are represented as circles and stars.]{\label{carmdis} \textit{Top panel}: CARMENES measured planetary masses against the planetary orbital periods are represented as color-coded circles using the stellar metallicity ([Fe/H]). The two planetary systems discussed in this paper are represented as stars. Other discoveries by CARMENES are depicted as circles. The scattered grey squares and black crosses are all known exoplanets\footnotemark ~and planets orbiting M stars, respectively. The horizontal lines represent the mass of solar system planets. \textit{Bottom panel}: Eccentricity against the planetary orbital period of exoplanets around M stars (crosses). CARMENES planets are represented as circles and stars. } 
\end{center}
\end{figure}
\footnotetext{\url{http://exoplanets.org}}

Finally, we also modeled the tentative long period signal found in the periodogram of residuals 
of the combined RV dataset after removing the planetary signal and the rotation period. We do that by simultaneously fitting two Keplerian orbits and an activity term modeled with a GP. We constrained the prior to periods longer than 100\,d. This yielded a stable solution at a period of $1161^{+53}_{-81}$\,d and a semi-amplitude of $1.44^{+0.44}_{-0.54}$\,m\,s$^{-1}$, which would be produced by a planet of at least 13.4\,M$_{\oplus}$ at 1.65\,au. Nevertheless, adding a second planet is not statistically significant ($\Delta \ln{L}$=4.9), and the signal could also be produced by the different offsets for each instrument or by a long-period activity cycle. Further observations are required to fully characterize this tentative long period signal.

\subsection{Exoplanets from CARMENES and habitability}

According to \cite{Kopparapu_2013}, both stars have very similar habitable-zone locations, with optimistic inner limits  at 0.13 and 0.14\,au from the host stars LSPM~J2116+0234 and GJ~686, respectively. 
The planets are closer to the host star than the conservative habitable-zone limits\footnote{http://depts.washington.edu/naivpl/sites/default/files/hz.shtml} of 0.16--0.32\,au \citep{Kopparapu_2013}.
With semi-major axes of 0.087\,au for the mini-Neptune around LSPM~J2116+0234 and 0.092\,au for the super-Earth around GJ~686, the two planets 
receive almost three times the flux received at the Earth by the Sun.

Both LSPM~J2116+0234~b and GJ~686~b are in the lower part of the planetary mass vs orbital period diagram represented as star symbols in  
Fig. \ref{carmdis} (top panel).  
We also show the known exoplanets and the planets orbiting M dwarfs. 
This demonstrates the capability of CARMENES as an instrument to discover low-mass planets on both short and longer orbits around M dwarfs. Nearly $\sim$75\% of CARMENES discoveries are super-Earths or mini-Neptunes at a wide range of periods. We also note the lack of close-in massive planets around M stars. However, the two small planets reported in this paper with periods $>$10\,d are part of a rather large population of planets with similar characteristic. The host stars of these planets have the lowest metallicity among the CARMENES discoveries. In Fig. \ref{carmdis} (bottom panel), we show the distribution of the eccentricity of M-dwarf exoplanets as a function of orbital period. The plot also shows the CARMENES discoveries including systems discussed in the current paper. 
We note that the majority of the super-Earths or mini-Neptunes have an eccentricity $e<0.2$.

\section{Summary}\label{sec:summary}
In this study, we analyzed 72 and 57 RV measurements of the M3.0\,V star LSPM~J2116+0234 taken with the visible and NIR channels of the high-resolution CARMENES \'echelle spectrograph, respectively. We also confirmed and refined the orbital parameters of the super-Earth around the M1.0\,V star GJ~686 reported in \citet{affer2019} with the addition of 100 new RV CARMENES measurements. 

The analysis of the RVs from LSPM~J2116+0234 revealed a signal stable in wavelength at 14.44\,d not present in activity indicators, which we interpret as being caused by a planet with a minimum mass of 13.3\,M$_{\oplus}$ and a semi-major axis of 0.087\,au. 

To obtain better constraints on the properties of GJ~686~b derived in \citet{affer2019}, who used the available RVs from HIRES, HARPS and HARPS-N, we combined these data with the CARMENES-VIS RVs. We derived a slightly smaller and more precise RV semi-amplitude of 3.02\,m\,s$^{-1}$, resulting in a lower minimum mass of the planet, of 6.64\,M$_{\oplus}$. The orbital period of 15.5314\,d and a semi-major axis of 0.092\,au are quite similar. Contrary to the best-fit model in \citet{affer2019}, ours suggests a non-zero eccentricity, obtaining a value of 0.077. 

We used the photometric measurements and the activity indices to estimate of the rotation period of both LSPM~J2116+0234 and GJ~686. For both targets, a non-parametric stellar variability model was adopted to account for correlated noise caused by stellar magnetic activity. We simultaneously modeled the stellar variability and the planetary signals to obtain a self-consistent planetary solution. From this model, we determined the stellar rotation period to be 42.0\,d for LSPM~J2116+0234 and 38.4\,d for GJ~686. With the data currently available, the RV time series favor a single planet model for both LSPM~J2116+0234 and GJ~686. However, an additional longer period signal may be present in the GJ~686 data, 
whose nature and properties need to be characterized with more measurements.

\begin{acknowledgements}
 
 L.S. acknowledges support from the Deutsche Forschungsgemeinschaft under DFG DR 281/32-1. CARMENES is an instrument for the Centro Astron\'omico Hispano-Alem\'an de Calar Alto (CAHA, Almer\'ia, Spain).  CARMENES is funded by the German Max-Planck-Gesellschaft (MPG), the Spanish Consejo Superior de Investigaciones Científicas (CSIC), the European Union through FEDER/ERF FICTS-2011-02 funds, and the members of the CARMENES Consortium (Max-Planck-Institut für Astronomie, Instituto de Astrof\'isica de Andaluc\'ia, Landessternwarte K\"onigstuhl, Institut de Ci\'encies de l'Espai, Insitut für Astrophysik G\"ottingen, Universidad Complutense de Madrid, Th\"uringer Landessternwarte Tautenburg, Instituto de Astrof\'isica de Canarias, Hamburger Sternwarte, Centro de Astrobiolog\'ia and Centro Astron\'omico Hispano-Alem\'an), with additional contributions by the Spanish Ministry of Economy, the German Science Foundation through the Major Research Instrumentation Programme and DFG Research Unit FOR2544 "Blue Planets around Red Stars", the Klaus Tschira Stiftung, the states of Baden-W\"urttemberg and Niedersachsen, and by the Junta de Andaluc\'ia.  Data were partly obtained with the MONET/South telescope of the MOnitoring NEtwork of Telescopes, funded by the Alfried Krupp von Bohlen und Halbach Foundation, Essen, and operated by the Georg-August-Universit\"at G\"ottingen, the McDonald Observatory of the University of Texas at Austin, and the South African Astronomical Observatory. Data were partly collected with the 90-cm telescope at 
Sierra Nevada Observatory (SNO) operated by the Instituto de Astrof\'\i fica de Andaluc\'\i a (IAA). We acknowledge financial support from the Spanish Ministry for Science, Innovation and Universities (MCIU) AYA2015-69350-C3-2-P, ESP2016-80435-C2-1-R, ESP2016-80435-C2-2-R, AYA2016-79425-C3-1/2/3-P, ESP2017-87676-C05-02-R, ESP2017-87143-R, BES-2017-082610, SEV-2015-0548-17-2, Generalitat de Catalunya/CERCA programme; Agència de Gestió d’Ajuts Universitaris i de Recerca of the Generalitat de Catalunya through grant 2018FI$\_$B$\_$00188, and  the Israel Science Foundation through grant 848/16. This work makes use of data from the HARPS-N Project, a collaboration between the Astronomical Observatory of the Geneva University (lead), the CfA in Cambridge, the Universities of St. Andrews and Edinburgh, the Queens University of Belfast, and the TNG-INAF Observatory; from observations obtained at the W. M. Keck Observatory, which is operated as a scientific partnership among the California Institute of Technology, the University of California and the National Aeronautics and Space Administration. The Observatory was made possible by the generous financial support of the W. M. Keck Foundation; from observations collected at the European Organisation for Astronomical Research in the Southern Hemisphere under ESO programmes 183.C-0437(A) and 072.C-0488(E); from the European Space Agency (ESA) mission {\it Gaia} (\url{https://www.cosmos.esa.int/gaia}), processed by the {\it Gaia} Data Processing and Analysis Consortium (DPAC, \url{https://www.cosmos.esa.int/web/gaia/dpac/consortium}). Funding for the DPAC has been provided by national institutions, in particular the institutions participating in the {\it Gaia} Multilateral Agreement. 
\end{acknowledgements}


\bibliographystyle{aa} 
\bibliography{paper} 

\begin{appendix}
\section{Additional periodograms, data table, and MCMC posterior distributions}

\begin{figure}[!h]
\begin{center}
\includegraphics[width=0.95\columnwidth,clip]{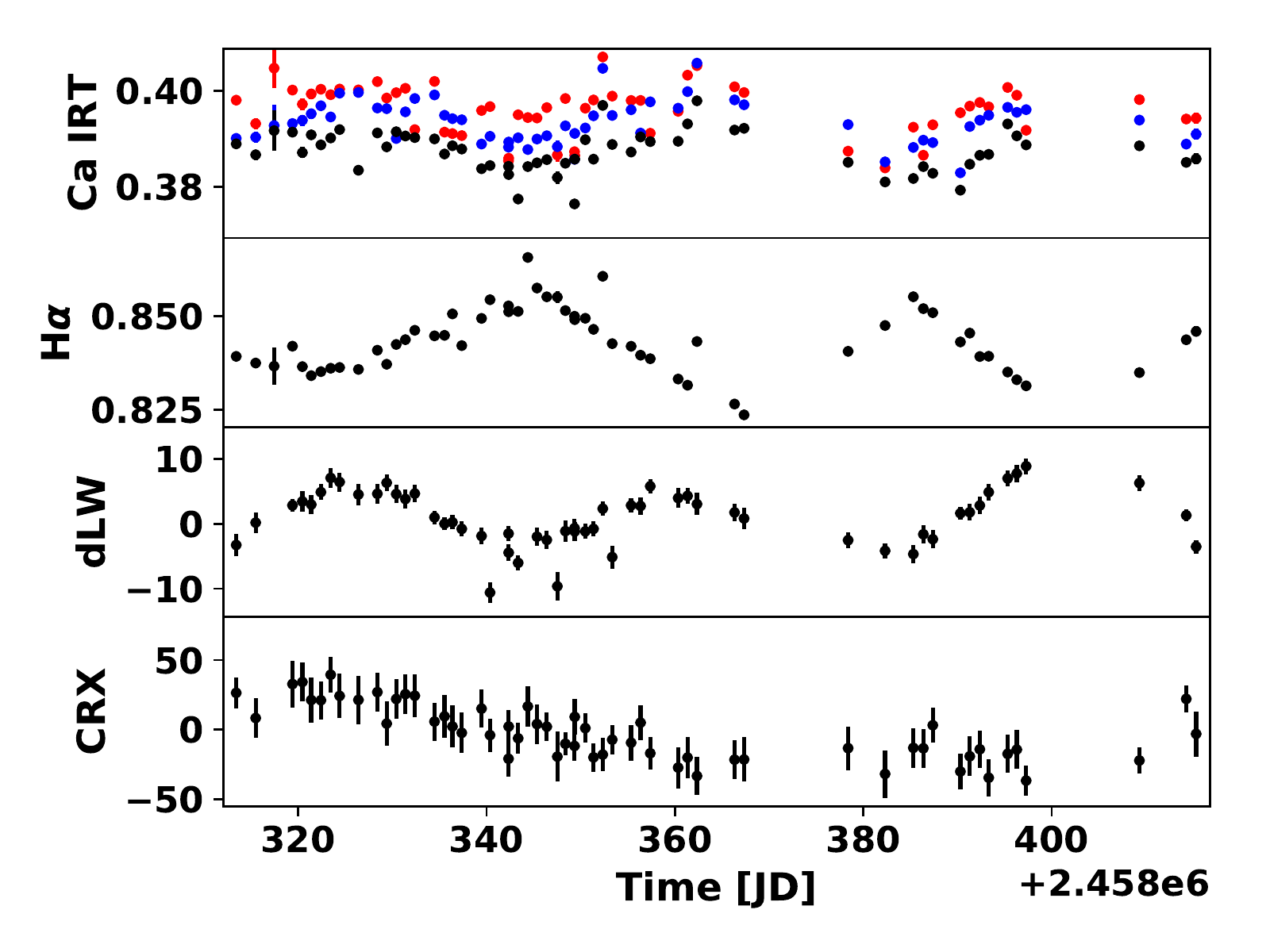}
\caption{\label{gj686activitytimeseries} Time series of activity indicators, dLW, and CRX for the last 100 days of observations of GJ~686.}
\end{center}
\end{figure}

\begin{figure}[!h]
\begin{center}
\includegraphics[width=\columnwidth,clip]{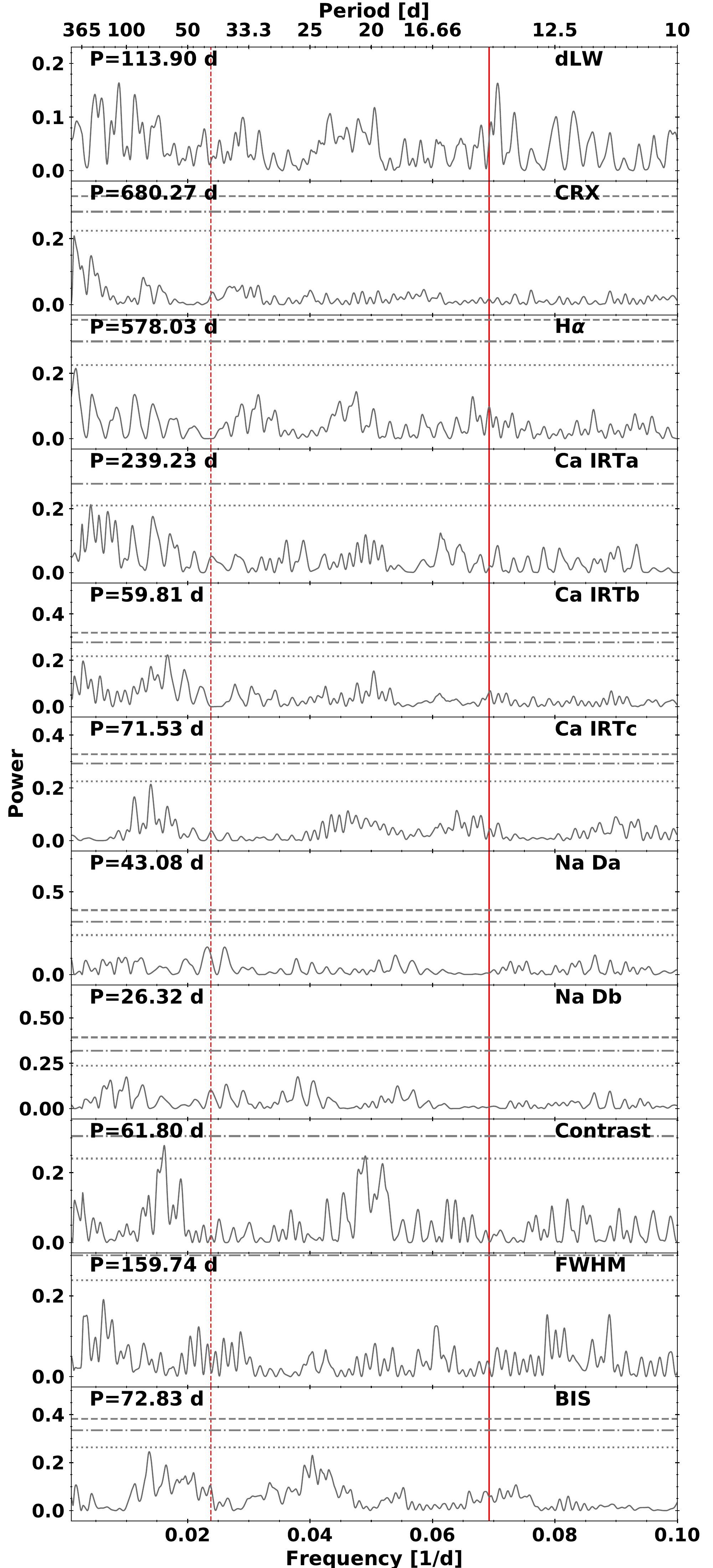}
\caption{\label{LSPMGLSresidualsactivity} Periodograms of the residuals after subtracting the highest signal of the activity indicators of LSPM~J2116+0234. The vertical solid line indicates the period of the suggested planet, while the vertical red dotted line denotes the period attributed to the rotation period. The periods reported in each panel refer to the highest peak. The horizontal lines represent the bootstrapped 10, 1, and 0.1$\%$ FAP levels.}
\end{center}
\end{figure}

\begin{figure}[!h]
\begin{center}
\includegraphics[width=0.95\columnwidth,clip]{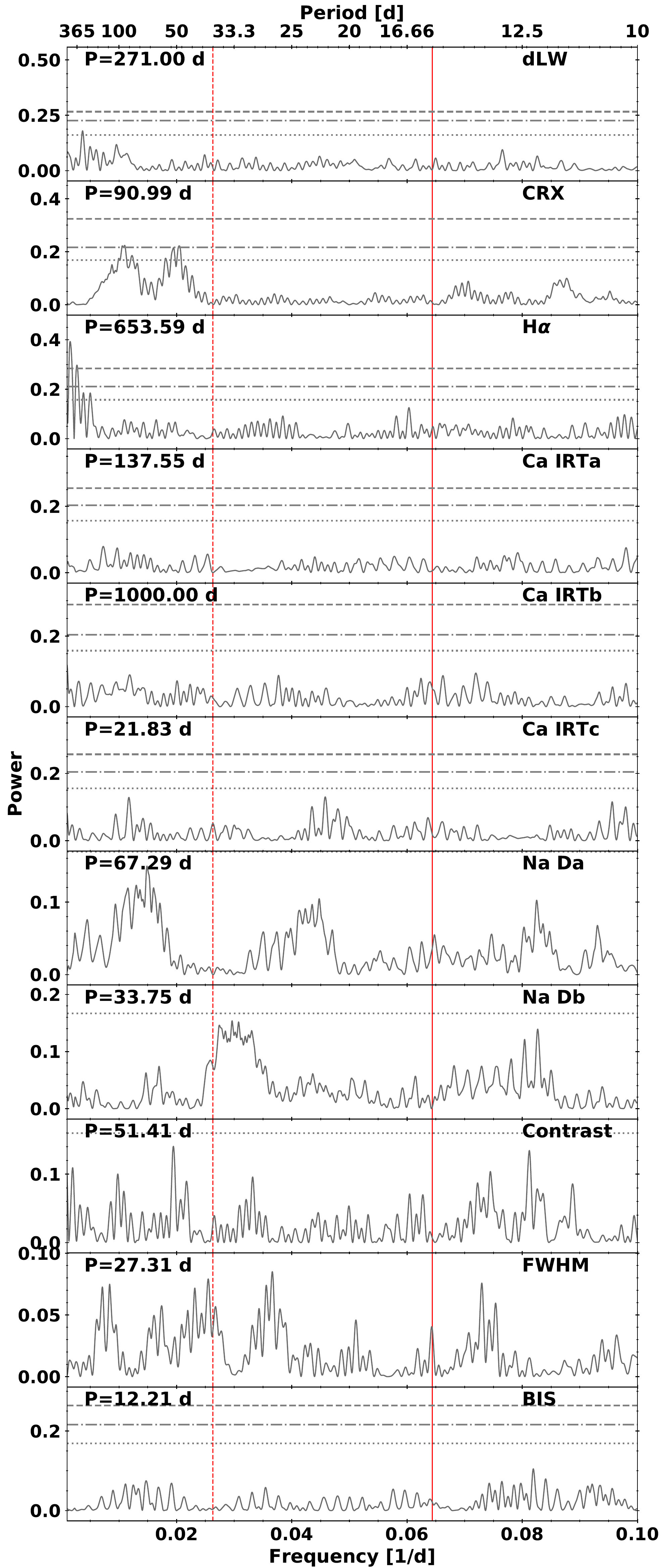}
\caption{\label{GJ686GLSresidualsactivity} Periodograms of the residuals after subtracting the highest signal of the activity indicators of GJ~686. The vertical solid line indicates the period of the suggested planet, while the vertical red dotted line denotes the period attributed to the rotation period. The periods reported in each panel refer to the highest peak. The horizontal lines represent the bootstrapped 10, 1, and 0.1$\%$ FAP levels.}
\end{center}
\end{figure}

\begin{table}[!h]
\centering
\caption{Radial velocities of LSPM~J2116+0234.}
\label{tab:lspmrvs}
\begin{tabular}{lccc} 
\hline\hline
BJD & RV [m\,s$^{-1}$] & $\sigma_{RV}$ [m\,s$^{-1}$] & Instrument\\
\hline
2457569.6342 & 1.49 & 9.89 & CARMENES-NIR \\ 
2457569.6343 & -5.32 & 2.53 & CARMENES-VIS \\ 
2457584.6234 & -4.09 & 8.92 & CARMENES-NIR \\ 
2457584.6238 & -5.36 & 1.80 & CARMENES-VIS \\ 
2457593.5641 & 4.47 & 7.31 & CARMENES-NIR \\ 
2457593.5656 & -1.92 & 1.90 & CARMENES-VIS \\ 
2457608.5397 & -3.37 & 8.65 & CARMENES-NIR \\ 
2457608.5403 & -2.58 & 1.80 & CARMENES-VIS \\ 
2457617.4723 & 9.24 & 2.01 & CARMENES-VIS \\ 
2457617.4751 & 9.64 & 4.73 & CARMENES-NIR \\ 
2457642.4391 & -6.18 & 8.04 & CARMENES-NIR \\ 
2457642.4415 & -0.08 & 1.48 & CARMENES-VIS \\ 
2457647.4671 & 5.68 & 10.92 & CARMENES-NIR \\ 
2457647.4678 & 9.74 & 2.01 & CARMENES-VIS \\ 
2457656.4869 & -2.32 & 9.86 & CARMENES-NIR \\ 
2457656.4893 & -1.62 & 1.57 & CARMENES-VIS \\ 
2457676.3731 & 9.23 & 3.85 & CARMENES-VIS \\ 
2457676.3754 & 15.26 & 10.77 & CARMENES-NIR \\ 
2457692.3954 & 5.46 & 4.37 & CARMENES-NIR \\ 
2457692.3963 & -0.30 & 1.26 & CARMENES-VIS \\ 
2457956.5190 & -7.31 & 8.75 & CARMENES-NIR \\ 
2457956.5192 & -8.58 & 1.44 & CARMENES-VIS \\ 
2457964.5469 & -1.43 & 7.84 & CARMENES-NIR \\ 
2457964.5477 & 6.55 & 1.30 & CARMENES-VIS \\ 
2457971.5622 & -5.36 & 7.71 & CARMENES-NIR \\ 
2457971.5638 & -4.49 & 1.38 & CARMENES-VIS \\ 
2458026.3456 & -3.77 & 5.41 & CARMENES-NIR \\ 
2458026.3482 & -7.17 & 1.39 & CARMENES-VIS \\ 
2458029.3942 & -12.90 & 6.48 & CARMENES-NIR \\ 
2458029.3943 & -6.23 & 1.60 & CARMENES-VIS \\ 
2458032.3668 & -2.57 & 5.95 & CARMENES-NIR \\ 
2458032.3674 & -2.40 & 1.88 & CARMENES-VIS \\ 
2458035.3032 & -6.91 & 5.47 & CARMENES-NIR \\ 
2458043.4445 & 10.37 & 7.09 & CARMENES-NIR \\ 
2458051.2813 & 8.24 & 7.25 & CARMENES-NIR \\ 
2458051.2827 & 5.78 & 1.41 & CARMENES-VIS \\ 
2458056.3113 & -5.72 & 6.35 & CARMENES-NIR \\ 
2458056.3121 & -3.04 & 1.43 & CARMENES-VIS \\ 
2458065.3831 & 12.29 & 10.34 & CARMENES-NIR \\ 
2458065.3847 & 3.71 & 1.30 & CARMENES-VIS \\ 
2458244.6746 & -4.20 & 1.74 & CARMENES-VIS \\ 
2458244.6757 & -7.23 & 6.49 & CARMENES-NIR \\ 
2458289.6045 & -6.63 & 7.88 & CARMENES-NIR \\ 
2458289.6052 & -8.30 & 1.63 & CARMENES-VIS \\ 
2458291.5331 & -16.15 & 7.91 & CARMENES-NIR \\ 
2458291.5338 & -2.93 & 2.27 & CARMENES-VIS \\ 
2458301.5964 & -1.27 & 1.66 & CARMENES-VIS \\ 
2458303.5421 & 3.35 & 11.35 & CARMENES-NIR \\ 
2458303.5436 & -0.03 & 3.05 & CARMENES-VIS \\ 
2458309.5670 & 20.14 & 7.79 & CARMENES-NIR \\ 
2458309.5685 & 9.95 & 1.95 & CARMENES-VIS \\ 
2458316.5615 & -0.76 & 1.39 & CARMENES-VIS \\ 
2458316.5621 & -1.85 & 5.81 & CARMENES-NIR \\ 
2458322.5278 & 7.17 & 8.98 & CARMENES-NIR \\ 
2458322.5278 & 1.92 & 1.31 & CARMENES-VIS \\ 
2458324.5333 & 3.43 & 5.60 & CARMENES-NIR \\ 
2458324.5343 & 7.05 & 1.56 & CARMENES-VIS \\ 
2458326.4999 & 5.89 & 5.05 & CARMENES-NIR \\ 
2458326.5011 & 3.17 & 1.31 & CARMENES-VIS \\ 
2458329.5600 & -2.83 & 1.21 & CARMENES-VIS \\ 
2458329.5608 & -9.63 & 5.39 & CARMENES-NIR \\ 
2458332.5170 & 0.53 & 5.45 & CARMENES-NIR \\ 
\hline
\end{tabular}
\end{table}

\addtocounter{table}{-1}

\begin{table}[!h]
\centering
\caption{Continued.}
\label{tab:lspmrvs}
\begin{tabular}{lccc} 
\hline\hline
BJD & RV [m\,s$^{-1}$] & $\sigma_{RV}$ [m\,s$^{-1}$] & Instrument\\
\hline
2458332.5189 & -7.82 & 1.29 & CARMENES-VIS \\ 
2458334.4985 & -2.93 & 8.16 & CARMENES-NIR \\ 
2458334.4985 & -5.42 & 1.62 & CARMENES-VIS \\ 
2458336.5145 & -3.46 & 5.47 & CARMENES-NIR \\ 
2458336.5146 & -0.84 & 1.63 & CARMENES-VIS \\ 
2458340.4775 & 7.21 & 6.31 & CARMENES-NIR \\ 
2458340.4791 & 7.33 & 1.32 & CARMENES-VIS \\ 
2458342.4909 & 5.47 & 6.08 & CARMENES-NIR \\ 
2458342.4920 & 3.07 & 1.51 & CARMENES-VIS \\ 
2458345.4754 & -1.01 & 4.88 & CARMENES-NIR \\ 
2458345.4765 & -0.81 & 1.31 & CARMENES-VIS \\ 
2458347.5284 & 3.35 & 12.40 & CARMENES-NIR \\ 
2458347.5294 & -3.24 & 3.00 & CARMENES-VIS \\ 
2458348.4407 & -2.73 & 1.33 & CARMENES-VIS \\ 
2458348.4408 & -6.28 & 4.79 & CARMENES-NIR \\ 
2458350.5600 & -3.67 & 4.88 & CARMENES-NIR \\ 
2458350.5604 & 0.84 & 1.47 & CARMENES-VIS \\ 
2458351.6279 & 7.89 & 6.06 & CARMENES-NIR \\ 
2458351.6293 & 2.61 & 1.78 & CARMENES-VIS \\ 
2458353.4635 & 10.04 & 1.59 & CARMENES-VIS \\ 
2458355.5021 & 5.32 & 6.21 & CARMENES-NIR \\ 
2458355.5027 & 6.84 & 1.46 & CARMENES-VIS \\ 
2458359.4942 & 3.35 & 7.14 & CARMENES-NIR \\ 
2458359.4965 & -3.41 & 1.81 & CARMENES-VIS \\ 
2458361.4588 & -7.97 & 10.86 & CARMENES-NIR \\ 
2458361.4594 & -8.24 & 1.43 & CARMENES-VIS \\ 
2458365.4355 & -1.78 & 9.03 & CARMENES-NIR \\ 
2458365.4361 & -3.40 & 1.99 & CARMENES-VIS \\ 
2458381.4792 & 0.26 & 6.54 & CARMENES-NIR \\ 
2458381.4797 & 2.87 & 1.48 & CARMENES-VIS \\ 
2458382.4046 & 1.31 & 5.73 & CARMENES-NIR \\ 
2458382.4051 & 6.72 & 1.26 & CARMENES-VIS \\ 
2458383.4052 & 3.60 & 5.31 & CARMENES-NIR \\ 
2458383.4060 & 7.21 & 1.33 & CARMENES-VIS \\ 
2458384.3978 & 1.07 & 5.38 & CARMENES-NIR \\ 
2458384.3988 & 4.91 & 1.27 & CARMENES-VIS \\ 
2458385.3888 & 2.94 & 5.55 & CARMENES-NIR \\ 
2458385.3897 & 4.00 & 1.34 & CARMENES-VIS \\ 
2458386.3893 & 3.72 & 1.34 & CARMENES-VIS \\ 
2458386.3894 & 4.50 & 7.01 & CARMENES-NIR \\ 
2458387.3877 & -0.50 & 10.09 & CARMENES-NIR \\ 
2458387.3895 & 0.91 & 1.44 & CARMENES-VIS \\ 
2458390.3754 & -3.00 & 7.69 & CARMENES-NIR \\ 
2458390.3790 & -3.17 & 1.34 & CARMENES-VIS \\ 
2458391.3786 & -3.01 & 1.17 & CARMENES-VIS \\ 
2458391.3791 & -9.94 & 5.87 & CARMENES-NIR \\ 
2458392.3461 & 0.88 & 5.74 & CARMENES-NIR \\ 
2458392.3467 & -2.30 & 1.22 & CARMENES-VIS \\ 
2458393.3536 & -0.32 & 1.43 & CARMENES-VIS \\ 
2458394.4332 & 3.13 & 1.74 & CARMENES-VIS \\ 
2458395.3578 & 8.04 & 2.03 & CARMENES-VIS \\ 
2458396.3535 & 8.58 & 1.35 & CARMENES-VIS \\ 
2458398.3438 & 10.13 & 1.37 & CARMENES-VIS \\ 
2458399.3576 & 3.30 & 1.42 & CARMENES-VIS \\ 
2458405.3408 & -7.80 & 2.43 & CARMENES-VIS \\ 
2458409.3285 & -3.68 & 2.35 & CARMENES-VIS \\ 
2458415.3673 & -4.39 & 1.73 & CARMENES-VIS \\ 
2458417.3118 & -4.83 & 1.29 & CARMENES-VIS \\ 
2458419.3963 & -7.27 & 3.00 & CARMENES-VIS \\ 
2458427.3437 & 5.83 & 1.79 & CARMENES-VIS \\ 
2458433.3611 & -1.35 & 2.71 & CARMENES-VIS \\ 
2458434.3013 & -3.51 & 1.56 & CARMENES-VIS \\ 
\hline
\end{tabular}
\end{table}

\begin{table}
\centering
\caption{Radial velocities of GJ~686.}
\label{tab:gj686rvs}
\begin{tabular}{lccc} 
\hline\hline
BJD & RV [m\,s$^{-1}$] & $\sigma_{RV}$ [m\,s$^{-1}$] & Instrument\\
\hline
2458451.3183 & -1.45 & 1.37 & CARMENES-VIS \\ 
2450604.9470 & -7.34 & 1.74 & HIRES \\ 
2450955.0818 & 3.77 & 1.91 & HIRES \\ 
2450956.9854 & -4.27 & 1.86 & HIRES \\ 
2450981.8611 & 1.94 & 2.01 & HIRES \\ 
2451050.8022 & -5.27 & 1.86 & HIRES \\ 
2451313.0310 & -2.36 & 1.69 & HIRES \\ 
2451367.8307 & -4.80 & 2.19 & HIRES \\ 
2451410.8169 & -3.08 & 2.00 & HIRES \\ 
2451703.9460 & -2.95 & 1.87 & HIRES \\ 
2452004.1131 & 1.90 & 1.95 & HIRES \\ 
2452097.9279 & 0.29 & 2.07 & HIRES \\ 
2452445.9041 & 9.18 & 2.01 & HIRES \\ 
2452446.8876 & 7.95 & 2.23 & HIRES \\ 
2452538.7482 & 1.52 & 1.87 & HIRES \\ 
2452777.9912 & -3.41 & 2.22 & HIRES \\ 
2452803.9463 & -0.52 & 2.10 & HIRES \\ 
2452849.8560 & 3.02 & 2.21 & HIRES \\ 
2453159.7492 & 4.40 & 0.71 & HARPS \\ 
2453180.8524 & 0.74 & 1.94 & HIRES \\ 
2453181.8583 & -4.01 & 1.92 & HIRES \\ 
2453478.9613 & 3.16 & 1.65 & HIRES \\ 
2453550.9287 & -1.70 & 1.88 & HIRES \\ 
2453574.6225 & 1.46 & 0.69 & HARPS \\ 
2453602.8651 & -4.32 & 2.01 & HIRES \\ 
2453817.8682 & -0.31 & 0.64 & HARPS \\ 
2453926.9486 & -3.31 & 1.80 & HIRES \\ 
2453984.8251 & -6.92 & 1.74 & HIRES \\ 
2454174.8713 & -1.10 & 0.65 & HARPS \\ 
2454194.9076 & 3.18 & 0.69 & HARPS \\ 
2454247.0275 & -1.66 & 1.54 & HIRES \\ 
2454248.0407 & 2.18 & 1.83 & HIRES \\ 
2454249.9630 & -4.35 & 1.70 & HIRES \\ 
2454251.9845 & -4.11 & 1.75 & HIRES \\ 
2454255.8363 & -0.62 & 1.73 & HIRES \\ 
2454255.8422 & -2.59 & 1.63 & HIRES \\ 
2454277.7856 & 4.17 & 1.73 & HIRES \\ 
2454278.7980 & 0.15 & 1.84 & HIRES \\ 
2454279.7981 & -6.26 & 1.88 & HIRES \\ 
2454285.8093 & -3.68 & 2.12 & HIRES \\ 
2454294.9811 & -2.07 & 1.75 & HIRES \\ 
2454300.6448 & -2.03 & 0.63 & HARPS \\ 
2454304.9433 & 2.02 & 1.92 & HIRES \\ 
2454305.9445 & 2.31 & 1.70 & HIRES \\ 
2454306.9197 & 0.89 & 1.78 & HIRES \\ 
2454307.9699 & 1.71 & 1.65 & HIRES \\ 
2454308.9402 & 1.93 & 1.88 & HIRES \\ 
2454309.9321 & 0.11 & 1.93 & HIRES \\ 
2454310.9276 & -1.81 & 1.62 & HIRES \\ 
2454311.9158 & 0.45 & 1.86 & HIRES \\ 
2454312.9226 & 0.42 & 1.78 & HIRES \\ 
2454313.9200 & -1.80 & 1.70 & HIRES \\ 
2454314.9628 & -2.34 & 1.81 & HIRES \\ 
2454335.8435 & -1.28 & 1.74 & HIRES \\ 
2454335.8636 & -0.57 & 1.74 & HIRES \\ 
2454343.8038 & -9.32 & 1.73 & HIRES \\ 
2454396.6967 & -2.12 & 1.79 & HIRES \\ 
2454397.6978 & -2.22 & 1.81 & HIRES \\ 
2454633.9259 & 3.73 & 2.01 & HIRES \\ 
2454666.8964 & 7.83 & 1.93 & HIRES \\ 
2454671.8867 & -2.59 & 1.88 & HIRES \\ 
2454671.8942 & -5.63 & 1.65 & HIRES \\ 
2454673.9248 & -4.95 & 1.68 & HIRES \\ 
2454686.9891 & 4.89 & 2.12 & HIRES \\ 
2454701.9167 & 2.04 & 2.00 & HIRES \\ 
2454702.8773 & -3.40 & 2.02 & HIRES \\ 
2454704.8849 & -4.37 & 1.80 & HIRES \\ 
\hline
\end{tabular}
\end{table}

\addtocounter{table}{-1}

\begin{table}
\centering
\caption{Continued.}
\label{tab:gj686rvs}
\begin{tabular}{lccc} 
\hline\hline
BJD & RV [m\,s$^{-1}$] & $\sigma_{RV}$ [m\,s$^{-1}$] & Instrument\\
\hline
2454704.8924 & -2.94 & 1.84 & HIRES \\ 
 2454948.8615 & -0.73 & 0.51 & HARPS \\ 
2454950.8621 & -2.02 & 0.55 & HARPS \\ 
2454956.8307 & -0.08 & 0.77 & HARPS \\ 
2454984.8828 & -2.04 & 1.91 & HIRES \\ 
2455024.0372 & -1.81 & 2.02 & HIRES \\ 
2455024.0446 & -0.22 & 1.76 & HIRES \\ 
2455025.0074 & -2.69 & 1.76 & HIRES \\ 
2455025.0151 & -0.67 & 1.64 & HIRES \\ 
2455042.8976 & 3.48 & 2.04 & HIRES \\ 
2455052.9485 & 8.36 & 1.75 & HIRES \\ 
2455052.9555 & 2.80 & 2.05 & HIRES \\ 
2455053.8513 & 4.51 & 2.05 & HIRES \\ 
2455053.8578 & 3.44 & 1.78 & HIRES \\ 
2455259.0848 & -0.19 & 1.90 & HIRES \\ 
2455259.0972 & 0.62 & 1.78 & HIRES \\ 
2455260.0874 & -2.35 & 1.72 & HIRES \\ 
2455372.0622 & -5.74 & 1.90 & HIRES \\ 
2455390.6409 & 1.69 & 0.79 & HARPS \\ 
2455392.6327 & 2.70 & 0.55 & HARPS \\ 
2455407.5859 & 1.53 & 0.63 & HARPS \\ 
2455408.9670 & 2.30 & 1.94 & HIRES \\ 
2455408.9744 & 5.26 & 1.94 & HIRES \\ 
2455409.5880 & 2.88 & 0.77 & HARPS \\ 
2455409.9415 & 1.22 & 1.71 & HIRES \\ 
2455409.9490 & 0.68 & 1.92 & HIRES \\ 
2455412.5785 & 0.98 & 0.84 & HARPS \\ 
2455437.5293 & -2.33 & 0.59 & HARPS \\ 
2455438.5335 & 0.78 & 0.58 & HARPS \\ 
2455446.5219 & -5.71 & 0.72 & HARPS \\ 
2455450.5081 & -2.88 & 0.80 & HARPS \\ 
2455458.5053 & 1.32 & 0.75 & HARPS \\ 
2455462.8181 & -6.02 & 1.53 & HIRES \\ 
2455637.1135 & -0.47 & 1.82 & HIRES \\ 
2455638.0598 & -0.75 & 1.69 & HIRES \\ 
2455638.0673 & 0.70 & 1.51 & HIRES \\ 
2455639.1080 & 3.29 & 1.50 & HIRES \\ 
2455639.1154 & 5.85 & 1.67 & HIRES \\ 
2455664.9911 & -2.91 & 1.78 & HIRES \\ 
2455664.9986 & -3.70 & 1.56 & HIRES \\ 
2455670.1260 & -1.07 & 1.66 & HIRES \\ 
2455670.1335 & -1.06 & 1.70 & HIRES \\ 
2455720.0090 & -2.53 & 1.99 & HIRES \\ 
2455720.0165 & -2.21 & 1.91 & HIRES \\ 
2455749.9152 & 8.61 & 2.00 & HIRES \\ 
2455749.9226 & 9.61 & 2.08 & HIRES \\ 
2455824.7659 & -0.21 & 1.84 & HIRES \\ 
2455824.7733 & 1.19 & 1.88 & HIRES \\ 
2455879.7285 & 2.79 & 1.93 & HIRES \\ 
2455971.1533 & 6.37 & 2.12 & HIRES \\ 
2455971.1605 & 6.69 & 2.11 & HIRES \\ 
2456027.0881 & 6.72 & 2.04 & HIRES \\ 
2456116.9645 & 0.51 & 2.07 & HIRES \\ 
2456116.9720 & -3.90 & 2.12 & HIRES \\ 
2456141.9782 & 2.64 & 1.90 & HIRES \\ 
2456141.9857 & 8.15 & 1.85 & HIRES \\ 
2456168.7913 & -0.89 & 1.88 & HIRES \\ 
2456329.1408 & 6.23 & 1.94 & HIRES \\ 
2456329.1483 & 6.14 & 1.81 & HIRES \\ 
2456433.0728 & 2.67 & 1.71 & HIRES \\ 
2456433.0802 & 3.50 & 1.83 & HIRES \\ 
2456548.7875 & -4.75 & 1.55 & HIRES \\ 
2456548.7949 & -6.39 & 1.63 & HIRES \\ 
2456551.7804 & -5.93 & 1.71 & HIRES \\ 
2456551.7876 & -7.40 & 1.56 & HIRES \\ 
2456700.7495 & 3.68 & 0.98 & HARPS-N \\ 
\hline
\end{tabular}
\end{table}

\addtocounter{table}{-1}

\begin{table}
\centering
\caption{Continued.}
\label{tab:gj686rvs}
\begin{tabular}{lccc} 
\hline\hline
BJD & RV [m\,s$^{-1}$] & $\sigma_{RV}$ [m\,s$^{-1}$] & Instrument\\
\hline

2456702.7585 & 1.36 & 1.14 & HARPS-N \\ 
2457440.7355 & -4.59 & 1.57 & CARMENES \\ 
2457444.7459 & -0.13 & 1.60 & CARMENES \\ 
2457472.7215 & -2.31 & 1.30 & CARMENES \\ 
2457490.6715 & 1.60 & 1.75 & CARMENES \\ 
2457493.6837 & -1.21 & 1.62 & CARMENES \\ 
2457504.6632 & -1.42 & 1.40 & CARMENES \\ 
2457508.5930 & -0.27 & 0.72 & HARPS-N \\ 
2457510.6084 & -0.33 & 0.84 & HARPS-N \\ 
2457536.6013 & 1.12 & 0.76 & HARPS-N \\ 
2457537.5723 & 1.16 & 0.58 & HARPS-N \\ 
2457537.6285 & -0.80 & 2.38 & CARMENES \\ 
2457538.5682 & -0.37 & 0.63 & HARPS-N \\ 
2457542.6223 & -7.16 & 1.67 & CARMENES \\ 
2457553.5607 & -0.58 & 1.96 & CARMENES \\ 
2457606.4629 & -6.78 & 0.81 & HARPS-N \\ 
2457608.5295 & -6.59 & 0.74 & HARPS-N \\ 
2457609.4813 & -5.52 & 0.88 & HARPS-N \\ 
2457610.4682 & -7.50 & 1.28 & HARPS-N \\ 
2457828.6905 & 1.12 & 1.44 & CARMENES \\ 
2457829.6992 & 2.37 & 1.42 & CARMENES \\ 
2457830.7226 & 3.51 & 1.36 & CARMENES \\ 
2457857.6418 & -2.92 & 0.80 & HARPS-N \\ 
2457858.6792 & -6.34 & 1.43 & CARMENES \\ 
2457860.5900 & 0.63 & 1.00 & HARPS-N \\ 
2457877.6513 & 0.48 & 1.34 & CARMENES \\ 
2457879.6773 & 1.39 & 3.16 & CARMENES \\ 
2457881.6194 & -0.43 & 0.72 & HARPS-N \\ 
2457887.5839 & -5.97 & 1.56 & CARMENES \\ 
2457893.6267 & 2.74 & 0.48 & HARPS-N \\ 
2457905.5417 & -0.82 & 0.73 & HARPS-N \\ 
2457907.6007 & -1.06 & 1.51 & CARMENES \\ 
2457909.5146 & 1.17 & 1.83 & CARMENES \\ 
2457913.4278 & 0.80 & 0.66 & HARPS-N \\ 
2457914.5686 & 0.91 & 1.71 & CARMENES \\ 
2457915.4820 & -0.50 & 0.64 & HARPS-N \\ 
2457922.5677 & -0.21 & 1.54 & CARMENES \\ 
2457928.6255 & 4.19 & 1.36 & HARPS-N \\ 
2457929.5786 & 7.43 & 1.11 & HARPS-N \\ 
2457931.6433 & 0.82 & 1.64 & HARPS-N \\ 
2457933.5701 & -5.17 & 0.71 & HARPS-N \\ 
2457935.4824 & -5.49 & 1.54 & CARMENES \\ 
2457935.5527 & -3.85 & 0.63 & HARPS-N \\ 
2457936.5023 & -2.29 & 0.71 & HARPS-N \\ 
2457937.6160 & 0.00 & 0.68 & HARPS-N \\ 
2457943.5627 & 1.68 & 0.94 & HARPS-N \\ 
2457944.5028 & 0.50 & 0.57 & HARPS-N \\ 
2457950.5135 & -2.56 & 0.86 & HARPS-N \\ 
2457954.4868 & 5.37 & 0.51 & HARPS-N \\ 
2457956.4337 & 6.53 & 0.61 & HARPS-N \\ 
2457961.4873 & 1.02 & 0.70 & HARPS-N \\ 
2457971.4244 & 1.76 & 0.62 & HARPS-N \\ 
2457972.3954 & 3.58 & 0.74 & HARPS-N \\ 
2457973.3935 & 1.48 & 0.69 & HARPS-N \\ 
2457974.5443 & -0.21 & 0.94 & HARPS-N \\ 
2457975.4754 & -1.64 & 0.74 & HARPS-N \\ 
2457976.4298 & -3.34 & 0.67 & HARPS-N \\ 
2457978.4013 & -2.62 & 0.61 & HARPS-N \\ 
2457981.4653 & -2.18 & 0.55 & HARPS-N \\ 
2457983.5109 & -1.81 & 0.69 & HARPS-N \\ 
2457984.4986 & 0.30 & 0.59 & HARPS-N \\ 
2457989.4081 & 3.28 & 0.73 & HARPS-N \\ 
\hline
\end{tabular}
\end{table}

\addtocounter{table}{-1}

\begin{table}
\centering
\caption{Continued.}
\label{tab:gj686rvs}
\begin{tabular}{lccc} 
\hline\hline
BJD & RV [m\,s$^{-1}$] & $\sigma_{RV}$ [m\,s$^{-1}$] & Instrument\\
\hline
2457992.4104 & 2.82 & 0.57 & HARPS-N \\ 
2457993.4198 & -0.40 & 0.93 & HARPS-N \\ 
2457994.4330 & 1.69 & 0.62 & HARPS-N \\ 
2457995.3917 & -0.83 & 0.59 & HARPS-N \\ 
2457996.4062 & -4.83 & 1.07 & HARPS-N \\ 
2457997.3939 & -1.83 & 0.81 & HARPS-N \\ 
2457999.4129 & 1.75 & 0.69 & HARPS-N \\ 
2458000.3640 & 0.41 & 1.02 & HARPS-N \\ 
2458007.4380 & 1.68 & 0.55 & HARPS-N \\ 
2458010.4463 & -2.55 & 0.77 & HARPS-N \\ 
2458020.3577 & 1.86 & 1.63 & HARPS-N \\ 
2458022.3410 & -1.04 & 0.60 & HARPS-N \\ 
2458024.3696 & -1.89 & 0.53 & HARPS-N \\ 
2458025.3542 & -1.75 & 0.53 & HARPS-N \\ 
2458027.3482 & -3.79 & 0.94 & HARPS-N \\ 
2458031.4052 & 1.56 & 0.59 & HARPS-N \\ 
2458037.3640 & 1.50 & 0.68 & HARPS-N \\ 
2458044.3621 & 0.23 & 0.67 & HARPS-N \\ 
2458047.3108 & 2.17 & 0.63 & HARPS-N \\ 
2458158.7338 & 6.07 & 2.50 & CARMENES \\ 
2458171.7159 & 3.08 & 1.94 & CARMENES \\ 
2458188.7101 & 0.93 & 2.07 & CARMENES \\ 
2458199.7161 & -0.33 & 2.17 & CARMENES \\ 
2458200.7063 & 2.23 & 1.84 & CARMENES \\ 
2458212.6323 & -1.55 & 1.72 & CARMENES \\ 
2458237.6302 & -0.19 & 2.54 & CARMENES \\ 
2458238.5125 & 1.67 & 2.00 & CARMENES \\ 
2458245.6203 & 6.60 & 2.48 & CARMENES \\ 
2458261.6011 & -3.43 & 2.00 & CARMENES \\ 
2458262.5657 & -2.63 & 3.18 & CARMENES \\ 
2458269.5744 & 0.36 & 1.76 & CARMENES \\ 
2458292.5373 & -3.19 & 1.60 & CARMENES \\ 
2458292.5664 & -2.36 & 1.71 & CARMENES \\ 
2458293.4054 & 0.27 & 1.56 & CARMENES \\ 
2458294.4627 & 0.09 & 1.57 & CARMENES \\ 
2458295.4348 & -0.25 & 1.43 & CARMENES \\ 
2458306.5580 & -4.15 & 1.78 & CARMENES \\ 
2458307.3833 & -3.33 & 1.84 & CARMENES \\ 
2458313.4281 & 3.59 & 1.52 & CARMENES \\ 
2458315.4950 & 5.45 & 2.07 & CARMENES \\ 
2458317.4518 & 5.29 & 3.73 & CARMENES \\ 
2458319.4005 & -0.18 & 2.22 & CARMENES \\ 
2458320.4524 & 2.63 & 1.91 & CARMENES \\ 
2458321.3915 & 0.89 & 1.84 & CARMENES \\ 
2458322.4181 & 1.15 & 1.66 & CARMENES \\ 
2458323.4559 & 2.68 & 1.69 & CARMENES \\ 
2458324.4056 & 2.18 & 2.03 & CARMENES \\ 
2458326.4024 & 4.55 & 1.90 & CARMENES \\ 
2458328.4109 & 3.75 & 1.67 & CARMENES \\ 
2458329.3987 & 5.86 & 1.71 & CARMENES \\ 
2458330.4128 & 4.34 & 1.64 & CARMENES \\ 
2458331.3875 & 3.53 & 1.85 & CARMENES \\ 
2458332.3847 & 1.91 & 1.72 & CARMENES \\ 
2458334.4765 & 0.37 & 1.64 & CARMENES \\ 
2458335.5404 & -4.51 & 1.75 & CARMENES \\ 
2458336.3807 & -1.16 & 1.95 & CARMENES \\ 
2458337.3720 & -2.32 & 1.74 & CARMENES \\ 
2458339.4705 & -2.44 & 1.74 & CARMENES \\ 
2458340.3701 & -3.20 & 1.44 & CARMENES \\ 
2458342.3492 & 1.72 & 1.51 & CARMENES \\ 
2458343.3563 & -0.42 & 1.45 & CARMENES \\ 
2458344.3802 & 6.28 & 1.91 & CARMENES \\ 
\hline
\end{tabular}
\end{table}
\addtocounter{table}{-1}

\begin{table}
\centering
\caption{Continued.}
\label{tab:gj686rvs}
\begin{tabular}{lccc} 
\hline\hline
BJD & RV [m\,s$^{-1}$] & $\sigma_{RV}$ [m\,s$^{-1}$] & Instrument\\
\hline
2458345.3552 & 1.80 & 1.57 & CARMENES \\ 
2458346.3949 & 0.62 & 1.33 & CARMENES \\ 
2458347.5135 & 2.21 & 3.11 & CARMENES \\ 
2458348.3681 & -1.45 & 1.21 & CARMENES \\ 
2458349.3622 & 0.17 & 1.50 & CARMENES \\ 
2458350.4837 & -2.96 & 1.36 & CARMENES \\ 
2458351.3562 & -3.64 & 1.40 & CARMENES \\ 
2458352.3370 & -0.24 & 1.44 & CARMENES \\ 
2458353.3404 & -4.09 & 1.62 & CARMENES \\ 
2458355.3459 & -1.71 & 1.58 & CARMENES \\ 
2458356.3502 & 0.37 & 1.54 & CARMENES \\ 
2458357.3786 & -0.76 & 1.40 & CARMENES \\ 
2458360.3393 & 2.58 & 1.87 & CARMENES \\ 
2458361.3358 & 3.45 & 1.77 & CARMENES \\ 
2458362.3361 & 3.03 & 1.73 & CARMENES \\ 
2458366.3262 & -2.29 & 1.70 & CARMENES \\ 
2458367.3357 & -2.06 & 2.50 & CARMENES \\ 
2458382.3043 & -6.62 & 1.80 & CARMENES \\ 
2458385.3046 & -7.11 & 1.54 & CARMENES \\ 
2458386.3668 & -4.15 & 1.63 & CARMENES \\ 
2458387.3683 & -4.88 & 1.58 & CARMENES \\ 
2458390.2989 & 0.17 & 1.55 & CARMENES \\ 
2458391.2936 & 0.08 & 1.52 & CARMENES \\ 
2458392.3587 & -0.78 & 1.51 & CARMENES \\ 
2458393.3048 & -0.76 & 1.56 & CARMENES \\ 
2458395.3148 & -0.62 & 2.21 & CARMENES \\ 
2458396.2826 & -1.26 & 1.63 & CARMENES \\ 
2458397.2773 & -1.53 & 1.90 & CARMENES \\ 
2458409.2996 & 3.74 & 2.26 & CARMENES \\ 
2458414.2706 & -2.59 & 2.38 & CARMENES \\ 
2458415.3223 & -3.86 & 1.90 & CARMENES \\ 
2458427.2976 & -1.57 & 1.97 & CARMENES \\ 
 
\hline
\end{tabular}
\end{table}

\addtocounter{table}{-1}
\begin{figure*}[!hb]
\begin{center}
\includegraphics[width=\textwidth,clip]{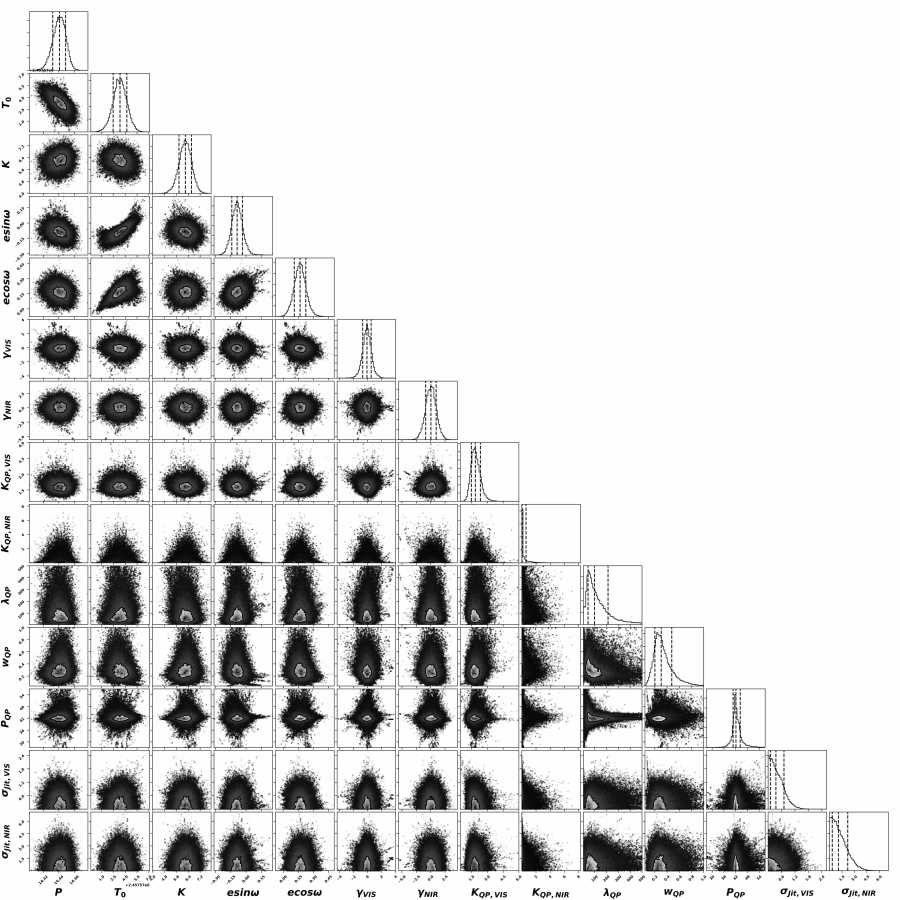}
\caption{\label{cornerlspm}  Posterior distributions from the MCMC analysis on LSPM~J2116+0234~b. Plotted are the planetary parameters, instrumental offsets ($\gamma_{\rm VIS}$, $\gamma_{\rm NIR}$), GP hyper-parameters ($K_{\rm QP}$, $\lambda_{\rm QP}$, $w_{\rm QP}$, $P_{\rm QP}$), and additional data jitters ($\sigma_{\rm Jit, VIS}$, $\sigma_{\rm Jit, NIR}$). The vertical dashed lines indicate the mean and 1$\sigma$ uncertainties of the fitted parameters.}
\end{center}
\end{figure*}

\begin{figure*}[!hb]
\begin{center}
\includegraphics[width=\textwidth,clip]{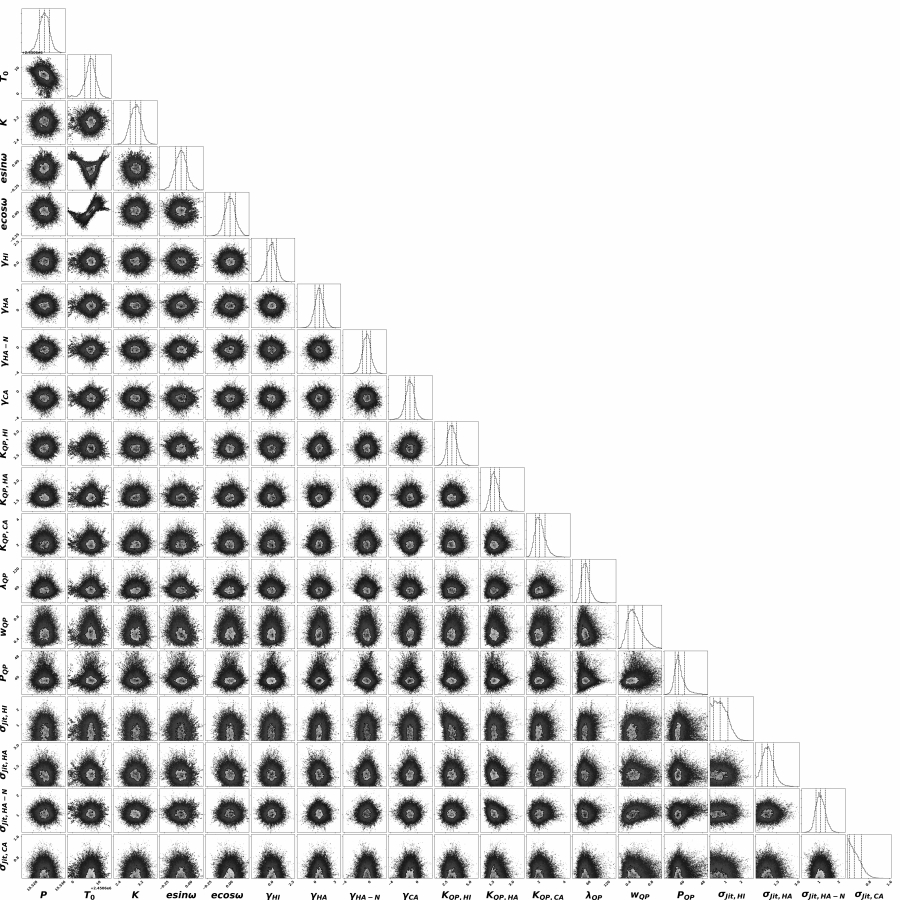}
\caption{\label{cornergj} Posterior distributions from the MCMC analysis on GJ~686~b. Plotted are the planetary parameters, instrumental offsets ($\gamma_{\rm HIRES}$, $\gamma_{\rm HARPS}$, $\gamma_{\rm HARPS-N}$, $\gamma_{\rm CARMENES}$), GP hyper-parameters ($K_{\rm QP}$, $\lambda_{\rm QP}$, $w_{\rm QP}$, $P_{\rm QP}$), and additional data jitters ($\sigma_{\rm Jit, HIRES}$, $\sigma_{\rm Jit, HARPS}$, $\sigma_{\rm Jit, HARPS-N}$, $\sigma_{\rm Jit, CARMENES}$). The vertical dashed lines indicate the mean and 1$\sigma$ uncertainties of the fitted parameters.}
\end{center}
\end{figure*}

\begin{figure*}[!hb]
\begin{center}
\includegraphics[width=\textwidth,clip]{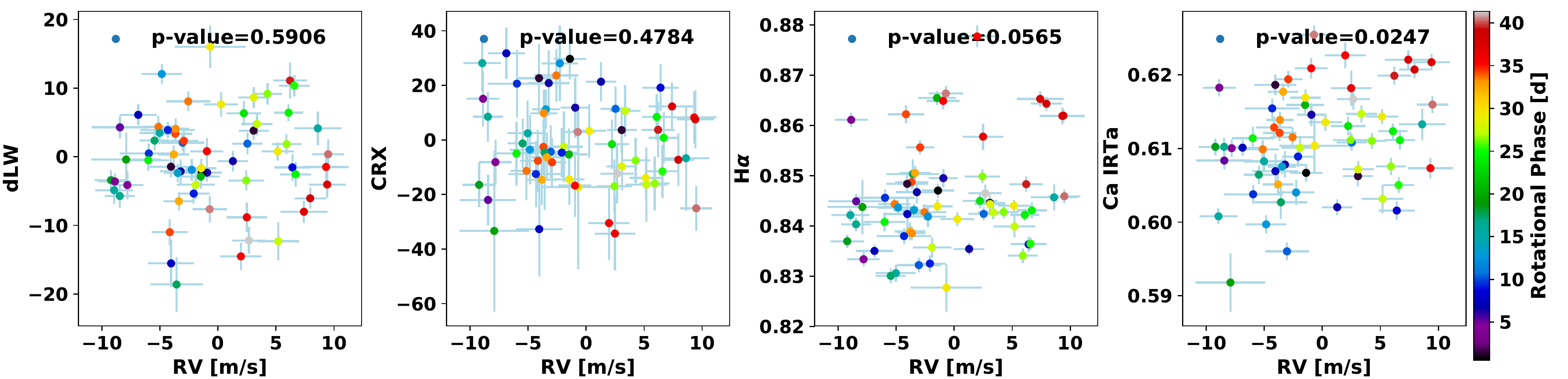}
\includegraphics[width=\textwidth,clip]{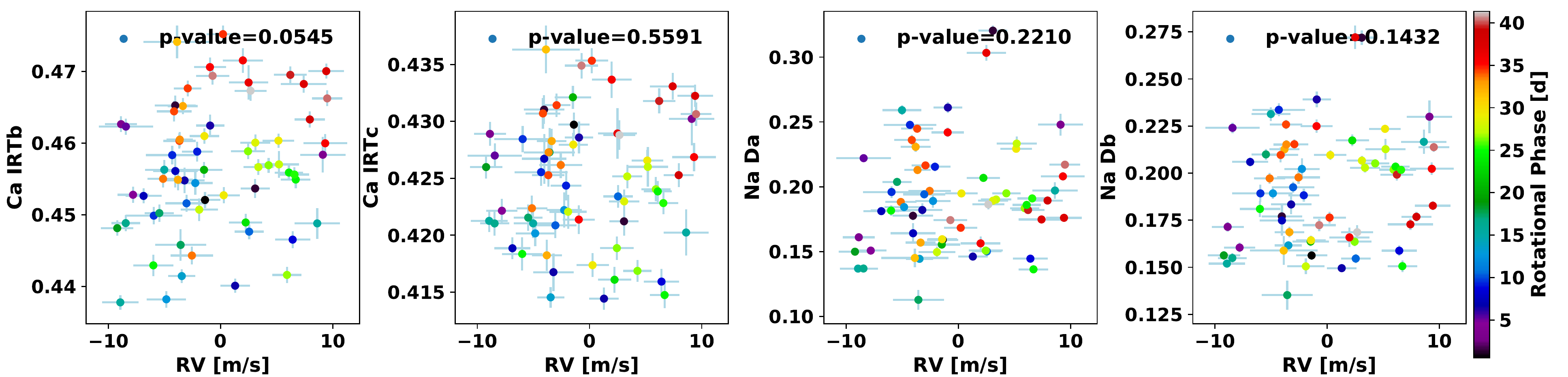}
\caption{\label{correlation_activity_lspm}  Correlation plots between the activity indices and radial velocities of LSPM~J2116+0234. Color code represents the phase with the estimated rotation period of 42 d. The p-value of a linear fit is shown.}
\end{center}
\end{figure*}

\begin{figure*}[!hb]
\begin{center}
\includegraphics[width=\textwidth,clip]{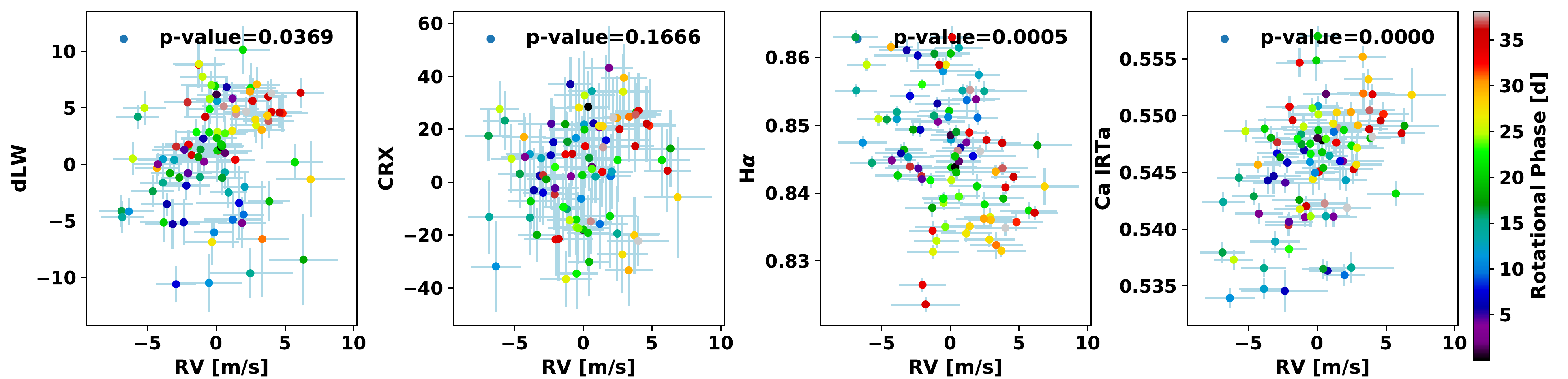}
\includegraphics[width=\textwidth,clip]{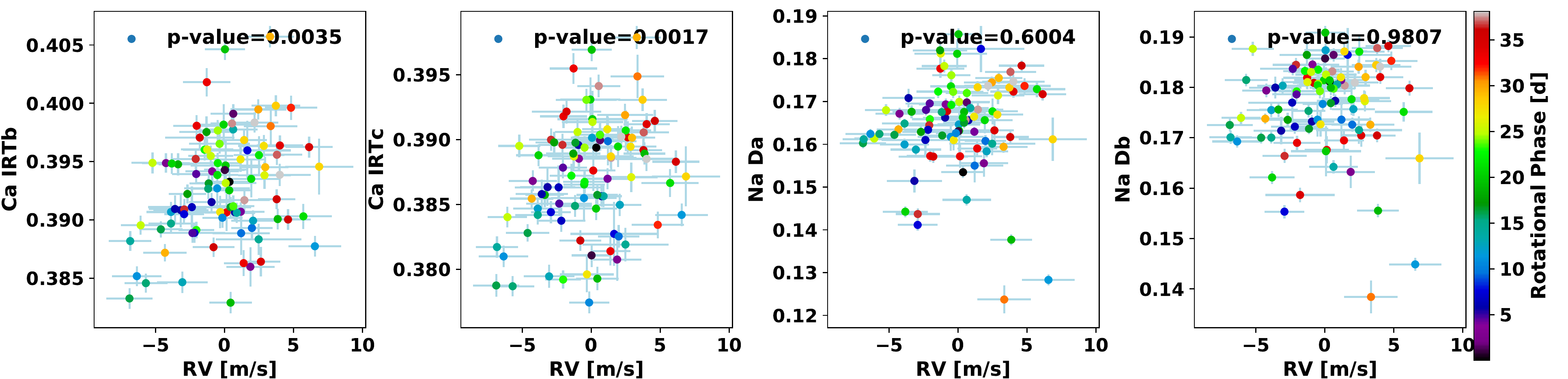}
\caption{\label{correlation_activity_gj}  Top panels: Correlation plots between the activity indices and radial velocities of GJ~686. Color code represents the phase with the estimated rotation period of 38.4 d. The p-value of a linear fit is shown. }
\end{center}
\end{figure*}

\end{appendix}

\end{document}